\documentclass[12pt,ascmac]{report}
\usepackage{eclepsf}
\usepackage{tascmac}
\usepackage{url}
\usepackage{array,tabularx}
\usepackage{amsmath}
\usepackage{mathtools}
\usepackage{caption}
\usepackage{physics}
\usepackage{subfig}
\usepackage[htt]{hyphenat}
\usepackage{here}
\usepackage{braket}
\usepackage{cite}
\usepackage{graphicx}
\usepackage{booktabs}
\usepackage{algorithm}
\usepackage{algpseudocode}
\usepackage{blkarray}
\usepackage[T1]{fontenc}
\usepackage{tabularx}
\usepackage{tablefootnote}
\usepackage{times}

\begin{document}
	\pagenumbering{roman}

\begin{titlepage}
\begin{center}
	\begin{large}
		Master's Thesis (Academic Year 2019)\\
		\vspace{32pt}
		\fontsize{24pt}{32pt}\selectfont
		Simulation of a Dynamic,\\ RuleSet-based Quantum Network\\
					~\\
		\fontsize{18pt}{32pt}\selectfont
					\end{large}
\end{center}
\vspace{30em}
\begin{flushleft}
	\LARGE Keio University \\
	\large Graduate School of Media and Governance\\
\end{flushleft}
\begin{flushright}
	\LARGE Takaaki Matsuo
\end{flushright}
\end{titlepage}


\begin{center}
\large Master's Thesis (Academic Year 2019)
\end{center}

\begin{center}
\begin{large}
	\vspace{32pt}
	\fontsize{24pt}{32pt}\selectfont
	 Simulation of a Dynamic,\\ RuleSet-based Quantum Network\\
	\end{large}
\end{center}

~ \bigskip ~ \\
~ \bigskip ~ \\

Similar to the classical Internet, the quantum Internet will require knowledge regarding link qualities used for purposes such as optimal route selection.
This is commonly accomplished by performing link-level tomography with or without purification -- a.k.a. quantum link bootstrapping.
Meanwhile, the gate selection and the resource (Bell pair) selection for a task must be coordinated beforehand.
This thesis introduces the RuleSet-based communication protocol aimed for supporting the autonomous coordination of quantum operations among distant nodes, with minimal classical packet transmission.
This thesis also discusses the RuleSet-based quantum link bootstrapping protocol, which consists of recurrent purifications and link-level tomography,
evaluated over a Markov-Chain Monte-Carlo simulation with noisy systems modeled on real world quality hardware.
Given a 10km MeetInTheMiddle based two-node system, each with 100 memory qubits ideally connected to the optical fiber,
the Recurrent Single selection - Single error purification (RSs-Sp) protocol is capable of improving the fidelity from an average input $F_{r}=0.675$ to approximately $F_{r}=0.865$.
The system gets noisier with longer channels, in which case errors may develop faster than the purification gain.
For a noisier system with a longer channel length, the double selection-based purification shows an advantage for improving the fidelity.

~ \bigskip ~ \\
~ \bigskip ~ \\

Keywords : \\
\underline{1. Quantum purification},
\underline{2. RuleSet},
\underline{3. Quantum link bootstrapping},
\\ \underline{4. Quantum repeater networks},
\underline{5. Coordinated quantum operations}

~ \bigskip ~ \\
\begin{flushleft}
\large Graduate School of Media and Governance\\
\large Keio University
\end{flushleft}
\begin{flushright}
\LARGE Takaaki Matsuo
\end{flushright}

	\clearpage

	\tableofcontents
	\clearpage
	\listoffigures
	\clearpage
	\listoftables
	\clearpage

	\pagenumbering{arabic}

\cleardoublepage
\chapter{Introduction}
\label{introduction}

This chapter focuses on the research background, novelty/contribution of this thesis and the structure of the remaining chapters\footnote{Portions of this thesis has been presented in \cite{2019arXiv190408605M}}.

\section{Background}

The quantum Internet, interconnecting distinct quantum repeater networks~\cite{Briegel1998},
 is a promising technology that can provide us new capabilities not reproducible by classical systems~\cite{Kimble2008,VanMeter:2014:QN:2683776,Wehnereaam9288,irtf-qirg-principles-00}.
One of the best known examples is quantum key distribution~\cite{BENNETT20147, PhysRevLett.68.557, PhysRevLett.67.661, TGW},
 which is designed to utilize quantum mechanics to securely share strings of random bits suitable for use as encryption keys.
Conducting a statistical experiment, such as Bell's inequality~\cite{PhysicsPhysiqueFizika.1.195}, allows participants to detect the presence of an eavesdropper.
Other example apllications include quantum blind computing~\cite{Broadbent2010, 2013arXiv1306.3664C} and quantum clock synchronization~\cite{PhysRevLett.85.2010, PhysRevLett.85.2006}.

Similar to the classical Internet, establishing a multi-hop quantum connection requires quantified link characteristics for purposes such as optimal route selection.
Such characterization is commonly achieved by performing tomography over generated Bell pairs -- a.k.a. quantum link bootstrapping.
In any situation, distant nodes need to agree the selected operation and the targeted resource for completing the task.
The simplest solution is to exchange classical messages at each attempt for guaranteeing synchronization but a more practical solution with minimal message transmission has not yet been discussed carefully in the field.

\section{Research Contribution}
This thesis introduces RuleSets, which can be used to autonomously coordinate quantum operations among multiple nodes.
This thesis also adapts the above protocol to quantum link bootstrapping, and simulates the process based on a carefully designed Markov-Chain Monte-Carlo simulation modeled on currently available hardware.
The network simulator dynamically executes the Physical Layer protocol, the Data Link Layer protocol and the Networking Layer protocol to realistically assess the capability of the available technology set.
The simulation results successfully showed the importance of the selected purification method within the recurrence purification -- the selected purification method may work the best for a particular case, but not for others.
This thesis also introduces a technique to switch the purification method in the middle of the recurrence purification, which can be beneficial in terms of fidelity and throughput.
The bootstrapping process, therefore, also needs to check which purification works the best for the particular link.

\section{Thesis structure}

The remaining of the thesis is constructed as follows.
In Chapter 2, basics of quantum information are provided as a preliminary to support the readers with minimal knowledge.
In Chapter 3, current status of experimental work is introduced.
In Chapter 4, several related publications are briefly explained.
In Chapter 5, the protocol design of the RuleSet-based communication protocol is explained.
In Chapter 6, the some details regarding the simulation is provided.
In Chapter 7, main quantum link bootstrapping results of Markov-Chain Monte-Carlo simulations are discussed with various settings over multiple protocols.
Finally in Chapter 7, this thesis is concluded with some discussions regarding the future work.

\chapter{Theory of Quantum Information\protect\footnote{Adapted from my bachelor's thesis, "Analysis of Measurement-based Quantum Network Coding over Repeater Networks under Noisy Conditions",
Faculty of Environment and Information Studies, Keio University, 2017.}}
\label{Preliminaries}

\section{Historical background of Quantum Computation}


In 1982, a physicist named Richard Feynman noted \cite{Feynman1982} that it is generally not feasible to represent the results of quantum mechanics with a classical universal device.
The newly introduced concept, quantum simulation,
took the advantage of the puzzling quantum effects to effectively simulate physics,
which cannot be handled by ordinary computers regardless of their computation power.

The term "Quantum Computer" was officially used in print for the first time in 1985,
by physicist David Deutsch \cite{Deutsch1985}.
He proposed a mathematical concept of a strictly modeled universal computer based on quantum mechanics with many properties not reproducible by classical Turing machines,
and generalized computing methods for quantum computers.

Contemporaneously, quantum networking appeared as a subfield of quantum computing.
The algorithm named BB84, Quantum key distribution (QKD), was firstly proposed by Bennett and Brassard \cite{Bennet1984} in 1984, and came into the experimental forefront in 1989 \cite{Bennet1989}.
The algorithm was widely recognized across the globe, as a result of its promising security by exploiting quantum mechanics compared to the classical technology.

In 1994, Peter W. Shor \cite{Shor1994} at Bell Laboratories introduced a quantum algorithm that has the capability of factoring large numbers within polynomial time -- known as Shor's algorithm.
The algorithm essentially showed that quantum computers have the ability to break commonly used classical cryptography techniques based on prime number factorization such as the RSA cryptosystem.

The theoretical proposal of quantum teleportation, which is a technique to map an arbitrary state of qubit to another, was introduced by Bennett et al. \cite{Bennet1993} in 1993.
Some years later, it was successfully demonstrated experimentally and became one of the essential ingredients for quantum networking \cite{Bouwmeester1997,Furusawa1998}.
In order to achieve long distance quantum communication, intermediate nodes called "quantum repeaters" were introduced by Briegel and D\"ur \cite{Briegel1998,Briegel1999} in the late 1990s, as a tool for managing errors, creating entanglement and enabling multi-hop communications.
Nevertheless, establishing a stable quantum communication over long distances still remains an outstanding challenge due to technical problems such as the operation errors and qubit degradation known as decoherence.


\section{Qubit}
The indivisible unit of a classical information is known as {\it binary digit} or {\it bit}. A single bit has a single binary value that is not limited to, but in general expressed with {\it 0} and {\it 1}.
The two values of a bit in a classical computer may be represented by the electric charge stored in a capacitor, the direction of a magnetic field, or anything else that is capable of physically representing two values.
Similarly, the smallest unit of a quantum information is known as {\it quantum bit} or {\it qubit} \footnote{The term qubit just had its 25th anniversary (2017).}.
Unlike a bit, a single qubit can be in a {\it superposition state} of two states, simultaneously representing {\it 0} and {\it 1}.
The two-level system of a quantum computer may be represented by the vertical polarization  and the horizontal polarization of a photon, the spin up and the spin down of an electron, or any other proposed state variables.

\subsection{Dirac notation}
The simplest quantum system is a two-state system, and the single qubit pure state can be expressed by using the {\it Dirac notation}:

\begin{eqnarray}
	\label{dirac_qubit_one}
\ket{\psi} =  \alpha \ket{0} + \beta \ket{1} \nonumber \\
\mid \alpha \mid^2 + \mid \beta \mid^2  =  1
\end{eqnarray}

Coefficients $\alpha$ and $\beta$ are arbitrary complex numbers representing the probability amplitudes.
The $\ket{0}$ and $\ket{1}$ are called {\it kets} and denote the two possible states, where the probability of the state being $\ket{0}$ can be found by the quantity $\alpha . \alpha^* = \mid \alpha \mid^{2}$, and $\beta . \beta^* = \mid \beta \mid^{2}$ for $\ket{1}$. As an example, $ \ket{\psi} =  \frac{1}{\sqrt{2}} \ket{0} + \frac{1}{\sqrt{2}} \ket{1} $ indicates that the state $\ket{\psi}$ is in a superposition of two states $\ket{0}$ and $\ket{1}$ with equally weighted probabilities.

A quantum state can also be described by a vector in a two dimensional complex Hilbert space.
State vectors of a single qubit may be:

\begin{align} \nonumber
\label{Z-basis state}
\intertext{Z-basis state}
	\ket{0} &\equiv \begin{bmatrix}
	1\\
	0
	\end{bmatrix} &   \ket{1} &\equiv \begin{bmatrix}
		0\\
		1
		\end{bmatrix}
\end{align}

Therefore, the superposition state of a single qubit can be expressed by the following two-dimensional vector.

\begin{equation}
\ket{\psi}  =  \alpha \ket{0} + \beta \ket{1}= \alpha \begin{bmatrix}1 \\0 \end{bmatrix} + \beta \begin{bmatrix}0 \\1 \end{bmatrix}= \begin{bmatrix} \alpha \\ \beta  \end{bmatrix}
\end{equation}

The computational vectors introduced in equation \ref{Z-basis state} are called the Z-basis states, and a linear combination of them can be used to express any pure quantum state of one qubit.
The alternative computational bases, the {\it X-basis states} and the {\it Y-basis states} are equally important. The X-basis states are based on the superposition states:

\begin{align}
\label{X-basis state}
\intertext{X-basis state}
\ket{+} &\equiv \frac{1}{\sqrt{2}} \begin{bmatrix}
	1\\
	1
\end{bmatrix} & \ket{-} &\equiv \frac{1}{\sqrt{2}}\begin{bmatrix}
	1\\
	-1
	\end{bmatrix}.
\end{align}

Similarly, the Y-basis states uses the complex bases:

\begin{align}
\label{Y-basis state}
\intertext{Y-basis state}
\ket{+i} &\equiv \frac{1}{\sqrt{2}} \begin{bmatrix}
	1\\
	i
\end{bmatrix} & \ket{-i} &\equiv \frac{1}{\sqrt{2}}\begin{bmatrix}
	1\\
	-i
	\end{bmatrix}.
\end{align}

Contrasting to {\it ket}, $\bra{\psi}$ is called {\it bra} which represents the conjugate transpose of $\ket{\psi}$.

\begin{align}
\ket{\alpha} &= \begin{bmatrix}
	\alpha_{1}\\
	\alpha_{2}\\
	\alpha_{3}\\
	\vdots\\
	\alpha_{n}\\
\end{bmatrix} & \bra{\alpha} &= \begin{bmatrix}
	\alpha^{*}_{1} & \alpha^{*}_{2} & \alpha^{*}_{3} & \cdots & \alpha^{*}_{n}\\
	\end{bmatrix}
\end{align}

The inner product and the outer product of states are:

\begin{align}
\ket{\alpha}\bra{\alpha} &= \begin{bmatrix}
	\alpha_{1}\\
	\alpha_{2}\\
	\alpha_{3}\\
	\vdots\\
	\alpha_{n}\\
\end{bmatrix}.
\begin{bmatrix}
	\alpha^{*}_{1} & \alpha^{*}_{2} & \alpha^{*}_{3} & \cdots & \alpha^{*}_{n}\\
\end{bmatrix}=\begin{bmatrix}
\alpha_{1}\alpha^{*}_{1} & \alpha_{1}\alpha^{*}_{2} & \alpha_{1}\alpha^{*}_{3} & \cdots & \alpha_{1}\alpha^{*}_{n}  \\
\alpha_{2}\alpha^{*}_{1} & \alpha_{2}\alpha^{*}_{2} & \alpha_{2}\alpha^{*}_{3} & \cdots & \alpha_{2}\alpha^{*}_{n}  \\
\alpha_{3}\alpha^{*}_{1} & \alpha_{3}\alpha^{*}_{2} & \alpha_{3}\alpha^{*}_{3} & \cdots & \alpha_{3}\alpha^{*}_{n}  \\
\vdots & \vdots & \vdots & \vdots & \vdots \\
\alpha_{n}\alpha^{*}_{1} & \alpha_{n}\alpha^{*}_{2} & \alpha_{n}\alpha^{*}_{3} & \cdots & \alpha_{n}\alpha^{*}_{n}  \\
\end{bmatrix}
\end{align}

\begin{align}
\braket{\alpha\mid\alpha} &= \begin{bmatrix}
	\alpha^{*}_{1} & \alpha^{*}_{2} & \alpha^{*}_{3} & \cdots & \alpha^{*}_{n}\\
\end{bmatrix}.
\begin{bmatrix}
	\alpha_{1}\\
	\alpha_{2}\\
	\alpha_{3}\\
	\vdots\\
	\alpha_{n}\\
\end{bmatrix}=\alpha^{*}_{1}\alpha_{1}  + \alpha^{*}_{2}\alpha_{2}  + \alpha^{*}_{3}\alpha_{3}  + \cdots + \alpha^{*}_{n}\alpha_{n}
\end{align}

Note that the diagonal elements of $\ket{\alpha}\bra{\alpha} $ are real and non-negative.

\subsection{Bloch sphere}
\label{blochsphere}
The Bloch sphere is a geometric representation of a single qubit pure state as a unit vector pointing on the surface of a unit sphere.

\begin{figure}[!hbt]
\center
\includegraphics[keepaspectratio,scale=0.7]{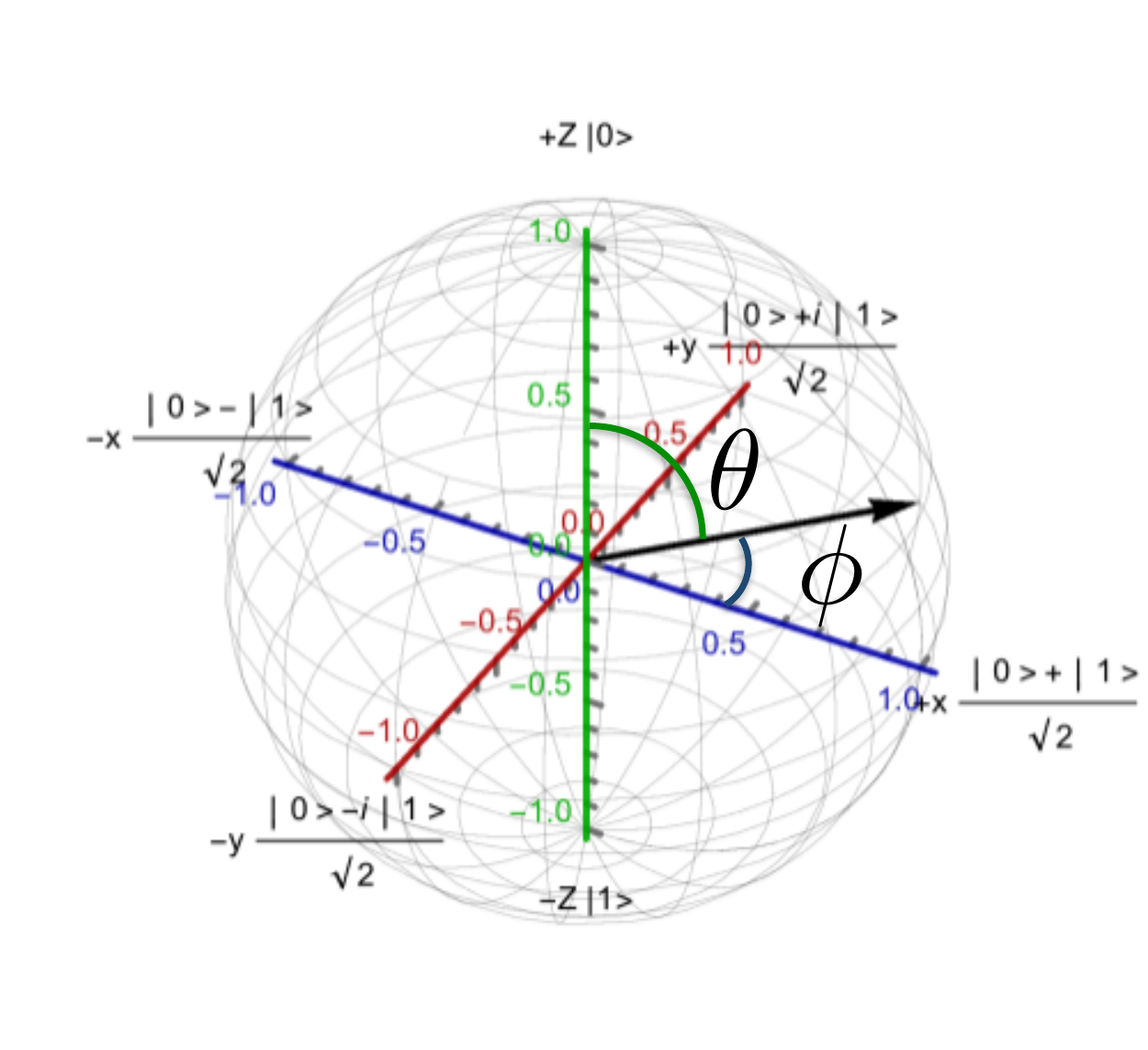}
\caption{Quantum State described on a Bloch Sphere}
\end{figure}

An arbitrary single qubit state can be written:

\begin{eqnarray}
\label{eq00}
\ket{\psi}  = e^{i\lambda} ( \cos{\frac{\theta}{2}} \ket{0} + e^{i\phi} \sin{\frac{\theta}{2}} \ket{1}) \\
0 \leq \theta \leq \pi \nonumber \\
0 \leq \phi \leq 2\pi \nonumber
\end{eqnarray}

where variables $\theta$, $\phi$ and $\lambda$ are real numbers.
The number $\theta$ represents the latitude with respect to the Z-axis and $\phi$ represents the longitude with respect to the Y-axis. Together they define a point on the Bloch sphere surface.
The variable $e^{i\lambda}$ is known as the global phase of a quantum state, and has no observable effects. Therefore, equation \ref{eq00} can simply be rewritten as:

\begin{eqnarray}
\ket{\psi} = \cos{\frac{\theta}{2}} \ket{0}+ e^{i\phi} \sin{\frac{\theta}{2}} \ket{1}
\end{eqnarray}

Hence, $ \cos{\frac{\theta}{2}} $ and $ e^{i\phi}\sin{\frac{\theta}{2}}$ correspond to $\alpha$ and $\beta$ in equation \ref{dirac_qubit_one} with $\alpha$ constrained to be real.

\subsection{Eigenvalue and Eigenvector}

The {\it eigenvalue} and the {\it eigenvector} are important concepts for quantum information.

As an example, if an operator U acts on a vector $\ket{\psi}$ and the result can be rearranged to a scalar $\lambda$ and the same vector $\ket{\psi}$:

\begin{equation}
U\ket{\psi} = \lambda \ket{\psi}
\end{equation}

Such $\lambda$ is called the {\it eigenvalue} and the corresponding vector $\ket{\psi}$ is called the {\it eigenvector}. In general, the eigenvalue is a complex number but here we often see it as a real number, $\pm 1$.
For example, performing a Pauli-Z gate on a state $\ket{1}$ will result in:

\begin{equation}
Z\ket{1} = -\ket{1}
\end{equation}

where in this case, the eigenvector of Z is $\ket{1}$ with an eigenvalue -1.
We will see eigenvalues and eigenvectors when we discuss measurement (see Sec.\ref{MEASUREMENTsection}).

\section{Composite quantum systems}
\label{compositesection}
In reality, a system may contain more than one qubit.
An independent two-qubit system may have states:

\begin{eqnarray}
\label{eq01}
	\ket{\psi}= \alpha \ket{0}+ \beta \ket{1}\\
\label{eq02}
	\ket{\phi}= \gamma \ket{0}+ \delta \ket{1}
\end{eqnarray}

Here, the joint system can be described by taking the tensor product of equation \ref{eq01} and equation \ref{eq02}.

\begin{equation}
	\ket{\psi}\otimes \ket{\phi}= \alpha \gamma \ket{00} + \alpha \delta \mid 01 \rangle + \beta \gamma \mid 10 \rangle + \beta \delta \ket{11} = \begin{bmatrix}
		\alpha \gamma\\
		\alpha \delta\\
		\beta \gamma\\
		\beta \delta
		\end{bmatrix}
\end{equation}
\begin{equation}
	\mid \alpha \gamma \mid^2 + \mid \alpha \delta \mid^2 + \mid \beta \gamma \mid^2 + \mid \beta \delta \mid^2 = 1 \nonumber
\end{equation}

where $\otimes$ is the tensor product of two vectors, and $\ket{00}$ equates to $\ket{0}\ket{0}$.
The tensor product of two vectors is:
\begin{equation}
\begin{bmatrix}
		a_{1}\\
		b_{1}
\end{bmatrix} \otimes
\begin{bmatrix}
		a_{2}\\
		b_{2}
\end{bmatrix} =
\begin{bmatrix}
		a_{1}  \begin{bmatrix}
					a_{2}\\
					b_{2}
			\end{bmatrix} \\
		b_{1}   \begin{bmatrix}
					a_{2}\\
					b_{2}
			\end{bmatrix}
\end{bmatrix} =
\begin{bmatrix}
		a_{1}a_{2}\\
		a_{1}b_{2}\\
		b_{1}a_{2}\\
		b_{1}b_{2}
\end{bmatrix}
\end{equation}

The vector representation of the two-qubit computational basis states is:

\begin{equation}
\ket{00}= \begin{bmatrix} 1\\0\\0\\0 \end{bmatrix},
\ket{01} = \begin{bmatrix} 0\\1\\0\\0 \end{bmatrix},
\ket{10} = \begin{bmatrix} 0\\0\\1\\0 \end{bmatrix},
\ket{11} = \begin{bmatrix} 0\\0\\0\\1 \end{bmatrix}
\end{equation}

Similarly, three qubits can be in eight states, and {\it n} qubits can be in a superposition of all $2^n$ states simultaneously.

\begin{equation}
\sum^{2^n-1}_{i=0}\alpha_{i}\ket{i}
\end{equation}
\begin{equation}
\sum^{2^n-1}_{i=0}\mid \alpha_{i} \mid^2 = 1 \nonumber
\end{equation}

\section{Quantum Gates}
As modern computers work based on Boolean logic gates, quantum computers performs similar gate operations to manipulate quantum information (see Figure \ref{fig01} for a simplified model of gate operations).
Such gates are often called unitary gates, as they give unitary transformation of the qubit states. Gates are unitary when $U^\dagger U=UU^\dagger=I$.
Quantum gates are reversible and can be represented as unitary matrices.

\begin{figure}[!hbt]
\center
\includegraphics[keepaspectratio,scale=0.7]{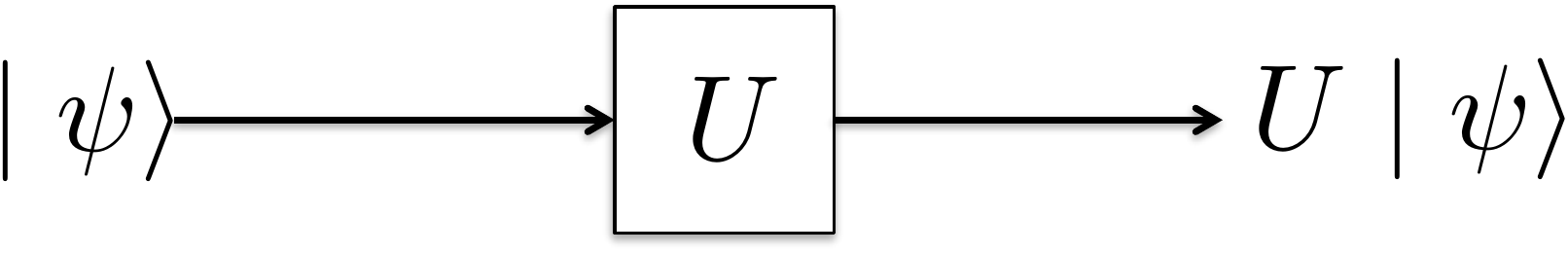}
\caption{Basic flow of quantum gate operation}
\label{fig01}
\end{figure}

\subsection{Single Qubit Gates}
\label{single_qubit_gates}

The most important operators for quantum computing are called the Pauli operators.

The Pauli-X gate is the equivalent of the classical NOT gate. The gate can be performed on a single qubit state and swaps the probability amplitude of $\ket{0}$ and $\ket{1}$.

\begin{equation}
	X = \begin{bmatrix} 0 & 1 \\ 1 & 0\end{bmatrix}
\end{equation}

\begin{figure}[!hbt]
\center
\includegraphics[keepaspectratio,scale=0.6]{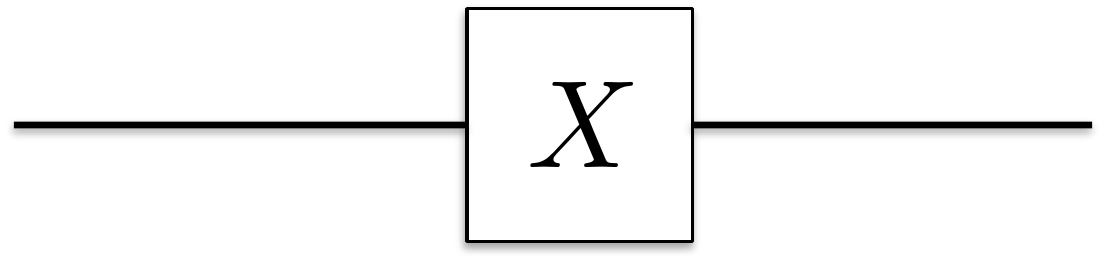}
\caption{Circuit representation of a Pauli-X gate}
\label{Xgate}
\end{figure}

Therefore, $\ket{0} = X\ket{1}$ and $\ket{1} = X\ket{0}$.

The Pauli-Y gate changes the quantum state $\ket{0}$ to $i\ket{1}$, and $\ket{1}$ to $-i\ket{0}$.

\begin{equation}
	Y = \begin{bmatrix} 0 & -i \\ i & 0\end{bmatrix}
\end{equation}

\begin{figure}[!hbt]
\center
\includegraphics[keepaspectratio,scale=0.6]{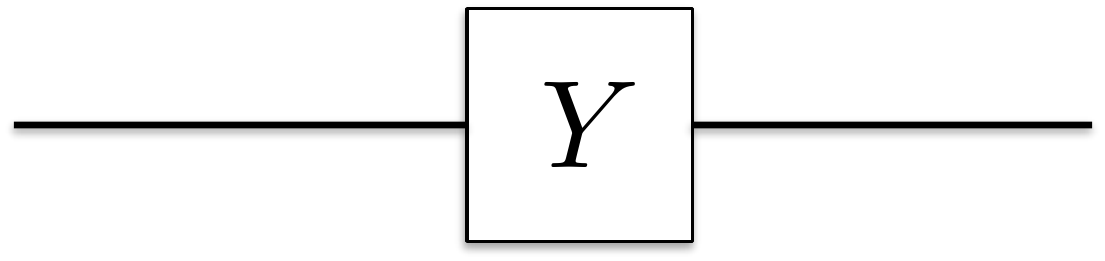}
\caption{Circuit representation of a Pauli-Y gate}
\label{Ygate}
\end{figure}

The Pauli-Z gate, or sometimes called the phase-flip gate, does not affect the basis state $\ket{0}$, but changes $\ket{1}$ to $-\ket{1}$, and $-\ket{1}$ to $\ket{1}$.

\begin{equation}
	Z = \begin{bmatrix} 1 & 0 \\ 0 & -1\end{bmatrix}
\end{equation}

\begin{figure}[!hbt]
\center
\includegraphics[keepaspectratio,scale=0.6]{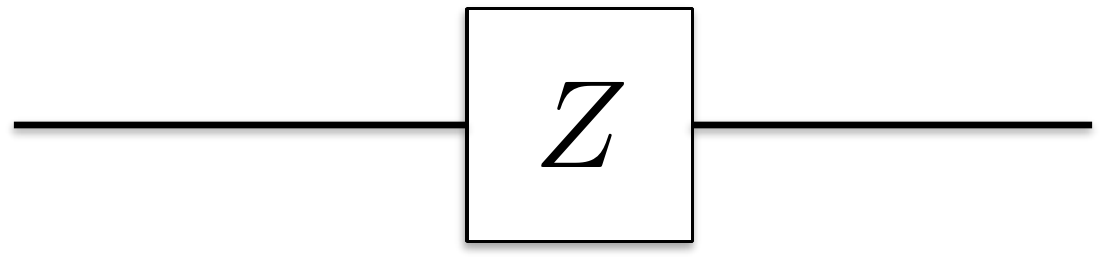}
\caption{Circuit representation of a Pauli-Z gate}
\label{Zgate}
\end{figure}

Besides Pauli operators, the Hadamard gate and the phase gate are equally important operators. These gates can be used to transform between the different basis states.

The Hadamard gate is as shown below.
\begin{equation}
	H = \frac{1}{\sqrt{2}}\begin{bmatrix} 1 & 1 \\ 1 & -1\end{bmatrix}
\end{equation}

\begin{figure}[!hbt]
\center
\includegraphics[keepaspectratio,scale=0.6]{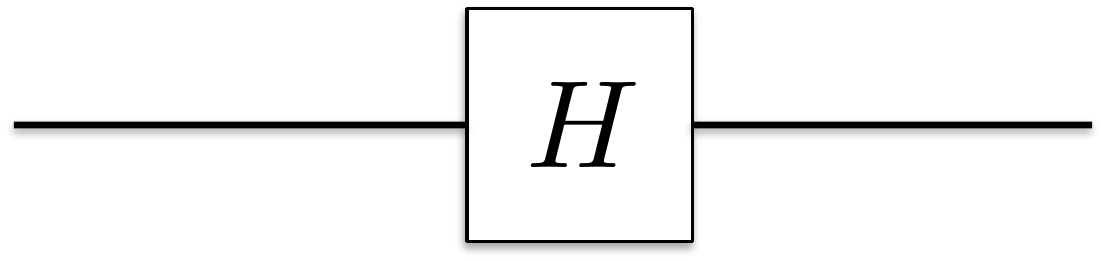}
\caption{Circuit representation of an Hadamard gate}
\label{Hgate}
\end{figure}

Applying an Hadamard gate to a qubit will result in:

\begin{equation}
H\ket{0}= \frac{1}{\sqrt{2}}\begin{bmatrix} 1 & 1 \\ 1 & -1\end{bmatrix}\begin{bmatrix} 1 \\ 0\end{bmatrix} = \frac{1}{\sqrt{2}} \begin{bmatrix} 1 \\ 1\end{bmatrix} = \ket{+}
\end{equation}
\begin{equation}
H\ket{1}= \frac{1}{\sqrt{2}}\begin{bmatrix} 1 & 1 \\ 1 & -1\end{bmatrix}\begin{bmatrix} 0 \\ 1\end{bmatrix} = \frac{1}{\sqrt{2}} \begin{bmatrix} 1 \\ -1\end{bmatrix} = \ket{-}
\end{equation}

\begin{equation}
H\ket{+}= \frac{1}{\sqrt{2}}\begin{bmatrix} 1 & 1 \\ 1 & -1\end{bmatrix}\begin{bmatrix} 1 \\ 1\end{bmatrix} = \frac{1}{\sqrt{2}} \begin{bmatrix} 1 \\ 0\end{bmatrix} = \ket{0}
\end{equation}

\begin{equation}
H\ket{-}= \frac{1}{\sqrt{2}}\begin{bmatrix} 1 & 1 \\ 1 & -1\end{bmatrix}\begin{bmatrix} 1 \\ -1\end{bmatrix} = \frac{1}{\sqrt{2}} \begin{bmatrix} 0 \\ 1\end{bmatrix} = \ket{1}
\end{equation}

Notice that an Hadamard gate can bring up a qubit into a superposition state from a basis state, or vice versa. Moreover, the X gate can be constructed by conjugating two Hadamard gates and a Z gate: $X = HZH$.

Similar to the Z gate, the phase shift gate changes the phase of a quantum state, from $\ket{1}$ to $e^{i\phi}\ket{1}$. Therefore, if $\phi = \pi$, the phase gate performs on a qubit the same way as the Z gate.
The T gate and the S gate are for specific, defined values of $\phi$.

\begin{eqnarray}
	Z_{\phi} = \begin{bmatrix} 1 & 0 \\ 0 & e^{i\phi}\end{bmatrix} \\
	S = Z_{\frac{\pi}{2}} = \begin{bmatrix} 1 & 0 \\ 0 & e^{i\frac{\pi}{2}}\end{bmatrix}\\
	T = Z_{\frac{\pi}{4}}=\begin{bmatrix} 1 & 0 \\ 0 & e^{i\frac{\pi}{4}}\end{bmatrix}
\end{eqnarray}

\begin{figure}[!hbt]
\center
\includegraphics[keepaspectratio,scale=0.6]{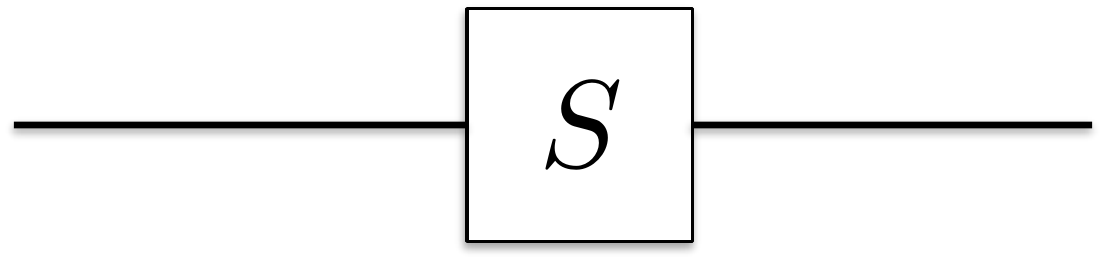}
\caption{Circuit representation of a S gate}
\label{Sgate}
\end{figure}

\begin{figure}[!hbt]
\center
\includegraphics[keepaspectratio,scale=0.6]{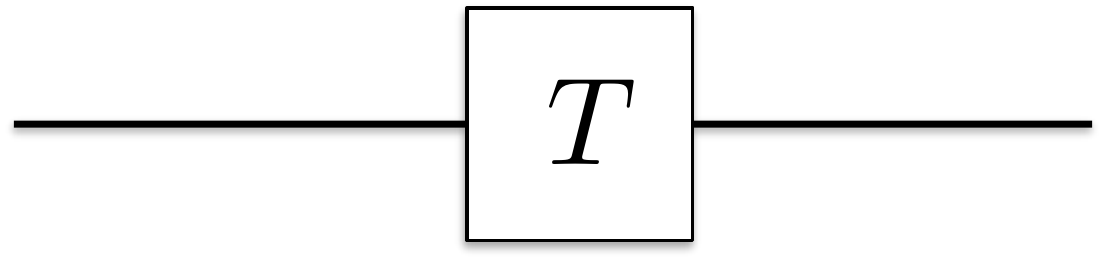}
\caption{Circuit representation of a T gate}
\label{Sgate}
\end{figure}

Two S gates and an X gate can be used to construct a Y gate: $Y = SXS^{\dagger}$.
A gate that rotates the quantum state by 180 degrees is not affected by the rotational direction; clockwise or anti-clockwise.
As an $S$ gate performs a rotation of less than 180 degrees, the direction of rotation matters.
The clockwise rotation is represented by the $S$ gate, and the anti-clockwise rotation is represented by the dagger of the gate $S^{\dagger}$ or sometimes $S^{-}$.

The generalized rotation operators, which rotate the Bloch vector about the X, Y and Z-axis by a given angle $\theta$, can be described as $R_{P}(\theta)=e^{-i\frac{\theta}{2}P}$, where P represents the axis.

\begin{eqnarray}
R_{X}(\theta) =e^{-i\frac{\theta}{2}X} = \cos\frac{\theta}{2}I - i\sin\frac{\theta}{2}X = \begin{bmatrix} \cos\frac{\theta}{2} & -i\sin\frac{\theta}{2} \\ -i\sin\frac{\theta}{2} & \cos\frac{\theta}{2} \end{bmatrix} \\
R_{Y}(\theta) =e^{-i\frac{\theta}{2}Y} = \cos\frac{\theta}{2}I - i\sin\frac{\theta}{2}Y = \begin{bmatrix} \cos\frac{\theta}{2} & -\sin\frac{\theta}{2} \\ \sin\frac{\theta}{2} & \cos\frac{\theta}{2} \end{bmatrix} \\
R_{Z}(\theta) =e^{-i\frac{\theta}{2}Z} = \cos\frac{\theta}{2}I - i\sin\frac{\theta}{2}Z = \begin{bmatrix} e^{-i\frac{\theta}{2}} & 0 \\ 0 & e^{i\frac{\theta}{2}} \end{bmatrix} \\
\end{eqnarray}

\subsection{Measurement in a circuit}
\label{MEASUREMENTsection}
\begin{figure}[!hbt]
\center
\includegraphics[keepaspectratio,scale=0.6]{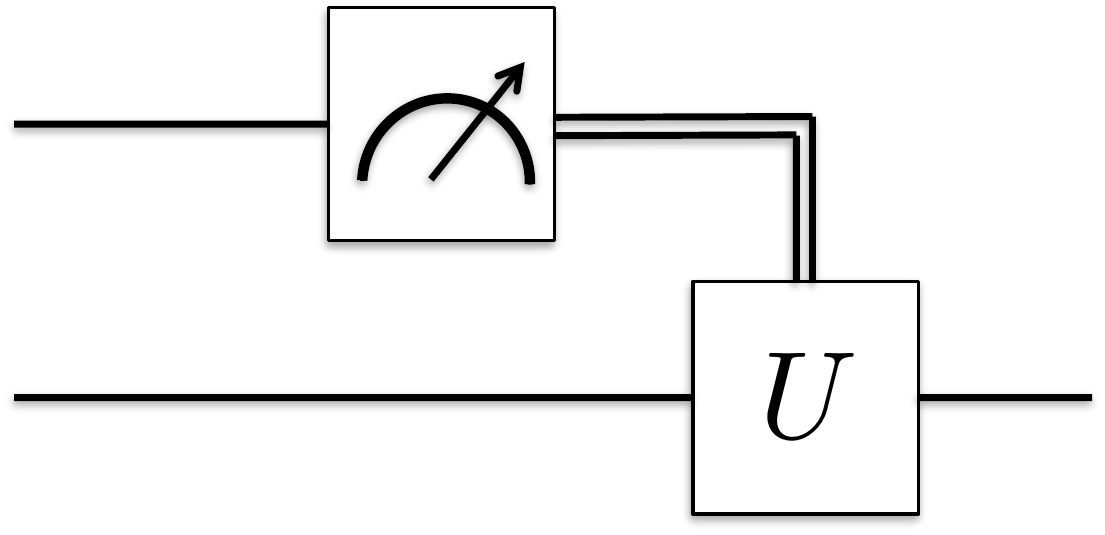}
\caption{Circuit representation of a Measurement gate and its byproduct operator}
\label{Mgate}
\end{figure}

The measurement result of an arbitrary qubit, if entangled with another, decides the residual quantum state.
That is to say, two post-measurement residue states may not equate depending on the measurement results, even with two identical pre-measurement quantum systems.
In some operations, measurement requires a classical feedforward operation to another qubit to fix the state to a wanted form.
Sometimes these operators are applied to more than one qubit.
Those operators are often called {\it byproduct operators} and are performed based on classically sent measurement results.
As shown in the circuit representation in Figure \ref{Mgate}, the classical message transmission is generally described with double lines interconnecting the measurement operator and the byproduct operator.

\subsubsection{Measuring a qubit}
Measuring a qubit will destroy the superposition state and determine the state in a classical fashion.
The measurement outcome of a quantum state is the eigenvalue of the eigenvector.

As an example, if the qubit state is:

\begin{equation}
\ket{\psi} =  \frac{1}{\sqrt{2}} \ket{0}+ \frac{1}{\sqrt{2}} \ket{1}= \begin{bmatrix} \frac{1}{\sqrt{2}}\\ \frac{1}{\sqrt{2}}  \end{bmatrix}
\end{equation}

the probability amplitude is equally weighted with respect to the Z-basis. Therefore, measuring the state in Z-basis will return the observer either $+Z$ or $-Z$ with a probability of $50\%$. Changing the measurement basis may affect the measurement result.
That is to say, if we use the X-basis for the measurement instead of Z, then the measurement result will always be $+X$.

If we measure the qubit $\ket{\psi}$ in the Z-basis, measurement outcomes 0 and 1 can be obtained with probabilities:

\begin{equation}
P(0) = \mid \bra{0} \ket{\psi} \mid^2 = Tr[\ket{0}\bra{0}\ket{\psi}\bra{\psi}]
\end{equation}

\begin{equation}
P(1) = \mid \bra{1} \ket{\psi} \mid^2 = Tr[\ket{1}\bra{1}\ket{\psi}\bra{\psi}]
\end{equation}

More generally, the above equation can be expressed using measurement operators $M_{i}$.
$P(i)$ is the probability of obtaining the measurement outcome $i$.

\begin{equation}
P(i) = \expval{M^{\dagger}_{i} M_{i}}{\psi} = Tr[M^{\dagger}_{i} M_{i}\ket{\psi}\bra{\psi}] = Tr[E_{i}\ket{\psi}\bra{\psi}]
\end{equation}

where $E_{i}$ is a set of positive operators such that $\Sigma_{i} E_{i} = I$.

The measurement operator can be obtained by:

\begin{equation}
M^{\pm}_{i} = \frac{1}{2}{(I \pm U)}
\end{equation}

where U in this case is an element of the matrix set $U =$ \{ $Z, Y, X$ \}.
As an example, the Z-basis measurement operators can be obtained by:

\begin{align}
M_{Z}^{+} = \frac{1}{2}(I+Z) = \frac{1}{2}(\begin{bmatrix}
1 & 0\\
0 & 1
\end{bmatrix} +  \begin{bmatrix}
1 & 0\\
0 & -1
\end{bmatrix}) = \begin{bmatrix}
	1 & 0\\
	0 & 0
	\end{bmatrix}
\end{align}

\begin{align}
M_{Z}^{-} = \frac{1}{2}(I-Z) = \frac{1}{2}(\begin{bmatrix}
1 & 0\\
0 & 1
\end{bmatrix} -  \begin{bmatrix}
1 & 0\\
0 & -1
\end{bmatrix}) = \begin{bmatrix}
	0 & 0\\
	0 & 1
	\end{bmatrix}
\end{align}

\subsection{Controlled Gates}
\begin{figure}[!hbt]
\center
\includegraphics[keepaspectratio,scale=0.5]{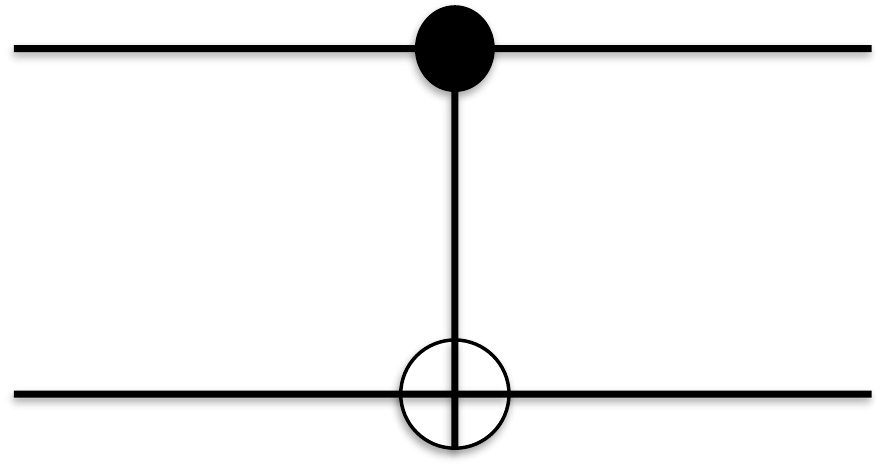}
\caption{Circuit representation of a Controlled-NOT gate}
\label{CNOTgate}
\end{figure}

Not all gates work on a single qubit; some act on two or more qubits.
One example of such a gate is the Controlled-NOT (CNOT) gate (see Fig.~\ref{CNOTgate}), which performs the X gate on one qubit ({\it the target qubit}), when the other qubit's ({\it the control qubit}) state is $\ket{1}$. The CNOT gate is defined by:

\begin{eqnarray}
\Lambda_{c,t}(X) \ket{i_{c}} \ket{j_{t}} = \ket{i_{c}} \ket {i \oplus j_{t}} \\
i,j = {0,1}
\end{eqnarray}

And the corresponding matrix is:

\begin{equation}
	CNOT = \begin{bmatrix} 1 & 0 & 0 & 0 \\ 0 & 1 & 0 & 0 \\ 0 & 0 & 0 & 1 \\ 0 & 0 & 1 & 0\end{bmatrix}
\end{equation}

Similarly, the Controlled-Z (CZ) gate performs the Z operation on the target qubit, when the controlled qubit's state is $\ket{1}$.
The CZ gate is defined by:

\begin{eqnarray}
\Lambda_{c,t}(Z) \ket{i_{c}} \ket{j_{t}} = (-1)^{ij}\ket{i_{c}} \ket{j_{t}} \\
i,j = {0,1}
\end{eqnarray}

The CZ gate is symmetric, $\Lambda_{c,t}(Z) =   \Lambda_{t,c}(Z)$ and the corresponding matrix is:

\begin{equation}
	CZ = \begin{bmatrix} 1 & 0 & 0 & 0 \\ 0 & 1 & 0 & 0 \\ 0 & 0 & 1 & 0 \\ 0 & 0 & 0 & -1\end{bmatrix}
\end{equation}

\begin{figure}[!hbt]
\center
\includegraphics[keepaspectratio,scale=0.5]{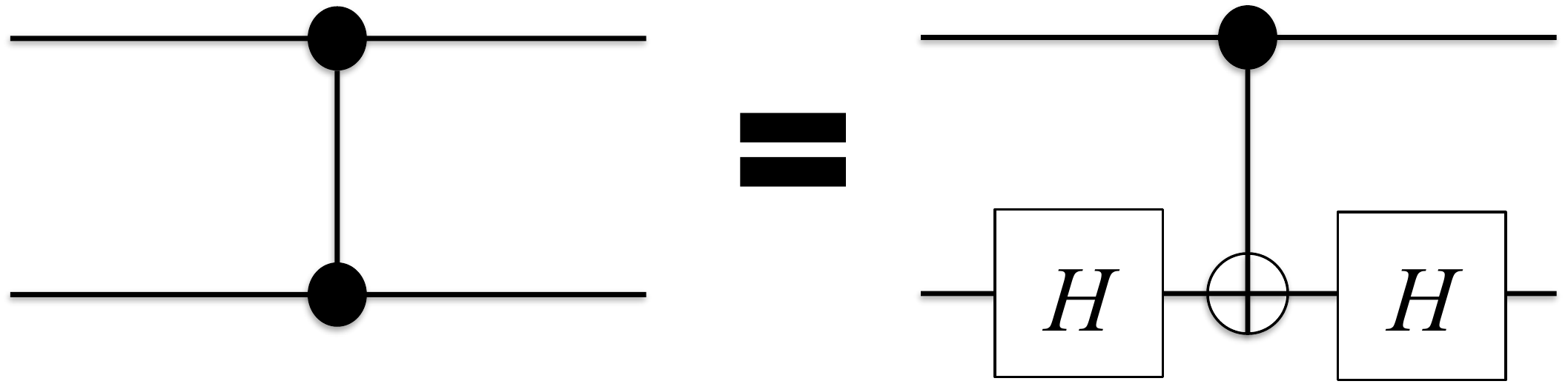}
\caption{Circuit representation of a Controlled-Z gate}
\label{CZgate}
\end{figure}

As shown in Fig.~\ref{CZgate}, the CZ gate can also be constructed by two Hadamard gates and one CNOT gate.

\section{Alternative state representations}

\subsection{Density matrix and Mixed States}

Quantum states are either mixed or pure. A pure state is in a closed system, with no interaction with the outside world.
In contrast, a state is mixed when a part of the quantum system becomes entangled with or is acted upon in unknown ways by the environment.
While any pure state can be written in state-vector form, mixed states can only be described using a {\it density matrix}.
The density matrix of a pure state $\ket{\psi}$ can be found by:

\begin{equation}
\rho = \ket{\psi} \bra{\psi}
\end{equation}

The corresponding probability of a state $\ket{j}$ can be found by the diagonal entries of $\rho$:

\begin{equation}
P(\ket{j}) = \rho_{j,j} \ket{j} \bra{j}
\end{equation}

If the state is pure, $\rho^2 = \rho$ and $Tr[\rho^2]=1$ and if the state is mixed $\rho^2 \neq \rho$ and $\mathrm{Tr}[\rho^2]<1$.
For example, the density matrix of a pure state Bell pair is:

\begin{eqnarray}
\rho = \ket{\psi} \bra{\psi} = \frac{1}{\sqrt{2}}(\ket{00}+\ket{11}) \frac{1}{\sqrt{2}}(\bra{00}+\bra{11}) \nonumber \\
=\frac{1}{2}(\ket{00}\bra{00} + \ket{00}\bra{11} + \ket{11}\bra{00} + \ket{11}\bra{11} ) =  \begin{bmatrix}
	\frac{1}{2} & 0 & 0 & \frac{1}{2} \\
	0 & 0 & 0 & 0\\
	0 & 0 & 0 & 0\\
	\frac{1}{2} & 0 & 0 & \frac{1}{2}
\end{bmatrix}
\end{eqnarray}

An example of a completely mixed state of 2 qubits, which represents the classical dependent probabilities is:

\begin{eqnarray}
\rho = \frac{1}{2}(\ket{00}\bra{00} + \ket{11}\bra{11} ) =  \begin{bmatrix}
	\frac{1}{2} & 0 & 0 & 0 \\
	0 & 0 & 0 & 0\\
	0 & 0 & 0 & 0\\
	0 & 0 & 0 & \frac{1}{2}
\end{bmatrix}
\end{eqnarray}

With a completely mixed state of 2 qubits, there is no entanglement between the two qubits -- each has a state, $\ket{0}$ or $\ket{1}$.
The off-diagonal elements are \emph{quantum coherences} and can be complex and the diagonal elements must be real.

Given a set of probabilities per state, we can take the sum of each weighted density matrix to construct the entire system's density matrix.
For example, the density matrix of a Bell pair (see Sec.\ref{entsection}) with 10\% bit flip error rate can be described as:
\begin{equation}
\rho = 0.9\ket{\Psi^+}\bra{\Psi^+} + 0.1\ket{\Phi^+}\bra{\Phi^+}.
\end{equation}

More generally, $\rho = \sum P(i)\ket{\psi_i}\bra{\psi_i}$ such that $\sum P(i) = 1$ where $P(i)$ is the probability of the system being in the state $\ket{\psi_i}$.
Note that the set \{$\ket{\psi_i}$\} may be but does not have to be a complete basis set.
\subsubsection{Calculating the density matrix of a left-over system after a measurement}

Qubit measurement and its outcome affect the state of the left-over system.
For example, a system may be composed of three qubits, entangled as a W state $\ket{\psi} = \frac{1}{\sqrt{3}}(\ket{0_A0_B1_C} + \ket{0_A1_B0_C} + \ket{1_A0_B0_C})$ (see section~\ref{WSTATEsection}).
The density matrix of the system is therefore:

\begin{eqnarray}
\rho = \ket{\psi}\bra{\psi}
= \begin{bmatrix}
0 \\
\frac{1}{\sqrt{3}} \\
\frac{1}{\sqrt{3}} \\
0 \\
\frac{1}{\sqrt{3}} \\
0 \\
0 \\
0
\end{bmatrix}.\begin{bmatrix}
0 &
\frac{1}{\sqrt{3}} &
\frac{1}{\sqrt{3}} &
0 &
\frac{1}{\sqrt{3}} &
0 &
0 &
0
\end{bmatrix} \nonumber \\
= \begin{bmatrix}
0 & 0 & 0 & 0 & 0 & 0 & 0 & 0\\
0 & 1/3 & 1/3 & 0 & 1/3 & 0 & 0 & 0\\
0 & 1/3 & 1/3 & 0 & 1/3 & 0 & 0 & 0\\
0 & 0 & 0 & 0 & 0 & 0 & 0 & 0\\
0 & 1/3 & 1/3 & 0 & 1/3 & 0 & 0 & 0\\
0 & 0 & 0 & 0 & 0 & 0 & 0 & 0\\
0 & 0 & 0 & 0 & 0 & 0 & 0 & 0\\
0 & 0 & 0 & 0 & 0 & 0 & 0 & 0
\end{bmatrix}
\end{eqnarray}

When qubit A is measured, say along the Z axis, the measurement outcome (or its state) will be either $\ket{0_A}$ or $\ket{1_A}$
and qubit A will no longer be entangled with qubit B and C.
As discussed in section~\ref{MEASUREMENTsection}, $P(\ket{0_A})$ and $P(\ket{1_A})$ can be derived as:

\begin{eqnarray}
P(\ket{0_A}) = \mbox{Tr}[ ({M_Z^+}_A \otimes I_B \otimes I_C)^\dagger . ({M_Z^+}_A \otimes I_B \otimes I_C) . \rho] = 2/3\\
P(\ket{1_A}) = \mbox{Tr}[ ({M_Z^-}_A \otimes I_B \otimes I_C)^\dagger . ({M_Z^-}_A \otimes I_B \otimes I_C) . \rho] = 1/3.
\end{eqnarray}

The overall density matrix of all three qibits, after performing the measurement on qubit A, can be derived as:

\begin{eqnarray}
\ket{\psi^\prime} = \begin{dcases}
(M_Z^+ \otimes I_B \otimes I_C) \rho (M_Z^+ \otimes I_B \otimes I_C)^\dagger ,& A \rightarrow \ket{0}\\
(M_Z^- \otimes I_B \otimes I_C) \rho (M_Z^- \otimes I_B \otimes I_C)^\dagger ,& A \rightarrow \ket{1}
\end{dcases}\\
\end{eqnarray}

The above density matrix still includes qubit A.
The reduced density matrix of the left-over state, $\ket{\psi^\prime_{B,C}}$, can also be calculated in a similar way:

\begin{eqnarray}
\ket{\psi^\prime_{B,C}} = \begin{dcases}
(\bra{0_A} \otimes I_B \otimes I_C) \rho (\bra{0_A} \otimes I_B \otimes I_C)^\dagger ,& A \rightarrow \ket{0}\\
(\bra{1_A} \otimes I_B \otimes I_C) \rho (\bra{1_A} \otimes I_B \otimes I_C)^\dagger ,& A \rightarrow \ket{1}
\end{dcases}\\
\end{eqnarray}

\subsection{Stabilizer}
\label{stabilizersection}
A quantum state may be described by a state vector or a density matrix.
If we use the ordinary vector notation, the size of the vector grows exponentially in the number of qubits, as we saw in Sec.\ref{compositesection}.
The bra-ket notation allows us to dispense with writing down the non-zero terms.
However, many states we create in a quantum computer have an exponentially large number of non-zero terms.
An alternative way to represent a pure state is to use a set of stabilizers. A stabilizer $S$ of a state $\ket{\psi}$ is:

\begin{equation}
S\ket{\psi} = \ket{\psi}
\end{equation}

Consequently, a stabilizer or a set of stabilizers uniquely determines a quantum state.
Below are examples of stabilizers of different states:

\begin{align}
	Z\ket{0} &= \ket{0} & -Z\ket{1} &= \ket{1} \\
	X\ket{+} &= \ket{+} & -X\ket{-} &= \ket{-} \\
	Y\ket{+i} &= \ket{+i} & -Y\ket{-i} &= \ket{-i}
\end{align}

Stabilizers of major entangled states are shown in Tab.\ref{table_stabilizer}.

\begin{table}[H]
\begin{center}
	\caption{Different entangled states and the corresponding stabilizer sets}
	\label{table_stabilizer}
	\begin{tabular}{|c||l|c|}
		\hline
		Quantum state & Stabilizer sets \\
		\hline
		\hline
		$\frac{1}{\sqrt{2}}(\ket{0_{1}0_{2}}+\ket{1_{1}1_{2}})$ & $X_{1}X_{2}, Z_{1}Z_{2}$ \\
		\hline
		$\frac{1}{\sqrt{2}}(\ket{0_{1}0_{2}}-\ket{1_{1}1_{2}})$ & $-X_{1}X_{2}, Z_{1}Z_{2}$ \\
		\hline
		$\frac{1}{\sqrt{2}}(\ket{0_{1}1_{2}}+\ket{1_{1}0_{2}})$ & $X_{1}X_{2}, -Z_{1}Z_{2}$ \\
		\hline
		$\frac{1}{\sqrt{2}}(\ket{0_{1}1_{2}}-\ket{1_{1}0_{2}})$ & $-X_{1}X_{2}, -Z_{1}Z_{2}$ \\
		\hline
		$\frac{1}{\sqrt{2}}(\ket{0_{1}0_{2}0_{3}}+\ket{0_{1}0_{2}0_{3}})$ & $X_{1}X_{2}X_{3}, Z_{1}Z_{2}, Z_{2}Z_{3}$ \\
		\hline
		$\frac{1}{\sqrt{2}}(\ket{0_{1}+_{2}}+\ket{1_{1}-_{2}})$ & $X_{1}Z_{2}, Z_{1}X_{2}$ \\
		\hline
		$\frac{1}{2}(\ket{0_{1}0_{2}+_{3}}+\ket{0_{1}1_{2}-_{3}}+\ket{1_{1}1_{2}+_{3}}+\ket{1_{1}1_{2}-_{3}})$ & $X_{1}Z_{2}, Z_{1}X_{2}Z_{3},Z_{2}X_{3}$ \\
		\hline
	\end{tabular}
\end{center}
\end{table}

Not all states can be stabilized by the Pauli operators.
So $\ket{\psi}$ is the eigenvector of all stabilizers in the set.
A quantum state with $n$ qubits can be fully described by $n$ stabilizers.
Stabilizers can be helpful when describing a state consisting of many qubits, as the number of stabilizers grows only linearly while other methods grow exponentially. On the other hand, finding the correct stabilizer set representing a particular complex quantum state is difficult.
Moreover, stabilizers may only be useful when representing states that are based on Clifford group operations (S gate, H gate and CZ gate).
A state with $n$ qubits has $n$ degree of freedom.
Each known stabilizer will decrease the degree of freedom linearly.

\section{Entanglement}
\label{entsection}
Two or more qubits can be in an {\it entangled} state.
When states are entangled, each qubit's state cannot be described independently.
That is to say, the collapse of the wave function of a qubit by measurement may immediately affect the other of the pair's state regardless of the physical distance between.

\subsection{Bell pair/Einstein-Podolsky-Rosen (EPR) pair}
One common example of an entangled state is called a {\it Bell pair} or sometimes called {\it EPR pair}:

\begin{equation}
	\ket{\psi}= \frac{1}{\sqrt{2}} (\ket{00} + \ket{11}) = \ket{\Phi^{+}}
\end{equation}

In the above example, note that each qubit state $\ket{00}$ and $\ket{11}$ has equally weighted probability,
so that each qubit has a 50/50 probability of being found in each state but not independently.
If one qubit's state is found to be 0, then the other qubit's state must be, and will be 0.
Therefore, measuring one qubit will also decide the other qubit's state. Specifically, the four entangled states that can be used as a basis set are known as {\it Bell states}:

\begin{equation}
\ket{\Phi^{+}}= \frac{1}{\sqrt{2}} (\ket{00} + \ket{11})
\end{equation}

\begin{equation}
\ket{\Phi^{-}}= \frac{1}{\sqrt{2}} (\ket{00} - \ket{11})
\end{equation}

\begin{equation}
\ket{\Psi^{+}}= \frac{1}{\sqrt{2}} (\ket{01} + \ket{10})
\end{equation}

\begin{equation}
\ket{\Psi^{-}}= \frac{1}{\sqrt{2}} (\ket{01} - \ket{10})
\end{equation}

\subsection{W state}
\label{WSTATEsection}
The W state is an entangled quantum state that consists of 3 qubits, which has a similar state to the $\ket{\Phi^+}$.
This state has an Hamming weight of 1.

\begin{equation}
\ket{W}= \frac{1}{\sqrt{3}} (\ket{001} + \ket{010} + \ket{100})
\end{equation}

Measuring one qubit will result in either state:

\begin{eqnarray}
\ket{W^{0}} = \frac{1}{\sqrt{2}}(\ket{01} + \ket{10}) \\
\ket{W^{1}} = \ket{00})
\end{eqnarray}

Depending on the measurement result, the residual system is entangled or unentangled.

\subsection{Greenberger-Horne-Zeilinger (GHZ) state}
\label{GHZsubsection}

The GHZ state is an entangled state of $N \geq 3$ qubits as in equation \ref{GHZ}.

\begin{equation}
\label{GHZ}
\ket{\psi}= \frac{1}{\sqrt{2}} (\ket{0}^{\otimes N} + \ket{1}^{\otimes N}) = \ket{GHZ}
\end{equation}

The simplest GHZ state includes 3 qubits:

\begin{equation}
\ket{\psi} = \frac{1}{\sqrt{2}} (\ket{0}^{\otimes 3} + \ket{1}^{\otimes 3}) =  \frac{1}{\sqrt{2}} (\ket{000} + \ket{111}) = \ket{GHZ}
\end{equation}

Similar to a Bell pair, measuring an arbitrary qubit of the GHZ state decides the overall state to $\ket{000}$ or $\ket{111}$ with equally weighted probabilities.
The GHZ state is not Local Operation and Classical Communication (LOCC) equivalent to the W state -- no local operation can convert the GHZ state to the W state or vice versa when qubits are physically distant.

\subsection{Cluster state}
The cluster state is an example of qubits that are maximally entangled -- a maximally entangled state has the maximum Von Neumann entropy obtainable by the number of qubits in the system. The Von Neumann entropy is a measure of entanglement:

\begin{equation}
S(\rho) = -\operatorname{Tr}{\rho \log_{2}(\rho)}
\end{equation}


A cluster state of $n$ vertices (qubits) can be defined by:
\begin{equation}
\ket{G} = \prod_{(a,b)\in E}\Lambda_{a,b}(Z)\ket{+}^{\otimes n}
\end{equation}

where E is the set of edges (entanglement) and $a,b$ are the corresponding vertices (qubits).

As an example, 3-qubit cluster state is:
\begin{equation}
\ket{\psi}= \frac{1}{2\sqrt{2}}(
\ket{0_{1}0_{2}0_{3}} +
\ket{0_{1}0_{2}1_{3}} +
\ket{0_{1}1_{2}0_{3}} -
\ket{0_{1}1_{2}1_{3}} +
\ket{1_{1}0_{2}0_{3}} +
\ket{1_{1}0_{2}1_{3}} -
\ket{1_{1}1_{2}0_{3}} +
\ket{1_{1}1_{2}1_{3}} )
\end{equation}

where the subscript denotes the labeled qubit for identification.
All states are equally weighted, and measuring qubit 1 will result in either one of the following system:

\begin{align}
\intertext{If the measurement result is 0}
\label{right_cluster}
\ket{\psi}= \frac{1}{2}(
\ket{0_{2}0_{3}} +
\mid 0_{2}1_{3} \rangle +
\mid 1_{2}0_{3} \rangle -
\ket{1_{2}1_{3}} )
\end{align}

\begin{align}
\intertext{If the measurement result is 1}
\label{wrong_cluster}
\ket{\psi}= \frac{1}{2}(
\ket{0_{2}0_{3}} +
\mid 0_{2}1_{3} \rangle -
\mid 1_{2}0_{3} \rangle +
\ket{1_{2}1_{3}} )
\end{align}

The equation \ref{wrong_cluster} equates to equation \ref{right_cluster} with an additional Z operation to qubit 2 as a byproduct, $\ket{\psi^{0}} = Z_{1}\ket{\psi^{1}}$.
Notice that measuring just one qubit does not fully decide the remaining state.
The 2-qubit cluster state is also LOCC equivalent to the Bell pair since $H_1 \ket{\Phi^+_{1,2}}=\ket{G_{1,2}}$ or  $H_2 \ket{\Phi^+_{1,2}}=\ket{G_{1,2}}$.

\section{Reading quantum circuits}

We can track the state through a computation step-by-step by writing down the bra-ket expression.
However, sometimes it is easier to understand the transition using a different mathematical representation.

\begin{figure}[!hbt]
\center
\includegraphics[keepaspectratio,scale=0.5]{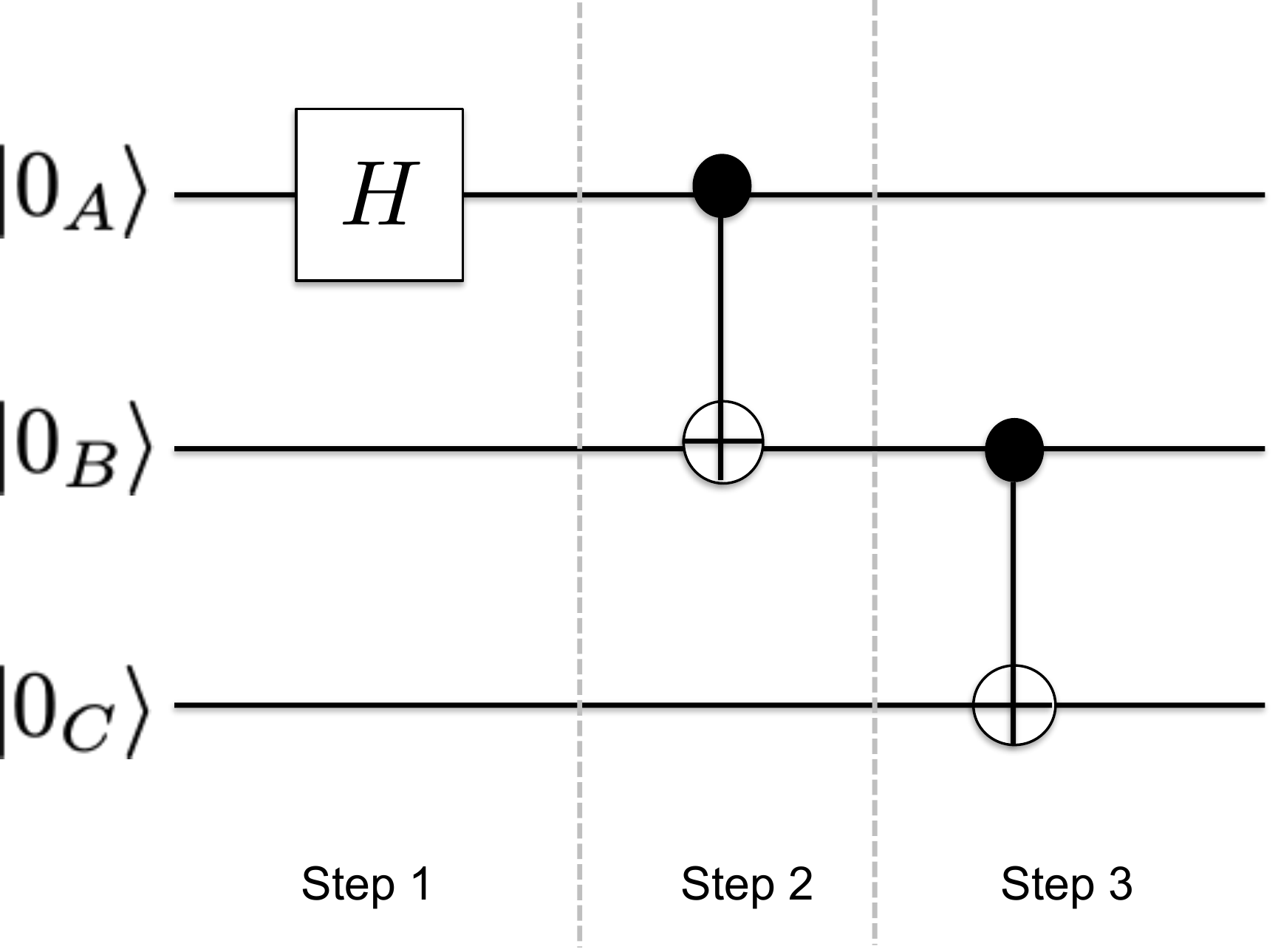}
\caption[A quantum circuit generating a GHZ state of three qubits]{A quantum circuit generating a GHZ state (refer to subsection~\ref{GHZsubsection}) of three qubits.}
\label{GHZcircuit}
\end{figure}

A common quantum circuit consists of multiple gate operations.
An example circuit that operates on three qubits (qubit A, qubit B and qubit C) initialized as $\ket{0}$ is shown in Fig.~\ref{GHZcircuit}.
This particular circuit consists of three steps, with one Hadamard gate and two CNOT gates.
This circuit uses three independent qubits as inputs, and outputs them as an entangled GHZ state.

\subsection{Understanding circuits using matrices}

It is sometimes easy to calculate the transition using matrices.

The input state $\ket{\psi_{0}} = \ket{0_A0_B0_C}$ in vector form is

\begin{eqnarray}
\ket{\psi_{0}}= \ket{0_A0_B0_C} =
\begin{bmatrix}
1\\
0\\
0\\
0\\
0\\
0\\
0\\
0
\end{bmatrix},
\end{eqnarray}

because each qubit has an initial state $\ket{0}$.
\subsubsection{Step 1}

The first step in Fig.\ref{GHZcircuit} performs Hadamard gate, which can be represented as a $2 \times 2$ matrix, on qubit A.
Other two qubits in Step 1 are simply waiting, in which case we commonly use an Identity matrix as a matrix representation.
The overall gate operation in Step 1 can be, therefore, seen as a set of three single qubit gate operations -- an Hadamard gate and two Identity gates.
The matrix representing the entire operation in Step 1, $U_{\mbox{1}}$, can be calculated by taking the tensor product of each matrix.

\begin{eqnarray}
U_{\mbox{1}} =
H \otimes I \otimes I \nonumber \\ =
\frac{1}{\sqrt{2}}
\begin{bmatrix}
1 & 1\\
1 & -1\\
\end{bmatrix} \otimes
\begin{bmatrix}
1 & 0\\
0 & 1\\
\end{bmatrix} \otimes
\begin{bmatrix}
1 & 0\\
0 & 1\\
\end{bmatrix} \nonumber \\  =
\begin{bmatrix}
\frac{1}{\sqrt{2}} & 0 & 0 & 0 & \frac{1}{\sqrt{2}} & 0 & 0 & 0\\
0 & \frac{1}{\sqrt{2}} & 0 & 0 & 0 & \frac{1}{\sqrt{2}} & 0 & 0\\
0 & 0 &\frac{1}{\sqrt{2}} & 0 & 0 & 0 & \frac{1}{\sqrt{2}} & 0\\
0 & 0 & 0 & \frac{1}{\sqrt{2}} & 0 & 0 & 0 & \frac{1}{\sqrt{2}}\\
\frac{1}{\sqrt{2}} & 0 & 0 & 0 & -\frac{1}{\sqrt{2}} & 0 & 0 & 0\\
0 & \frac{1}{\sqrt{2}} & 0 & 0 & 0 & -\frac{1}{\sqrt{2}} & 0 & 0\\
0 & 0 &\frac{1}{\sqrt{2}} & 0 & 0 & 0 & -\frac{1}{\sqrt{2}} & 0\\
0 & 0 & 0 & \frac{1}{\sqrt{2}} & 0 & 0 & 0 & -\frac{1}{\sqrt{2}}
\end{bmatrix}
\end{eqnarray}

The state after Step 1 ($\ket{\psi_1}$), of course, can be calculated by multiplying the acquired matrix to the input state as:

\begin{eqnarray}
\ket{\psi_1}=U_{1}\ket{\psi_{0}} =
\begin{bmatrix}
\frac{1}{\sqrt{2}} \\
0\\
0\\
0\\
\frac{1}{\sqrt{2}}\\
0\\
0\\
0
\end{bmatrix} \\ = \frac{1}{\sqrt{2}}(\ket{0_A0_B0_C}+\ket{1_A0_B0_C}).
\end{eqnarray}

\subsubsection{Step 2}

$U_{2}$ can be calculated in a similar way as in Step 1.
Hence, $U_{2}$ is:

\begin{eqnarray}
U_{\mbox{2}} =
CNOT \otimes I \nonumber \\ =
\frac{1}{\sqrt{2}}
\begin{bmatrix}
1 & 0 & 0 & 0\\
0 & 1 & 0 & 0\\
0 & 0 & 0 & 1\\
0 & 0 & 1 & 0
\end{bmatrix} \otimes
\begin{bmatrix}
1 & 0\\
0 & 1\\
\end{bmatrix} \nonumber \\ =
\begin{bmatrix}
1 & 0 & 0 & 0 & 0 & 0 & 0 & 0\\
0 & 1 & 0 & 0 & 0 & 0 & 0 & 0\\
0 & 0 & 1 & 0 & 0 & 0 & 0 & 0\\
0 & 0 & 0 & 1 & 0 & 0 & 0 & 0\\
0 & 0 & 0 & 0 & 0 & 0 & 1 & 0\\
0 & 0 & 0 & 0 & 0 & 0 & 0 & 1\\
0 & 0 & 0 & 0 & 1 & 0 & 0 & 0\\
0 & 0 & 0 & 0 & 0 & 1 & 0 & 0
\end{bmatrix}
\end{eqnarray}

The state after Step 2, $\ket{\psi_2}$, is therefore:

\begin{eqnarray}
\ket{\psi_2} = U_{\mbox{2}}\ket{\psi_1} = U_{\mbox{2}}U_{\mbox{1}}\ket{\psi_0} =
\begin{bmatrix}
\frac{1}{\sqrt{2}} \\
0\\
0\\
0\\
0\\
0\\
\frac{1}{\sqrt{2}}\\
0
\end{bmatrix} \\ = \frac{1}{\sqrt{2}}(\ket{0_A0_B0_C}+\ket{1_A1_B0_C}) \\ = \frac{1}{\sqrt{2}}(\ket{0_A0_B}+\ket{1_A1_B})\ket{0_C}.
\end{eqnarray}

\subsubsection{Step 3}

The last step in the circuit, $U_3$, is:

\begin{eqnarray}
U_{\mbox{3}} =
I \otimes CNOT \nonumber \\
= \frac{1}{\sqrt{2}}
\begin{bmatrix}
1 & 0\\
0 & 1\\
\end{bmatrix} \otimes
\begin{bmatrix}
1 & 0 & 0 & 0\\
0 & 1 & 0 & 0\\
0 & 0 & 0 & 1\\
0 & 0 & 1 & 0
\end{bmatrix} \nonumber \\  =
\begin{bmatrix}
1 & 0 & 0 & 0 & 0 & 0 & 0 & 0\\
0 & 1 & 0 & 0 & 0 & 0 & 0 & 0\\
0 & 0 & 0 & 1 & 0 & 0 & 0 & 0\\
0 & 0 & 1 & 0 & 0 & 0 & 0 & 0\\
0 & 0 & 0 & 0 & 1 & 0 & 0 & 0\\
0 & 0 & 0 & 0 & 0 & 1 & 0 & 0\\
0 & 0 & 0 & 0 & 0 & 0 & 0 & 1\\
0 & 0 & 0 & 0 & 0 & 0 & 1 & 0
\end{bmatrix}
\end{eqnarray}

The output state of the circuit, $\ket{\psi_3}$ , can be calculated using the acquired three matrices.

\begin{eqnarray}
U_{\mbox{3}}\ket{\psi_2} = U_{\mbox{3}}U_{\mbox{2}}\ket{\psi_1} = U_{\mbox{3}}U_{\mbox{2}}U_{\mbox{1}}\ket{\psi_0} =
\begin{bmatrix}
\frac{1}{\sqrt{2}} \\
0\\
0\\
0\\
0\\
0\\
0\\
\frac{1}{\sqrt{2}}
\end{bmatrix} \nonumber \\ = \frac{1}{\sqrt{2}}(\ket{0_A0_B0_C}+\ket{1_A1_B1_C}).
\end{eqnarray}

Notice that the entire circuit, composed of the discussed three steps, can be described using a single $8 \times 8$ matrix, which is:

\begin{eqnarray}
U = U_{\mbox{3}}.U_{\mbox{2}}.U_{\mbox{1}} =
\begin{bmatrix}
\frac{1}{\sqrt{2}} & 0 & 0 & 0 & \frac{1}{\sqrt{2}} & 0 & 0 & 0\\
0 & \frac{1}{\sqrt{2}} & 0 & 0 & 0 & \frac{1}{\sqrt{2}} & 0 & 0\\
0 & 0 & 0 & \frac{1}{\sqrt{2}} & 0 & 0 & 0 & \frac{1}{\sqrt{2}}\\
0 & 0 & \frac{1}{\sqrt{2}} & 0 & 0 & 0 & \frac{1}{\sqrt{2}} & 0\\
0 & 0 & \frac{1}{\sqrt{2}} & 0 & 0 & 0 & -\frac{1}{\sqrt{2}} & 0\\
0 & 0 & 0 & \frac{1}{\sqrt{2}} & 0 & 0 & 0 & -\frac{1}{\sqrt{2}}\\
0 & \frac{1}{\sqrt{2}} & 0 & 0 & 0 & -\frac{1}{\sqrt{2}} & 0 & 0\\
\frac{1}{\sqrt{2}} & 0 & 0 & 0 & -\frac{1}{\sqrt{2}} & 0 & 0 & 0
\end{bmatrix}.
\end{eqnarray}

\subsubsection{Other cases}

\begin{figure}[!hbt]
\center
\includegraphics[keepaspectratio,scale=0.45]{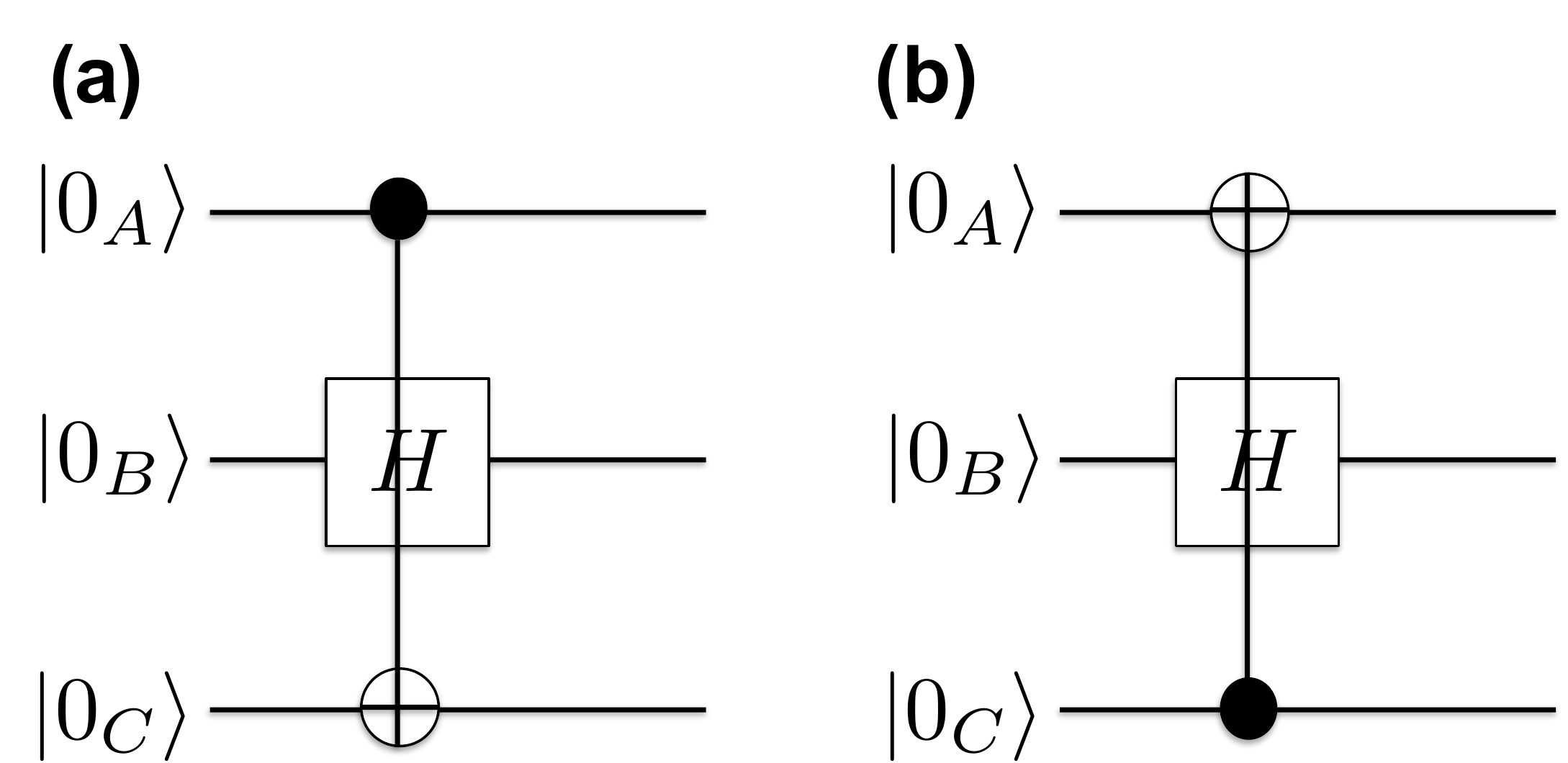}
\caption{Example circuits of an Hadamard gate and a CNOT gate.}
\label{MatrixCircuit}
\end{figure}

For a different quantum circuit, a two-qubit gate may be performed between non-adjacent qubits (see Fig.~\ref{MatrixCircuit}).

For such cases, we have to calculate the matrix separately based on the controlled qubit's state, $\ket{0}$ and $\ket{1}$, and take the sum as:

\begin{eqnarray}
U_{(a)} = (\ket{0_A}\bra{0_A} \otimes H_B \otimes I_C) + (\ket{1_A}\bra{1_A} \otimes H_B \otimes X_C) \nonumber \\
= ( \begin{bmatrix}
1 & 0 \\
0 & 0
\end{bmatrix} \otimes \begin{bmatrix}
\frac{1}{\sqrt{2}} & \frac{1}{\sqrt{2}} \\
\frac{1}{\sqrt{2}} & -\frac{1}{\sqrt{2}}
\end{bmatrix} \otimes \begin{bmatrix}
1 & 0 \\
0 & 1
\end{bmatrix}) + ( \begin{bmatrix}
0 & 0 \\
0 & 1
\end{bmatrix} \otimes \begin{bmatrix}
\frac{1}{\sqrt{2}} & \frac{1}{\sqrt{2}} \\
\frac{1}{\sqrt{2}} & -\frac{1}{\sqrt{2}}
\end{bmatrix} \otimes \begin{bmatrix}
0 & 1 \\
1 & 0
\end{bmatrix}) \nonumber \\
= \begin{bmatrix}
\frac{1}{\sqrt{2}} & 0 & \frac{1}{\sqrt{2}} & 0 & 0 & 0 & 0 & 0\\
0 & \frac{1}{\sqrt{2}} & 0 & \frac{1}{\sqrt{2}} & 0 & 0 & 0 & 0\\
\frac{1}{\sqrt{2}} & 0 & -\frac{1}{\sqrt{2}} & 0 & 0 & 0 & 0 & 0\\
0 & \frac{1}{\sqrt{2}} & 0 & -\frac{1}{\sqrt{2}} & 0 & 0 & 0 & 0\\
0 & 0 & 0 & 0 & 0 & \frac{1}{\sqrt{2}} & 0 & \frac{1}{\sqrt{2}}\\
0 & 0 & 0 & 0 & \frac{1}{\sqrt{2}} & 0 & \frac{1}{\sqrt{2}} & 0\\
0 & 0 & 0 & 0 & 0 & \frac{1}{\sqrt{2}} & 0 & -\frac{1}{\sqrt{2}}\\
0 & 0 & 0 & 0 & \frac{1}{\sqrt{2}} & 0 & -\frac{1}{\sqrt{2}} & 0
\end{bmatrix}
\end{eqnarray}

Similarly, for an inverted CNOT, we can calculate the matrix as:

\begin{eqnarray}
U_{(b)} = (I_A \otimes H_B \otimes \ket{0_C}\bra{0_C}) + (X_A \otimes H_B \otimes \ket{1_C}\bra{1_C}) \nonumber \\
= (  \begin{bmatrix}
1 & 0 \\
0 & 1
\end{bmatrix} \otimes \begin{bmatrix}
\frac{1}{\sqrt{2}} & \frac{1}{\sqrt{2}} \\
\frac{1}{\sqrt{2}} & -\frac{1}{\sqrt{2}}
\end{bmatrix} \otimes \begin{bmatrix}
1 & 0 \\
0 & 0
\end{bmatrix}) + (\begin{bmatrix}
0 & 1 \\
1 & 0
\end{bmatrix}
\otimes \begin{bmatrix}
\frac{1}{\sqrt{2}} & \frac{1}{\sqrt{2}} \\
\frac{1}{\sqrt{2}} & -\frac{1}{\sqrt{2}}
\end{bmatrix} \otimes \begin{bmatrix}
0 & 0 \\
0 & 1
\end{bmatrix}) \nonumber \\
= \begin{bmatrix}
\frac{1}{\sqrt{2}} & 0 & \frac{1}{\sqrt{2}} & 0 & 0 & 0 & 0 & 0\\
0 & 0 & 0 & 0 & 0 & \frac{1}{\sqrt{2}} & 0 & \frac{1}{\sqrt{2}}\\
\frac{1}{\sqrt{2}} & 0 & -\frac{1}{\sqrt{2}} & 0 & 0 & 0 & 0 & 0\\
0 & 0 & 0 & 0 & 0 & \frac{1}{\sqrt{2}} & 0 & -\frac{1}{\sqrt{2}}\\
0 & 0 & 0 & 0 & \frac{1}{\sqrt{2}} & 0 & \frac{1}{\sqrt{2}} & 0\\
0 & \frac{1}{\sqrt{2}} & 0 & \frac{1}{\sqrt{2}} & 0 & 0 & 0 & 0\\
0 & 0 & 0 & 0 & \frac{1}{\sqrt{2}} & 0 & -\frac{1}{\sqrt{2}} & 0\\
0 & \frac{1}{\sqrt{2}} & 0 & -\frac{1}{\sqrt{2}} & 0 & 0 & 0 & 0
\end{bmatrix}
\end{eqnarray}

The required dimension of matrices will increase exponentially to the number of involved qubits.

\subsection{Understanding circuits using stabilizers}
Not only does the number of terms in the vector increase exponentially, but naturally the size of a full matrix representing the unitary operation on a set of qubits does as well.
However, as discussed in section~\ref{stabilizersection}, the number of stabilizers only increases linearly.
The transition of stabilizers through a circuit can be understood in a similar way as error propagations (see section~\ref{ERRORsection}).
Here, we study the behavior based on the same GHZ state circuit shown in Fig.~\ref{GHZcircuit}.

The Z operator stabilizes the state $\ket{0}$.
Thus, the input state $\ket{\psi_{i}} = \ket{0_A0_B0_C}$ in stabilizer formalism is as shown in Tab.~\ref{stab1}.
Note that each empty space in the table is actually a stabilizer I.

\begin{table}[htb]
\center
\caption{Stabilizers of the input state.}
\begin{tabular}{|c||c|c|c|} \hline
	 & qubit A & qubit B & qubit C \\ \hline \hline
	1st stabilizer & Z &   &   \\ \hline
	2nd stabilizer &   & Z &   \\ \hline
	3rd stabilizer &   &   & Z  \\ \hline
\end{tabular}
\label{stab1}
\end{table}

\subsubsection{Step 1}
The first step is an Hadamard gate on qubit A.
Similar to the error propagation, the Hadamard gate converts the Z stabilizer of qubit A to X as in Tab.~\ref{stab2}.

\begin{table}[H]
\center
\caption{Stabilizers of the output state after Step 1.}
\begin{tabular}{|c||c|c|c|} \hline
	 & qubit A & qubit B & qubit C \\ \hline \hline
	1st stabilizer & X &   &   \\ \hline
	2nd stabilizer &   & Z &   \\ \hline
	3rd stabilizer &   &   & Z  \\ \hline
\end{tabular}
\label{stab2}
\end{table}

\subsubsection{Step 2}
The second step is a CNOT gate between qubit A and B.
The X stabilizer of qubit A propagates to qubit B, and the Z stabilizer of qubit B propagates to qubit A.

\begin{table}[H]
\center
\caption{Stabilizers of the output state after Step 2.}
\begin{tabular}{|c||c|c|c|} \hline
	 & qubit A & qubit B & qubit C \\ \hline \hline
	1st stabilizer & X & X  &   \\ \hline
	2nd stabilizer & Z & Z &   \\ \hline
	3rd stabilizer &   &   & Z  \\ \hline
\end{tabular}
\end{table}

\subsubsection{Step 3}
The third step is also a CNOT gate, but between qubit B and C.
The X stabilizer propagates from qubit B propagates to qubit C, and
the Z stabilizer propagates from qubit C propagates to qubit B.

\begin{table}[H]
\center
\caption{Stabilizers of the output state after Step 3.}
\begin{tabular}{|c||c|c|c|} \hline
	 & qubit A & qubit B & qubit C \\ \hline \hline
	1st stabilizer & X & X & X  \\ \hline
	2nd stabilizer & Z & Z &   \\ \hline
	3rd stabilizer &   & Z & Z  \\ \hline
\end{tabular}
\end{table}

\begin{figure}[!hbt]
\center
\includegraphics[keepaspectratio,scale=0.28]{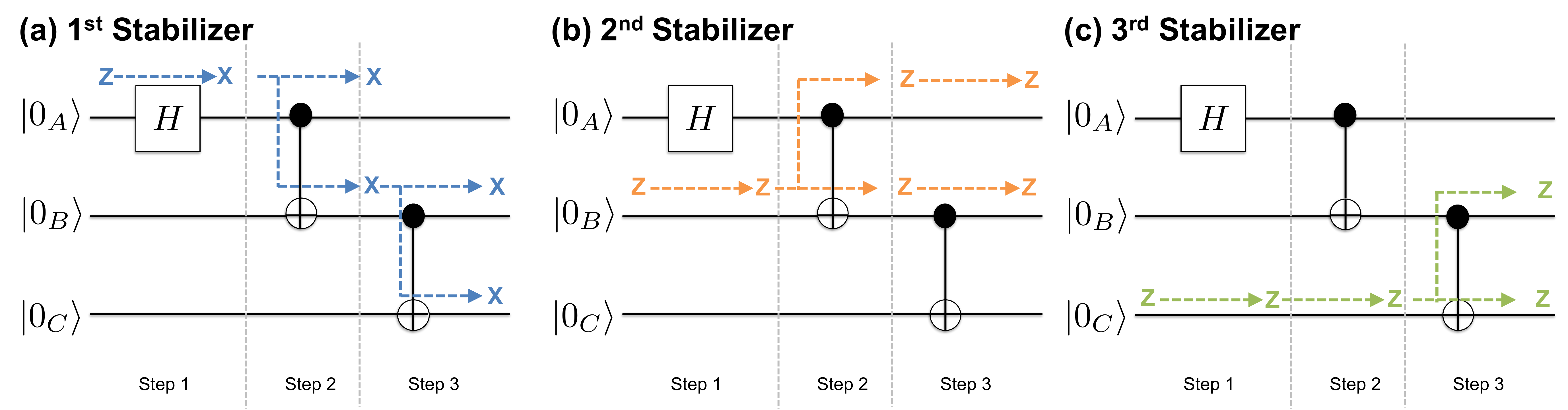}
\caption[Propagation of stabilizers in the creation of a 3-qubit GHZ state]{Propagation of stabilizers in the creation of a 3-qubit GHZ state. Stabilizer I is omitted in the figure for simplification. Stabilizer I will stay as I unless propagated from a different qubit through a multi-qubit gate operation.}
\label{StabilizerPropagation}
\end{figure}

Thus, the stabilizer set of a three qubit GHZ state is $\{X_{A}X_{B}X_{C},Z_{A},Z_{B},Z_{B}Z_{C}\}$.
A pictorial transition of the stabilizers is summarized in Fig.~\ref{StabilizerPropagation}.

\section{Quantum networking}

There are several important operations for accomplishing quantum networking.
The following subsections explain some of the major operations in detail.

\subsection{Quantum Teleportation}
\begin{figure}[H]
\center
\includegraphics[keepaspectratio,scale=0.7]{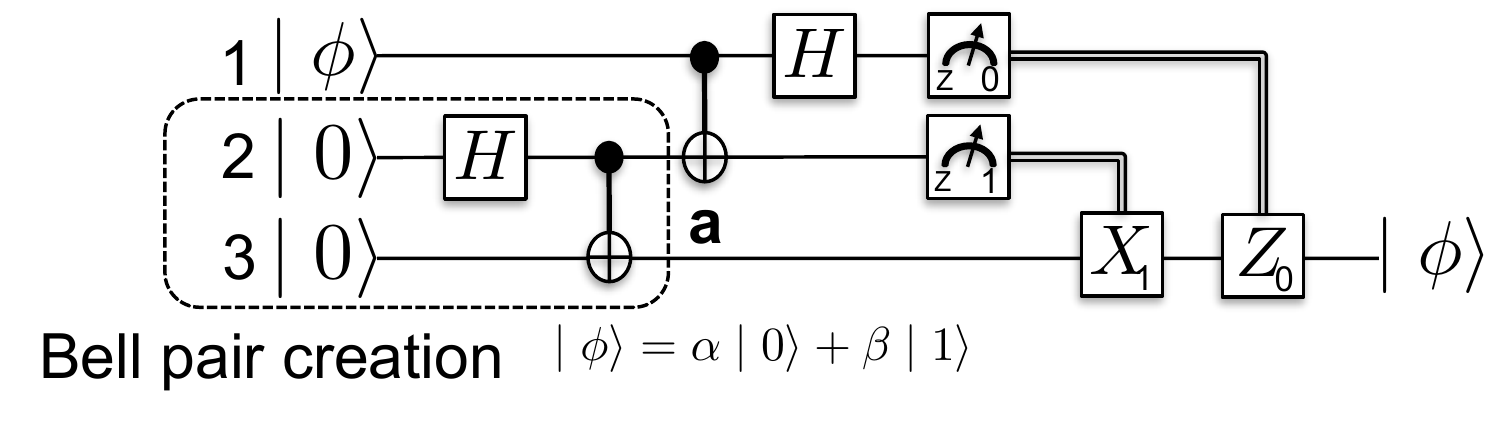}
\caption{Simple quantum circuit of quantum teleportation}
\label{Quantum_teleportation_circuit}
\end{figure}

Using gate operations, it is possible to teleport quantum information from one place to another. This is not only limited to close distance but also for long distances. The simplest circuit implementation for quantum teleportation is as shown in Figure \ref{Quantum_teleportation_circuit}.

After the Bell pair creation at \emph{a}, the two qubits may be separated physically.
As shown in the circuit, each measurement comes with a classical feedforward operation to the residual qubit, which is essential for completing the teleportation of an arbitrary quantum state.
Even though entangled particles always share physical properties regardless of the distance between them, the necessity of classical communication forbids the transmission of information from one place to another faster than the speed of light.

\begin{eqnarray}
H_{1} \Lambda_{1,2}(X)  \Lambda_{2,3}(X) H_{2} \ket{0_{3}} \ket{0_{2}} \ket{\Phi_{1}} =  H_{1} \Lambda_{1,2}(X)  \Lambda_{2,3}(X) H_{2} \ket{0_{3}0_{2}} \otimes \alpha \ket{0_{1}} + \beta \ket{1_{1}} \nonumber \\
= H_{1} \Lambda_{1,2}(X)  (\frac{1}{\sqrt{2}}(\ket{0_{3}0_{2}} + \ket{1_{2}1_{3}}) \otimes \alpha \ket{0_{1}} + \beta \ket{1_{1}}) \nonumber \\
= H_{1} (\frac{1}{\sqrt{2}}(\alpha \ket{0_{1}0_{2}0_{3}} + \alpha \ket{1_{1}1_{2}1_{3}} + \beta \ket{1_{1}0_{2}0_{3}} + \beta \ket{1_{1}1_{2}1_{3}}) \nonumber \\
= (\frac{1}{2}(\alpha \ket{0_{1}0_{2}0_{3}} + \alpha \ket{1_{1}0_{2}0_{3}} + \alpha \ket{0_{1}1_{2}1_{3}} + \alpha \ket{1_{1}1_{2}1_{3}} \nonumber \\
- \beta \ket{1_{1}1_{2}0_{3}} + \beta \ket{0_{1}1_{2}0_{3}} - \beta \ket{1_{1}0_{2}1_{3}} + \beta \ket{0_{1}0_{2}1_{3}}) \nonumber \\
\end{eqnarray}

Depending on the measurement results, byproduct operations are applied to the remaining qubit to complete the teleportation. For details, see Table \ref{table_quantum_teleportation}. Each outcome can be found with equal probability of 25\%.

\begin{table}[H]
\begin{center}
	\caption{Byproduct operations to complete quantum teleportation}
	\label{table_quantum_teleportation}
	\begin{tabular}{|c||l|c|}
		\hline
		Measurement result & Output state & Byproduct operation \\
		\hline
		\hline
		$0_{1}0_{2}$ & $\alpha \ket{0_{3}} + \beta \ket{1_{3}}$ & I \\
		\hline
		$0_{1}1_{2}$  & $\alpha \ket{1_{3}} + \beta \ket{0_{3}}$ & X \\
		\hline
		$1_{1}0_{2}$  & $\alpha \ket{0_{3}} - \beta \ket{1_{3}}$ & Z \\
		\hline
		$1_{1}1_{2}$  & $\alpha \ket{1_{3}} - \beta \ket{0_{3}}$ & XZ \\
		\hline
	\end{tabular}
\end{center}
\end{table}

\begin{figure}[H]
\center
\includegraphics[keepaspectratio,scale=0.7]{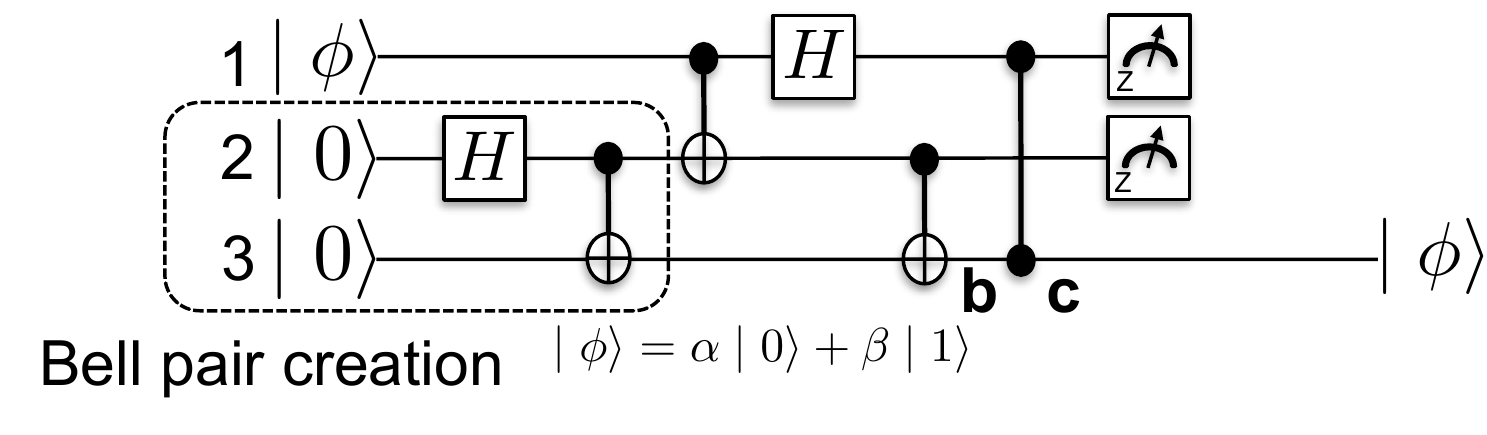}
\caption{Simple quantum circuit of quantum teleportation without byproduct operations}
\label{Quantum_teleportation_circuit_2}
\end{figure}

Notice that when the measurement result of qubit 2 is 1, there is always a bit-flip error on the remaining qubit.
Similarly, when the measurement result of qubit 1 is 1, Z gate is must be applied to qubit 3 as a byproduct to fix the phase.
Thus, qubit 3's state can be manipulated beforehand to avoid any byproduct operation after the measurement as in Figure \ref{Quantum_teleportation_circuit_2}. The communication speed is still not faster than the speed of light as qubit 3, which is a part of the Bell pair, needs to be sent to another node to establish a long distance communication.

\begin{figure}[!hbt]
\center
\includegraphics[keepaspectratio,scale=0.7]{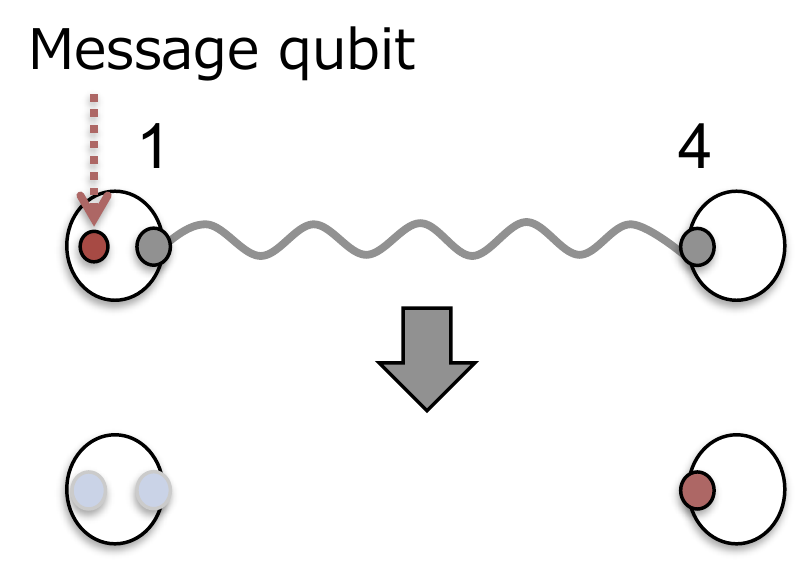}
\caption[Visualized model of quantum teleportation]{Visualized model of quantum teleportation. At the end of the operation, the red qubit has been recreated on the right.}
\label{quantum_teleportation_image}
\end{figure}

If the qubits of the Bell pair are separated, \emph{b} or \emph{c} will require long-distance multi-qubit operations.
The visualized model of quantum teleportation is shown in Figure \ref{quantum_teleportation_image}.

\subsection{Entanglement Swapping}

\begin{figure}[!hbt]
\center
\includegraphics[keepaspectratio,scale=0.7]{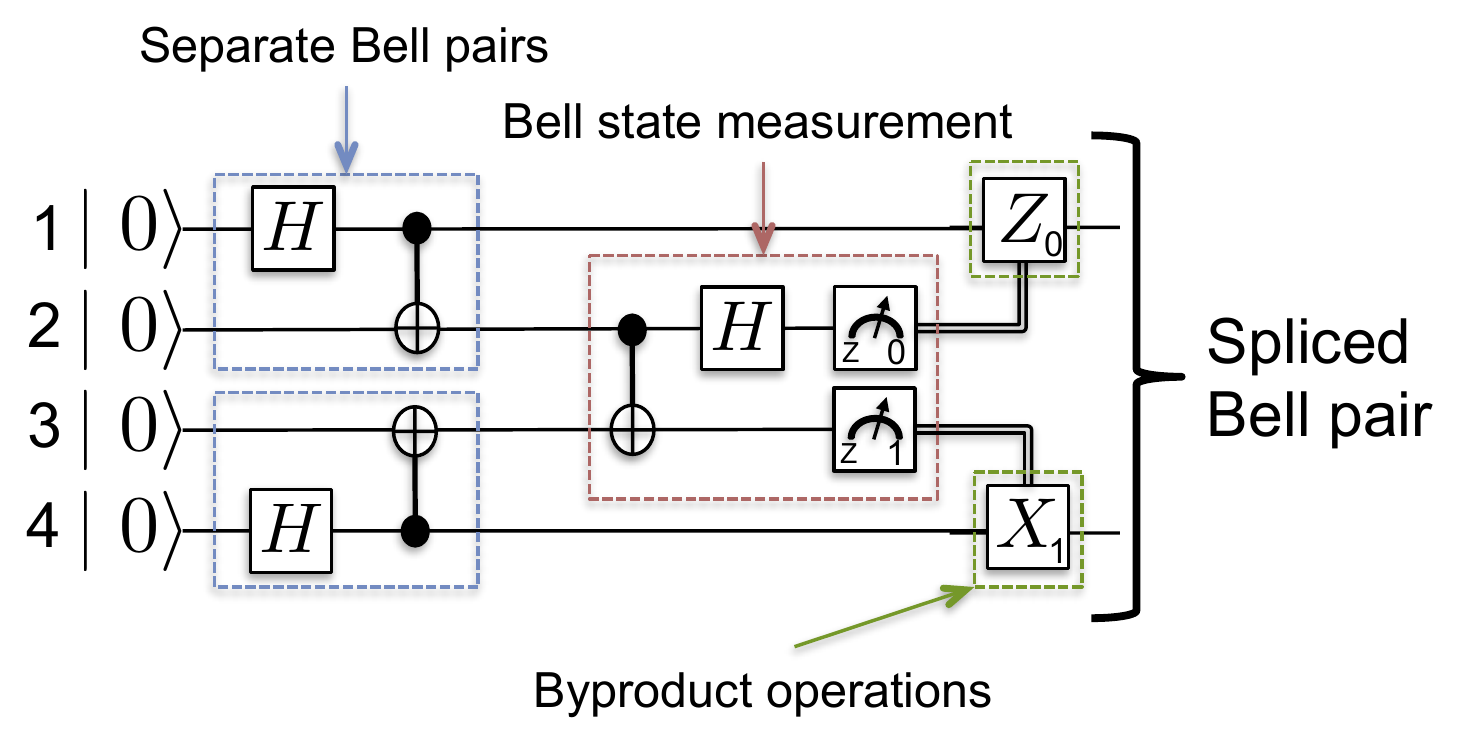}
\caption{Simple quantum circuit of entanglement swapping}
\label{2_Entanglement_swapping}
\end{figure}

The quantum teleportation technique introduced above is an example of transmitting one bit of quantum information from one place to another but networking often requires multi-hop communication. The simplest solution for such demand is to apply quantum teleportation hop-by-hop. Nevertheless, operating on the message qubit directly many times degrades the information. Another solution is the use of {\it entanglement swapping} \cite{Zukowski1993}, which is capable of lengthening the Bell pair to allow a direct teleportation of quantum information over multiple repeaters. As shown in Figure \ref{2_Entanglement_swapping}, entanglement swapping is based on the teleportation circuit that was introduced above and in Figure \ref{Quantum_teleportation_circuit}. In the below example, 2 Bell pairs are consumed to output 1 end-to-end Bell pair.

Subscripts of operators are identifiers for clarifying the interconnection between the measurement operator and its byproduct operator.

\begin{eqnarray}
		H_{2}\Lambda_{2,3}(X)\Lambda_{4,3}(X)\Lambda_{1,2}(X) H_{4}H_{1}\ket{0_{4}}\ket{0_{3}}\ket{0_{2}}\ket{0_{1}} \nonumber \\
		= H_{2}\Lambda_{2,3}(X) \frac{1}{\sqrt{2}}(\ket{0_{1}0_{2}}+\ket{1_{1}1_{2}}) \otimes \frac{1}{\sqrt{2}}(\ket{0_{3}0_{4}}+\ket{1_{3}1_{4}}) \nonumber \\
		=  H_{2}\frac{1}{2}(\ket{0_{1}0_{2}0_{3}0_{4}}+\ket{0_{1}0_{2}1_{3}1_{4}}+\ket{1_{1}1_{2}1_{3}0_{4}}+\ket{1_{1}1_{2}0_{3}1_{4}}) \nonumber \\
		\frac{1}{2\sqrt{2}}(\ket{0_{1}0_{2}0_{3}0_{4}}+\ket{0_{1}1_{2}0_{3}0_{4}}+\ket{0_{1}0_{2}1_{3}1_{4}}+\ket{0_{1}1_{2}1_{3}1_{4}} \nonumber \\ -\ket{1_{1}1_{2}1_{3}0_{4}}+\ket{1_{1}0_{2}1_{3}0_{4}}-\ket{1_{1}1_{2}0_{3}1_{4}}+\ket{1_{1}0_{2}0_{3}1_{4}})
\end{eqnarray}

\begin{table}[H]
\begin{center}
	\caption{Byproduct operations to complete entanglement swapping}
	\label{table_entanglement_swapping}
	\begin{tabular}{|c||l|c|}
		\hline
		Measurement result & Output state & Byproduct operation \\
		\hline
		\hline
		$0_{2}0_{3}$ & $\frac{1}{\sqrt{2}}(\ket{0_{1}0_{4}} + \ket{1_{1}1_{4}})$ & I \\
		\hline
		$0_{2}1_{3}$  & $\frac{1}{\sqrt{2}}(\ket{0_{1}1_{4}} + \ket{1_{1}0_{4}})$ & $X_{4}$ \\
		\hline
		$1_{2}0_{3}$  & $\frac{1}{\sqrt{2}}(\ket{0_{1}0_{4}} - \ket{1_{1}1_{4}})$ & $Z_{1}$\\
		\hline
		$1_{2}1_{3}$  & $\frac{1}{\sqrt{2}}(\ket{0_{1}1_{4}} - \ket{1_{1}0_{4}})$ & $X_{4}Z_{1}$ \\
		\hline
	\end{tabular}
\end{center}
\end{table}

\begin{figure}[H]
\center
\includegraphics[keepaspectratio,scale=0.7]{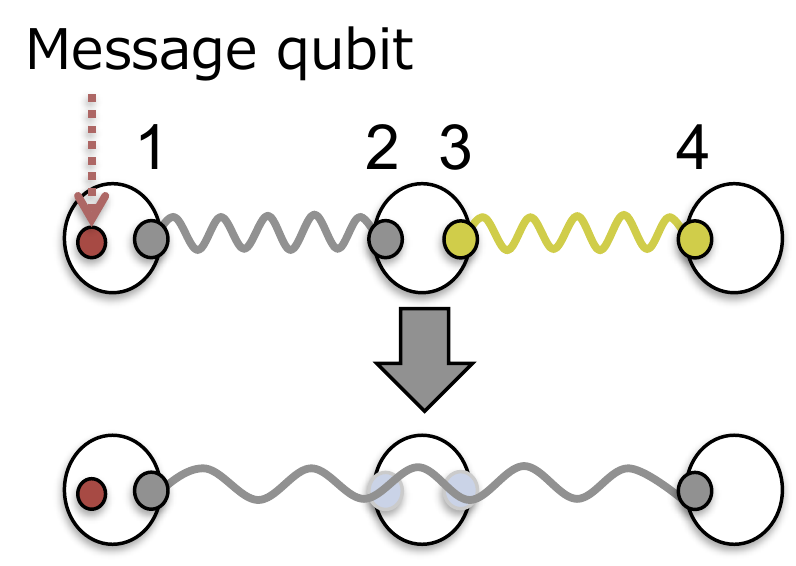}
\caption{Visualized model of entanglement swapping}
\label{entanglement_swapping_image}
\end{figure}

The byproduct operator, Z gate, can also be applied to qubit 4 instead of qubit 1. The visualized model is at Figure \ref{entanglement_swapping_image}.

\subsection{Entanglement Purification}
\label{PURIFICATIONsection}

\begin{figure}[H]
\center
\includegraphics[keepaspectratio,scale=0.7]{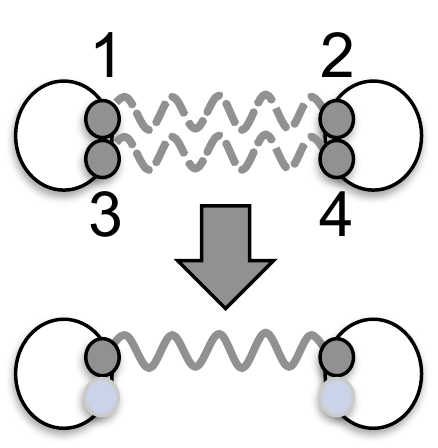}
\caption{Visualized model of entanglement purification}
\label{entanglement_purification}
\end{figure}

Using two or more less-entangled mixed pairs shared among nodes, it is possible to create one pair with a higher entanglement.
The easiest example of entanglement purification can be shown using 2 Bell pairs as in Figure \ref{entanglement_purification}.

In this example, two Bell pairs $\ket{\Phi_{1,2}^+}$, $\ket{\Phi_{3,4}^+}$ are shared between two nodes which are physically far apart, each holding half of the two pairs.
In order to create and check the parity of two qubits in each node, both nodes locally perform a CNOT operation and measure one of the pairs with respect to the Z-basis, which destroys one Bell pair.
If the measurement results agree, $0_{3}0_{4}$ or $1_{3}1_{4}$, the remaining pair will have a higher fidelity than the original pairs given $F_{input} > 50\%$. On the other hand, the whole resource needs to be discarded if the measurement results do not agree.

When Bell pairs are both perfect and gates have no noise $F_{input} = 1.00$ and $ F_{operation} = 1.00$:
\begin{eqnarray}
\Lambda_{2,4}(X)\Lambda_{1,3}(X)\ket{\Phi_{1,2}^+}\ket{\Phi_{3,4}^+} = \Lambda_{2,4}(X)\Lambda_{1,3}(X) \frac{1}{\sqrt{2}}(\ket{0_{1}0_{2}}) \otimes \frac{1}{\sqrt{2}}(\ket{0_{3}0_{4}}) \nonumber \\
=\Lambda_{2,4}(X)\Lambda_{1,3}(X)\frac{1}{2}(\ket{0_{1}0_{2}0_{3}0_{4}}+\ket{0_{1}0_{2}1_{3}1_{4}}+\ket{1_{1}1_{2}0_{3}0_{4}}+\ket{1_{1}1_{2}1_{3}1_{4}})\nonumber \\
=\Lambda_{2,4}(X)\frac{1}{2}(\ket{0_{1}0_{2}0_{3}0_{4}}+\ket{0_{1}0_{2}1_{3}1_{4}}+\ket{1_{1}1_{2}1_{3}0_{4}}+\ket{1_{1}1_{2}0_{3}1_{4}})\nonumber \\
=\frac{1}{2}(\ket{0_{1}0_{2}0_{3}0_{4}}+\ket{0_{1}0_{2}1_{3}1_{4}}+\ket{1_{1}1_{2}1_{3}1_{4}}+\ket{1_{1}1_{2}0_{3}0_{4}})\nonumber \\
= \ket{\Phi_{1,2}^+}\ket{\Phi_{3,4}^+}
\end{eqnarray}

When either Bell pair is imperfect, a bit-flip error on $\ket{\Phi^+}$ changes the state to $\ket{\Psi^+}$.
Tab.\ref{table_entanglement_swapping} is the summary of the behavior when $F_{output} > F_{input}$ and $F_{input} > 0.5$.

\begin{table}[H]
\begin{center}
	\caption{Entanglement swapping with noisy Bell pairs}
	\label{table_entanglement_swapping}
	\begin{tabular}{|c||l|c|c|}
		\hline
		Input state & Probability & Output state & Result\\
		\hline
		\hline
		$\ket{\Phi^+_{1,2}}\ket{\Phi^+_{3,4}}$ & $F^2$ & $\ket{\Phi^+_{1,2}}$ & True positive\\
		\hline
		$\ket{\Phi^+_{1,2}}\ket{\Psi^+_{3,4}}$ & $F(1-F)$ & - & True negative\\
		\hline
		$\ket{\Psi^+_{1,2}}\ket{\Phi^+_{3,4}}$ & $F(1-F)$ & - & False negative\\
		\hline
		$\ket{\Psi^+_{1,2}}\ket{\Psi^+_{3,4}}$ & $(1-F)^2$ & $\ket{\Psi^+_{1,2}}$ & False positive\\
		\hline
	\end{tabular}
\end{center}
\end{table}

With an assumption of two Bell pairs having the same fidelity, the probability of getting the same measurement results can be obtained by $F_{input}^2 + (1-F_{input})^2$, while the probability of it actually being right is $F_{input}^2$. Therefore, The relation between the input fidelity and the output fidelity after performing entanglement purification can be derived as:
\begin{equation}
 F_{output} = \frac{F_{input}^2}{F_{input}^2+(1-F_{input})^2}
\end{equation}

The iteration of the above steps can be used to create an entangled pair of arbitrarily high purity $F_{out} \rightarrow 1$.

\subsection{Quantum state tomography}
\label{TOMOGRAPHYsection}
Because measuring a qubit destroys the state, one cannot simply check the quality of a generated Bell pair.
Instead, a common methodology is to prepare many of the same states, measure them along different axes, and estimate the average quality of the measured states by reconstructing the density matrix from the measurement outcomes.
This process is called quantum tomography~\cite{2005AAMOP..52..105A}.

Given matrices $\sigma_i$, corresponding to our Pauli operators explained in subsection~\ref{single_qubit_gates}, as:

\begin{eqnarray}
\sigma_0 = \begin{bmatrix} 1 & 0 \\ 0 & 1 \end{bmatrix}\\
\sigma_1= \begin{bmatrix} 0 & 1 \\ 1 & 0 \end{bmatrix}\\
\sigma_2= \begin{bmatrix} 0 & -i \\ i & 0 \end{bmatrix}\\
\sigma_3= \begin{bmatrix} 1 & 0 \\ 0 & -1 \end{bmatrix},
\end{eqnarray}

an arbitrary n-qubit density matrix can be rewritten using Stokes parameters ($S$):

\begin{eqnarray}
\rho = \frac{1}{2^n} \sum_{i_1,i_2,...i_n=0}^{3} S_{i_1,i_2,...i_n} \sigma_{i_1} \otimes \sigma_{i_2} \otimes ...\sigma_{i_n} \\
s.t.~S_{i_1,i_2,...i_n} = \mathrm{Tr}[(\sigma_{i_1} \otimes \sigma_{i_2} \otimes ... \sigma_{i_n}) \rho].
\end{eqnarray}

We need a total of $4^n$ Stokes parameters for an n-qubit system.
Notice that each Stokes parameter requires the density matrix $\rho$ to be known for the calculation.
This is not accessible, unless for a theoretical calculation.
However, each of the parameters corresponds to the outcome of a specific pair of projective measurements.

\subsubsection{Single qubit tomography using projective measurements}

A single qubit density matrix, therefore, can be described as:

\begin{eqnarray}
\rho = \frac{1}{2} \sum_{i=0}^{3} S_{i} \sigma_{i}.
\end{eqnarray}

Stokes parameters for a single qubit tomography can be rewritten based on measurements as:

\begin{eqnarray}
S_0 = P(\ket{0}) + P(\ket{1})\\
S_1 = P(\frac{1}{\sqrt{2}}(\ket{0}+\ket{1})) - P(\frac{1}{\sqrt{2}}(\ket{0}-\ket{1}))\\
S_2 = P(\frac{1}{\sqrt{2}}(\ket{0}+i \ket{1})) - P(\frac{1}{\sqrt{2}}(\ket{0}-i \ket{1}))\\
S_3 = P(\ket{0}) - P(\ket{1}),
\end{eqnarray}

where $P(\ket{\Psi})$ is the probability to observe outcome state $\ket{\Psi}$, which mathematically is $P(\ket{\psi})=\bra{\Psi}\rho\ket{\Psi}=\mbox{Tr}[\ket{\Psi}\bra{\Psi}\rho]$.
In the real world, those probabilities can be directly accessed from the accumulated measurement results.
Notice that observing the state \{$\ket{0}, \ket{1}$\} requires a Z measurement.
Similarly, observing \{$\frac{1}{\sqrt{2}}(\ket{0}+\ket{1}),\frac{1}{\sqrt{2}}(\ket{0}-\ket{1})$\} requires an X measurement,
and \{$\frac{1}{\sqrt{2}}(\ket{0}+i \ket{1}), \frac{1}{\sqrt{2}}(\ket{0}-i \ket{1})$\} requires a Y measurement.
If one measured an unknown state along the Z axis 1000 times, in which 489 outcomes showed up as $\ket{0}$ and 511 as $\ket{1}$,
then $P(\ket{0})=0.489$ and $P(\ket{1})=0.511$. $S_0$ is, thus, always 1.
The notation used is similar to optics, although measurement process is different.

\subsubsection{Two-qubit tomography using projective measurements}

Similarly, a two-qubit density matrix can be described as:

\begin{eqnarray}
\rho = \frac{1}{4} \sum_{i_1,i_2=0}^{3} S_{i_1,i_2} \sigma_{i_1} \sigma_{i_2}.
\end{eqnarray}

Assuming that
\begin{eqnarray}
H = \ket{0}\\
V = \ket{1}\\
D = \frac{1}{\sqrt{2}}(\ket{0}+\ket{1})\\
A = \frac{1}{\sqrt{2}}(\ket{0}-\ket{1})\\
R = \frac{1}{\sqrt{2}}(\ket{0}+i\ket{1})\\
L = \frac{1}{\sqrt{2}}(\ket{0}-i\ket{1}),
\end{eqnarray}

then the 16 Stokes parameters can be expressed as:

\begin{eqnarray}
S_{0,0}=1\\
S_{0,1}=P(DD)-P(DA)+P(AD)-P(AA)\\
S_{0,2}=P(RR)-P(LR)+P(RL)-P(LL)\\
S_{0,3}=P(HH)-P(HV)+P(VH)-P(VV)\\
S_{1,0}=P(DD)+P(DA)-P(AD)-P(AA)\\
S_{1,1}=P(DD)-P(DA)-P(AD)+P(AA)\\
S_{1,2}=P(DR)-P(DL)-P(AR)+P(AL)\\
S_{1,3}=P(DH)-P(DV)-P(AH)+P(AV)\\
S_{2,0}=P(RR)+P(LR)-P(RL)-P(LL)\\
S_{2,1}=P(RD)-P(RA)-P(LD)+P(LA)\\
S_{2,2}=P(RR)-P(RL)-P(LR)+P(LL)\\
S_{2,3}=P(RH)-P(RV)-P(LH)+P(LV)\\
S_{3,0}=P(HH)+P(HV)-P(VH)-P(VV)\\
S_{3,1}=P(HD)-P(HA)-P(VD)+P(VA)\\
S_{3,2}=P(HR)-P(HL)-P(VR)+P(VL)\\
S_{3,3}=P(HH)-P(HV)-P(VH)+P(VV).
\end{eqnarray}

\subsection{Verification of quantum correlation (entangled states) via CHSH inequality}
\label{CHSHsection}
Whether qubits are entangled or not cannot be checked directly.
In stead, the presence of an entanglement can be verified via a statistical experiment called the \emph{CHSH inequality}, which is a similar technique to tomography.
The classical CHSH inequality assumes two nodes performing measurements {$\alpha=A(a,\lambda)$, $\alpha '=A(a',\lambda)$}, and {$B(b,\lambda)$, $B(b',\lambda)$} each with outcome {$\pm 1$}.
A and B are functions that take a variable \{a,a'\} and \{b,b'\} accordingly, which are adjustable parameters representing its orientation of measurement.
The variable $\lambda$ is the hidden variable.
The hidden variable is an unknown parameter determining the result of each of these experiments.
Given $p(\lambda )$ as the probability distribution determining the measurement outcome, the classical correlation between two measurements, $E$, can be gained as:

\begin{eqnarray}
E(a,b) = E(a)E(b) = \int \mathrm{d}\lambda p(\lambda) A(a,\lambda) B(b,\lambda) = \braket{\alpha \beta}.
\end{eqnarray}

Because each measurement outputs either +1 or -1,

\begin{eqnarray}
\begin{dcases}
\alpha+\alpha ' = A(a,\lambda) + A(a',\lambda) =0 \\ \alpha-\alpha '  = A(a,\lambda) - A(a',\lambda) =\pm 2 ,& \alpha,\alpha ' \rightarrow \{1,-1\} \mbox{{ }or{ }} \{-1,1\},\\
\\ \alpha+\alpha ' = A(a,\lambda) + A(a',\lambda) =\pm 2 \\  \alpha-\alpha '  = A(a,\lambda) - A(a',\lambda) =0 ,& \alpha,\alpha ' \rightarrow \{-1,-1\} \mbox{{ }or{ }} \{1,1\}
\end{dcases},\\
\end{eqnarray}

in which case the following satisfies for a single measurement:

\begin{eqnarray}
C \equiv  (\alpha+\alpha ')\beta + (\alpha-\alpha ')\beta ' = \pm 2.
\end{eqnarray}

The classical correlation, CHSH inequality, is:

\begin{eqnarray}
\mid \braket{\alpha \beta} - \braket{\alpha \beta '} \mid  + \mid \braket{\alpha ' \beta} + \braket{\alpha ' \beta '}\mid  \leq 2.
\end{eqnarray}

This inequality, however, can be violated in quantum mechanics using entanglement.
An example of an entangled state is:

\begin{eqnarray}
\ket{\psi} = \frac{1}{\sqrt{2}}(\ket{00}+\ket{11}).
\end{eqnarray}

The degree of its violation depends on the measurement orientations, where the settings for accomplishing the maximum violation are:

\begin{eqnarray}
\alpha = \frac{\pi}{2}\\
\alpha ' = 0 \\
\beta = \frac{\pi}{4}\\
\beta ' = \frac{-\pi}{4}\\.
\end{eqnarray}

Here, each angle is adjusted over the ZX plane of the Bloch sphere (see subsection~\ref{blochsphere}).
Hence, the quantum correlation is:

\begin{eqnarray}
E(a,b) = \bra{\psi}\hat{\sigma_{\alpha}}\otimes\hat{\sigma_{\beta}} \ket{\psi} = \cos{(\alpha - \beta )}= \braket{\alpha \beta}.
\end{eqnarray}

Each term of the CHSH inequality will therefore be:

\begin{eqnarray}
\braket{\alpha \beta} = \braket{\alpha ' \beta} = - \braket{\alpha \beta '} = \braket{\alpha ' \beta '} = \frac{1}{\sqrt{2}}.
\end{eqnarray}

Thus, the quantum correlation takes a value between:

\begin{eqnarray}
2 < \mid \braket{\alpha \beta} - \braket{\alpha \beta '} \mid  + \mid \braket{\alpha ' \beta} + \braket{\alpha ' \beta '}\mid  \leq 2\sqrt{2}.
\end{eqnarray}

The noise on qubits, of course, decreases the calculated value.
Because the probability of getting $\pm$ 1 measurement outcome depends on the measurement angles,
we must choose the measurement angle at random.
Classical correlation by pre-arranged photon states, therefore, can never violate the CHSH inequality.
But the measurement basis must not be chosen before the photons are emitted.
Those constrains that affect the validity of the experiment are known as the \emph{loopholes}.
Some of the major loopholes are summarized in the Wikipedia~\cite{LoopHole}.

\subsection{Quantum Repeaters}

\begin{figure}[!hbt]
\center
\includegraphics[keepaspectratio,scale=0.4]{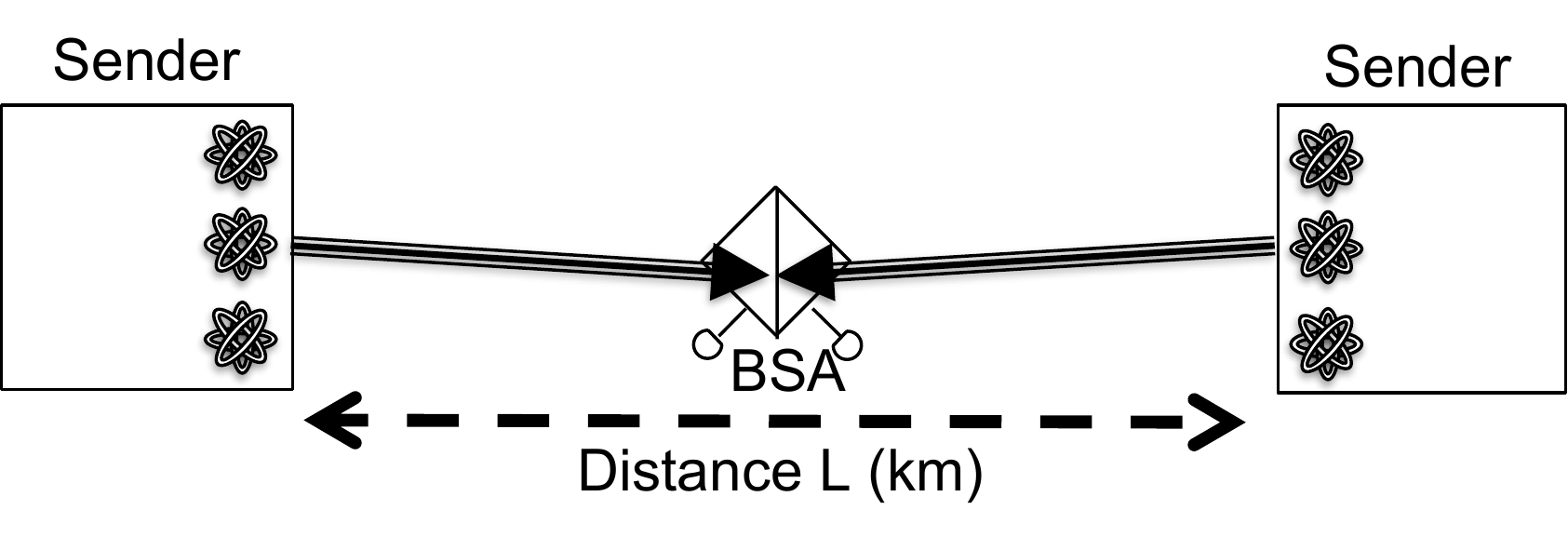}
\caption{Architecture of a simple two-node quantum repeater network.}
\label{Quantum_repeater_network}
\end{figure}

\begin{figure}[H]
\center
\includegraphics[keepaspectratio,scale=0.36]{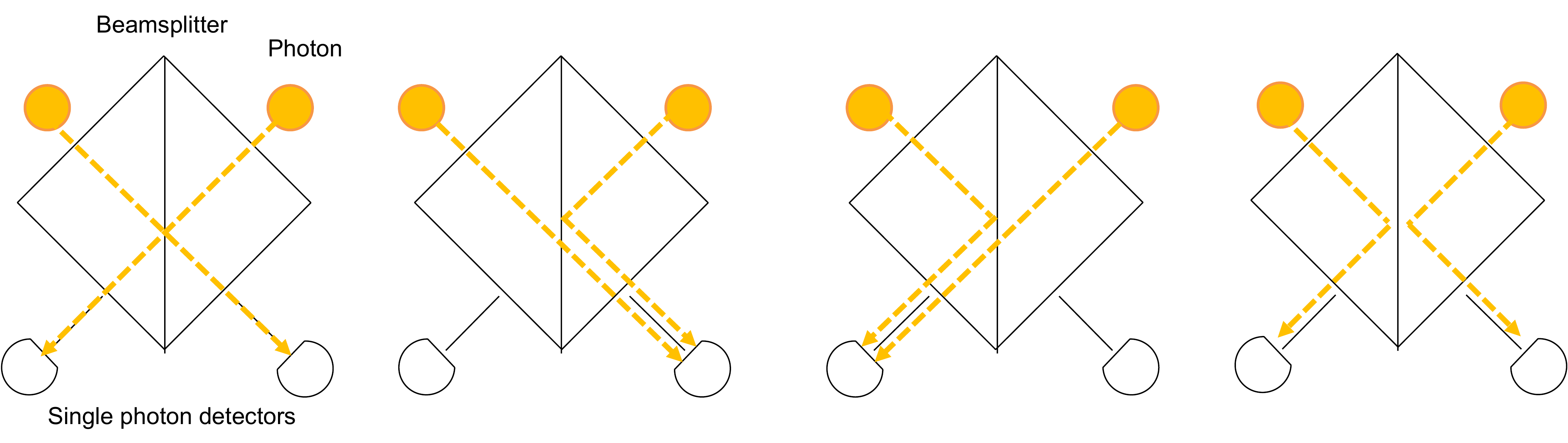}
\caption{Bell measurement using a beam splitter and photon detectors.}
\label{BellMeasurement}
\end{figure}

A quantum repeater network is a system of connected quantum repeater nodes (see Fig.~\ref{Quantum_repeater_network}).
The abstract architecture of a simple repeater network consists of 6 important elements.
The first element is the memory qubits, or sometimes called the stationary qubits.
Memory qubits are physically fixed at each node, and used for storing quantum information.
A quantum repeater network requires two types of channels.
The first channel is called the quantum channel, which is used for transmitting quantum information.
The other one is called the classical channel, which is used for transmitting ordinary classical packets.
A memory qubit emits an entangled optical qubit through the quantum channel, towards the Bell State Analyzer (BSA).
The BSA's role is to create an entangled state of memory qubits, by performing Bell measurements on received optical qubits (see Fig.~\ref{BellMeasurement}).
Entangled states shared between nodes are commonly, but not limited to, $\ket{\Phi^+}$ Bell pairs.


Nodes emit photons that are entangled with its source memory qubit.
For each attempt, two photons arrive at the BSA, one from the left and another from the right.
The polarizing beamsplitter functions in a similar way as a polarizer, except it reflects the photon instead of blocking.
When they arrive at the beamsplitter, they interfere in the middle.
Detector outcomes are summarized in Tab.~\ref{table_BellMeasurement}.
As shown, when two detectors click, you cannot tell whether the two photons had states $\ket{0_L 0_R}$ or $\ket{1_L 1_R}$, which in quantum mechanics, makes it a superposition state.
The state of the photon immediately gets destroyed by the measurements, but preserves the entanglement of the source memory qubits.

\begin{table}[H]
 \begin{center}
	 \caption{Bell measurement outcomes. The horizontal polarization $\ket{H}$ may be used for representing $\ket{0}$, and $\ket{V}$ for $\ket{1}$. The subscript \emph{L} denotes the \emph{Left photon}, and \emph{R} denotes the \emph{Right photon}.}
	 \label{table_BellMeasurement}
	 \begin{tabular}{|c|c|c|}
		 \hline
			Left detector count & Right detector count & State \\
		 \hline
		 \hline
		 True & False & $\ket{0_L}\ket{1_R}$ \\
		 \hline
		 False & True & $\ket{1_L}\ket{0_R}$\\
		 \hline
		 True & True & $\ket{0_L}\ket{0_R}$ || $\ket{1_L}\ket{1_R}$\\
		 \hline
	 \end{tabular}
 \end{center}
\end{table}

\subsection{Generations of Network Evolution}

Muralidharan et al. categorized quantum repeaters, based on capabilities, into three generations~\cite{Muralidharan2016}.
The 1st generation quantum repeater network works based on purification and entanglement swapping, for detecting and discarding erroneous resources and to connect non-adjacent nodes.
While this scheme is relatively simple and straightforward, its capability could strictly depend on the distance,
mainly limited by classical latencies for receiving acknowledgements regarding purification and entanglement swapping.
Its performance is also known to be limited by memory lifetime~\cite{hartmann06}.
The 2nd generation utilizes encoded Bell pairs prepared between adjacent nodes,
and performs quantum error correction~\cite{PhysRevA.79.032325, PhysRevLett.104.180503, 1367-2630-15-2-023012}.
Swapping is done at the logical level for Jiang's architecture~\cite{PhysRevA.79.032325}, via an extended surface for Fowler's~\cite{PhysRevLett.104.180503}.
In the 3rd generation, quantum states are directly encoded to a block of physical qubits that will be sent through the channel.
The receiver node can correct errors using the received physical qubits.
The 3rd generation is very similar to the 2nd generation, but requires a very high success rate of photon reception through all of the physical elements of the channel.
These two generations are semantically identical but the temporal behavior of the 3rd generation is more like classical packet forwarding networks.

\section{Errors on quantum systems and its measure}

\subsection{Error}
\label{ERRORsection}
Quantum gates are inherently noisy, and may cause errors while operating on qubits.

\subsubsection{Pauli errors}

In general, there are two types of Pauli errors that should be taken into consideration,
the bit-flip (X) error and the phase (Z) error -- the Y error is a combination of both errors.

An X error, for example, can be seen as an undesired X gate applied to the system.
For instance, assuming a single qubit ideal system in state $\ket{\psi_i}=\ket{0_A}$,
an X error on the actual system ($\ket{\psi_a}$) is:

\begin{eqnarray}
\ket{\psi_a}=X_{A}.\ket{\psi_i}=X_{A}.\ket{0_A} = \ket{1_A}
\end{eqnarray}

There are also cases where an error does not impact the state at all.
An example can be seen with the same ideal state, $\ket{\psi_i}=\ket{0_A}$, and a Z error on it.

\begin{eqnarray}
\ket{\psi_a}=Z_{A}.\ket{\psi_i}=Z_{A}.\ket{0_A} = \ket{0_A}
\end{eqnarray}

When a particular error (or an intentional unitary operation) does not influence the state,
the error is a \emph{stabilizer} of the state (for details regarding stabilizers, see section~\ref{stabilizersection}).
In this particular example, the Z operator is a stabilizer of state $\ket{0}$.

As mentioned above, the Y error is a combination of both X and Z error.
Thus, a Y error on $\ket{\psi_i}=\ket{0_A}$ is:

\begin{eqnarray}
\ket{\psi_a}=Y_{A}.\ket{\psi_i}=Y_{A}.\ket{0_A} = \ket{1_A}.
\end{eqnarray}

Naturally, the Z error part will not be physically observable, and what's left on the output is solely an X error.

\subsubsection{Propagation of Pauli errors}

Equally important, errors may propagate through quantum circuits throughout the operation.
The Hadamard gate converts bit-flip error to phase error, and phase error to bit-flip error.

\begin{figure}[H]
\center
\includegraphics[keepaspectratio,scale=0.6]{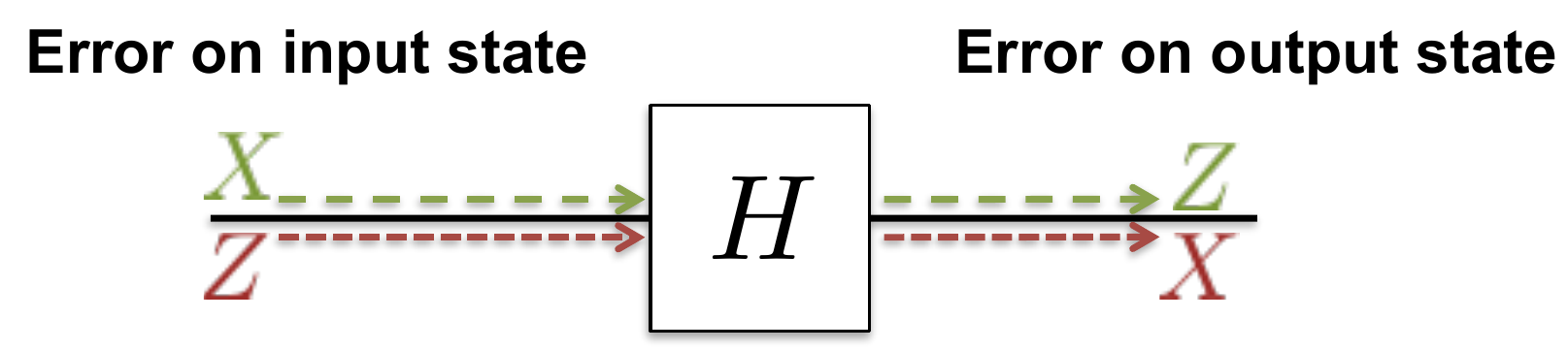}
\caption{Error propagation of Hadamard gate}
\label{H_prop}
\end{figure}

The CNOT operation will result in a propagation of the bit-flip error on the control qubit, ending up with bit-flip errors on both qubits. Similarly, a phase error on the target qubit will be transferred, after application of CNOT gate, to the control qubit.

\begin{figure}[H]
\center
\includegraphics[keepaspectratio,scale=0.6]{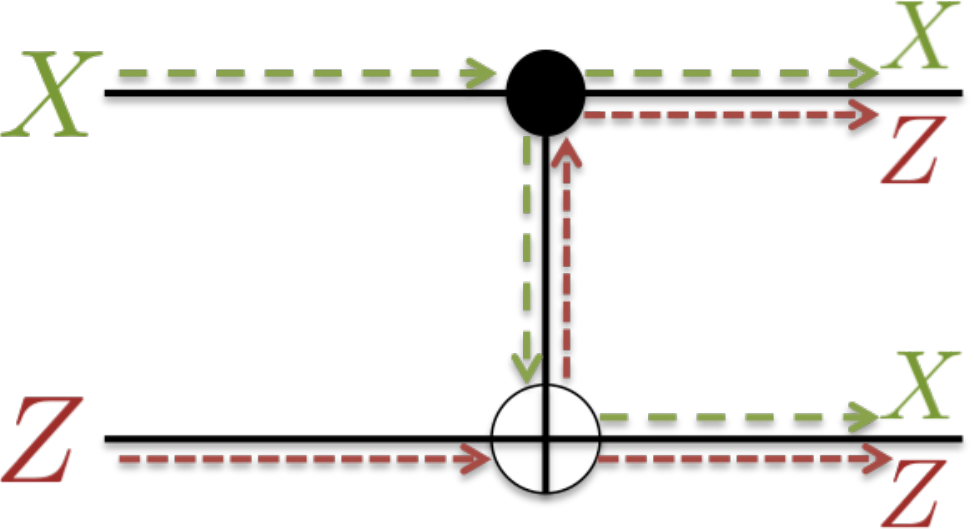}
\caption{Error propagation of Controlled-Not gate}
\label{CNOT_prop}
\end{figure}

Similarly, the bit-flip error on the control qubit will result in a bit-flip error on the target qubit and a phase error on the control qubit after the CZ operation. Furthermore, a bit-flip error on the target qubit will result in a bit-flip error on the target qubit and a phase error on the control qubit.

\begin{figure}[H]
\center
\includegraphics[keepaspectratio,scale=0.6]{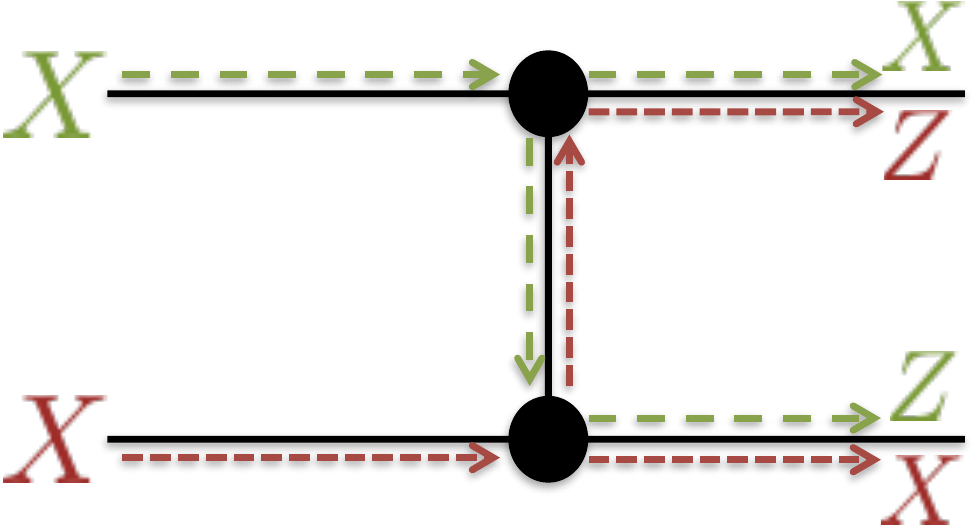}
\caption{Error propagation of Controlled-Z gate}
\label{CZ_prop}
\end{figure}

The measurement outcome depends on the quantum state, and thus a measurement error leads to an erroneous feedforward operation.
Therefore measurement errors in MBQC, explained in section~\ref{MBQCsection}, also propagate to other qubits through byproduct operations.

\begin{figure}[H]
\center
\includegraphics[keepaspectratio,scale=0.6]{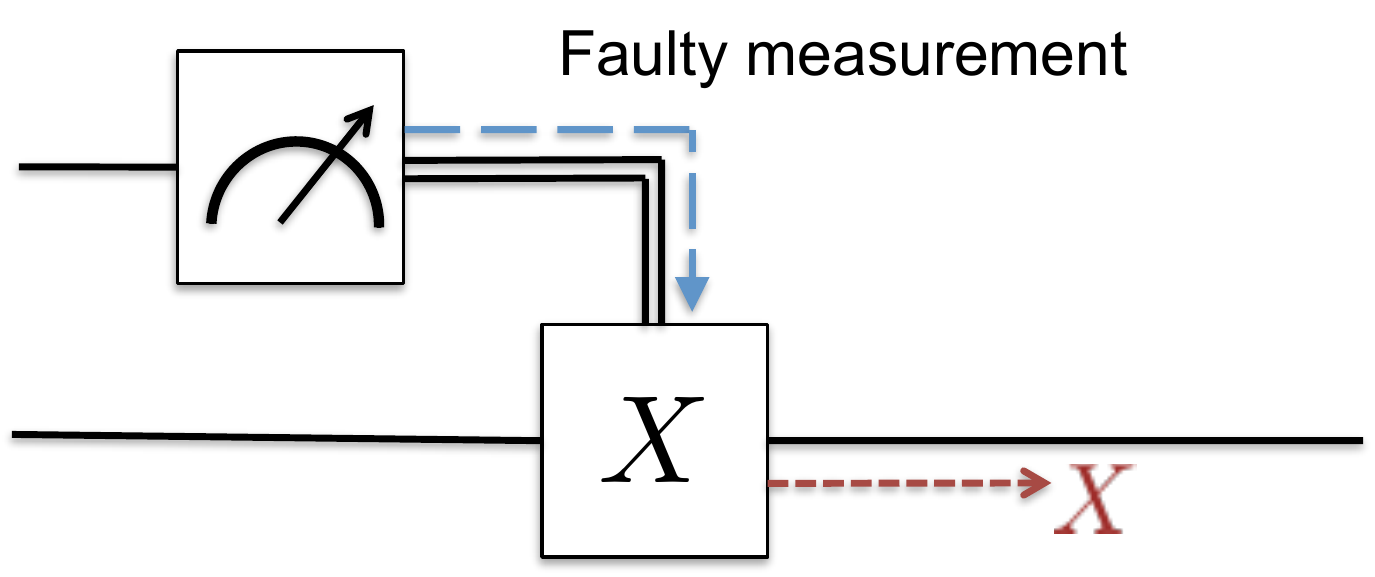}
\caption{Error propagation of measurement}
\label{CZ_prop}
\end{figure}

\subsubsection{Other unfixable errors}

In general, Pauli errors can be detected and discarded through purification (see section~~\ref{PURIFICATIONsection}), or corrected through error correction.
However, there exists a type of errors that cannot be fixed.
The completely mixed state is of course an example of this.
Other examples include qubit memory excitation/relaxation error.

A single qubit may hold an arbitrary quantum information $\ket{\psi} = \alpha \ket{0} + \beta \ket{0}$.
The memory excitation error is a type of error where the state $\alpha \ket{0} + \beta \ket{0}$ is forced to make a transition to $\ket{1}$ state due to external energy absorbed from the environment.
Similarly, the relaxation error happens when the qubit unintentionally loses energy, and is forced to move to the ground state ($\ket{0}$ state).

\subsection{Fidelity}

The imperfection of a quantum state can be described by the $fidelity$. The fidelity is often defined as~\cite{Jozsa1994}:

\begin{equation}
	F = \expval{\rho_a}{\psi} = \mbox{Tr}[\rho_a \rho_i]
\end{equation}

where $0 \leq F \leq 1$, $\ket{\psi}$ is the desired ideal state vector, $\rho_i$ is the desired ideal density matrix and $\rho_a$ is the density matrix of the actual state.
With $F=1$, the actual state is identical to the desired state.
A single qubit system has a fidelity of $50\%$ with a completely mixed state.
Similarly, in n-qubit system, a completely mixed state has a fidelity of $F=\frac{1}{2^n}$.

The same fidelity, however, may imply different states.
For instance, when the actual state ($\rho_a$) is completely mixed and the assuming ideal state is $\ket{\Phi^+}$ with a density matrix $\rho_i$, then the fidelity is:

\begin{eqnarray}
\rho_a = \begin{bmatrix}
		\frac{1}{2} & 0 & 0 & 0 \\
		0 & 0 & 0 & 0\\
		0 & 0 & 0 & 0\\
		0 & 0 & 0 & \frac{1}{2}
\end{bmatrix}\\
\rho_i =  \begin{bmatrix}
		\frac{1}{2} & 0 & 0 & \frac{1}{2} \\
		0 & 0 & 0 & 0\\
		0 & 0 & 0 & 0\\
		\frac{1}{2} & 0 & 0 & \frac{1}{2}
\end{bmatrix}\\
F = \mbox{Tr}[\rho_a \rho_i] = 1/2=0.5
\end{eqnarray}

But the same fidelity can be observed when the state is 50\% $\ket{\Phi^+}$ and 50\% $\ket{\Psi^+}$.
The former example of $F=0.5$ is a case where none of the systems are actually entangled but completely mixed.
The latter example with the same fidelity, however, is a case where all of the systems are entangled, but with a 50\% chance of having an X error.
Hence, $F=1-P(\mbox{error\ rate})$ is only true under a system with only Pauli errors.

\section{Measurement-based Quantum Computing}
\label{MBQCsection}

\begin{figure}[H]
\center
\includegraphics[keepaspectratio,scale=0.4]{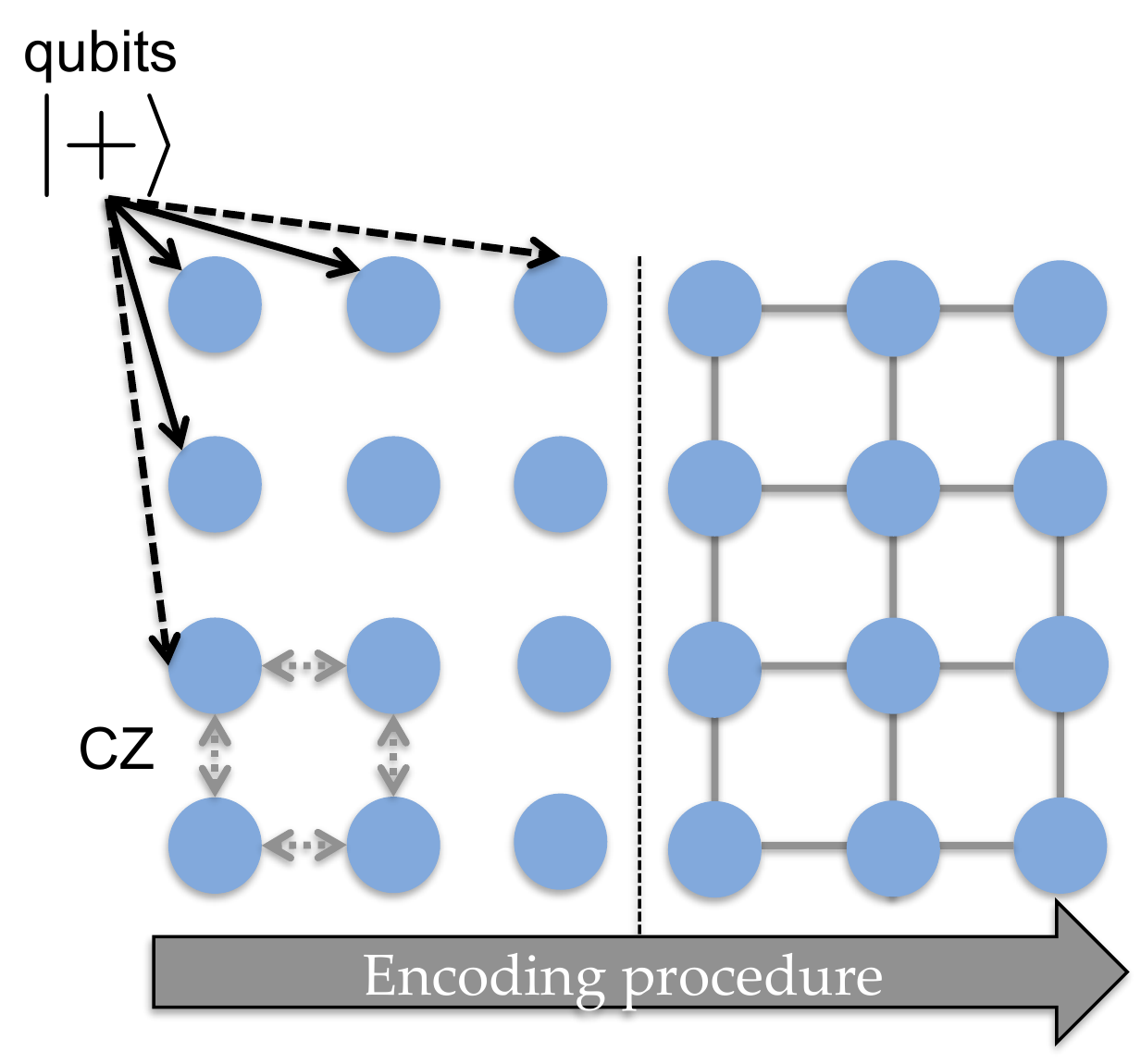}
\caption{2D resource state generation}
\label{grid}
\end{figure}

The Measurement-based Quantum Computing (MBQC) is the basis of Blind computing, which is an important application of the quantum Internet.
MBQC is an alternative universal computation method based on single qubit measurements that was proposed by Raussendorf, Browne and Briegel in 2003 \cite{Raussendorf2003} -- also known as one-way quantum computing.
Unlike the circuit model, the scheme of MBQC generally requires a two dimensional grid of qubits that are initialized as $\ket{+}$ and entangled with all neighboring qubits using CZ gates, as a cluster state (see Figure \ref{grid}).
The initialized cluster state for MBQC is also called the {\it resource state}.

The following set of abilities sufficiently provide enough capabilities for universal computing.
First is to be able to prepare qubits.
Second is to be able to perform Hadamard gates to each of them.
Third is to be able to perform CZ gates between neighboring qubits.
And last is to be able to measure each of the qubit along an arbitrary axis.

\begin{figure}[!hbt]
\center
\includegraphics[keepaspectratio,scale=0.6]{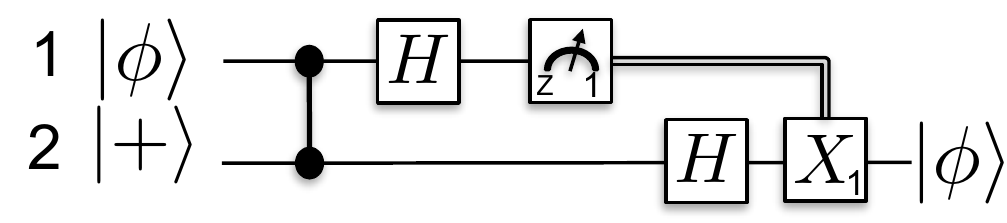}
\caption{Quantum circuit for one-bit teleportation}
\label{one-bit_teleportation}
\end{figure}

The technique to map an unknown quantum state from one qubit to another is known as {\it one-bit teleportation} (see Figure \ref{one-bit_teleportation}) -- introduced by Zhou, Leung and Chuang in 2000 \cite{Zhou2000}.

The calculation using state-vector form is:

\begin{eqnarray}
H_{1}\Lambda_{1,2}(Z)\ket{\phi_{1}}\ket{+_{2}} =  H_{1}\Lambda_{1,2}(Z)\alpha \ket{0}+\beta\ket{1} \otimes \ket{+_{2}} \nonumber \\
=H_{1} \alpha \ket{0_{1}+_{2}} + \beta \ket{1_{1}-_{2}} \nonumber \\
=\alpha \ket{+_{1}+_{2}} + \beta \ket{-_{1}-_{2}} \nonumber \\
\end{eqnarray}

After the measurement operations, the residual state ends up in either output state in Table \ref{table_entanglement_swapping}.

\begin{table}[H]
 \begin{center}
	 \caption{Byproduct operation to complete one-bit teleportation}
	 \label{table_entanglement_swapping}
	 \begin{tabular}{|c||l|c|}
		 \hline
		 Measurement result & Output state & Byproduct operation \\
		 \hline
		 \hline
		 $0_{1}$ & $\alpha \ket{+}+\beta \ket{-}$ & ($H_{2})I$ \\
		 \hline
		 $1_{1}$  & $\alpha \ket{+}-\beta \ket{-}$ & $(H_{2})X_{2}$ \\
		 \hline
	 \end{tabular}
 \end{center}
\end{table}

The X gate in this case behaves like a Z gate on $\ket{0}, \ket{1}$ basis.
The Hadamard gate is used to convert the state $\ket{+}$ to $\ket{0}$ and $\ket{-}$ to $\ket{1}$.

\begin{figure}[!hbt]
\center
\includegraphics[keepaspectratio,scale=0.6]{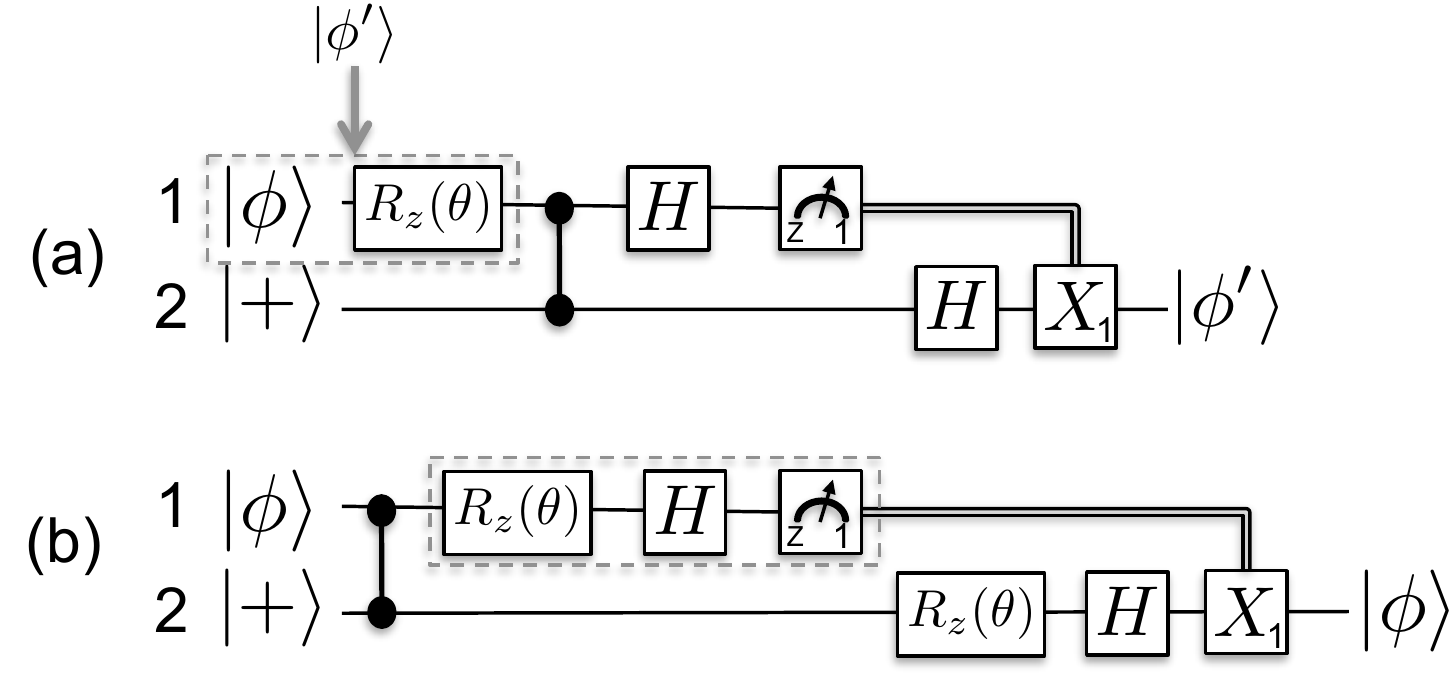}
\caption{One-bit teleportation to Measurement-based Quantum Computing}
\label{one-bit_teleportation_Z}
\end{figure}

Inserting a $R_{Z}(\theta)$ operation to the one-bit teleportation circuit can simply be accomplished by adding the gate just before the CZ gate (shown Figure \ref{one-bit_teleportation_Z}(a)). As $R_{Z}(\theta)\ket{\phi}$ can be considered as state $\ket{\phi'}$, the circuit can be directly related to the circuit in Figure \ref{one-bit_teleportation}.
However, as rotations about the Z-axis commute with the CZ operation, the circuit can be rewritten as Figure \ref{one-bit_teleportation_Z}(b).
The set of operations on qubit 1 after the CZ gate can be considered to be a single measurement operator in a particular basis, and the overall protocol can be simplified to a collection of CZ gates and measurement operations.
Finally, the computation depends on the measurement basis, which specifies the $\theta$.

An arbitrary single qubit unitary operation $U$ can be decomposed into Euler angles:

\begin{equation}
U= H e^{i\theta Z} e^{i\phi X} e^{i\xi Z}  = H e^{i\theta Z}H e^{i\phi Z} H e^{i\xi Z}
\end{equation}

e.g., an X gate can also be described by two Hadamard gates and a Z gate $X=HZH$. Therefore, an arbitrary single qubit unitary operation can be represented by a sequence of one-bit teleportations.

\begin{figure}[!hbt]
\center
\includegraphics[keepaspectratio,scale=0.6]{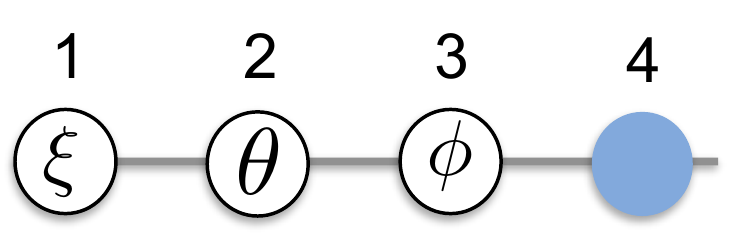}
\caption{Measurement-based single qubit unitary operation}
\label{single-unitary_MBQC}
\end{figure}

The realization of the CNOT operation can be accomplished by a 2-dimensional sequence of measurements as shown in Figure \ref{MBQCvsCM} below. Z-basis measurements are performed beforehand to omit unwanted qubits from the graph.

\begin{figure}[!hbt]
\center
\includegraphics[keepaspectratio,scale=0.4]{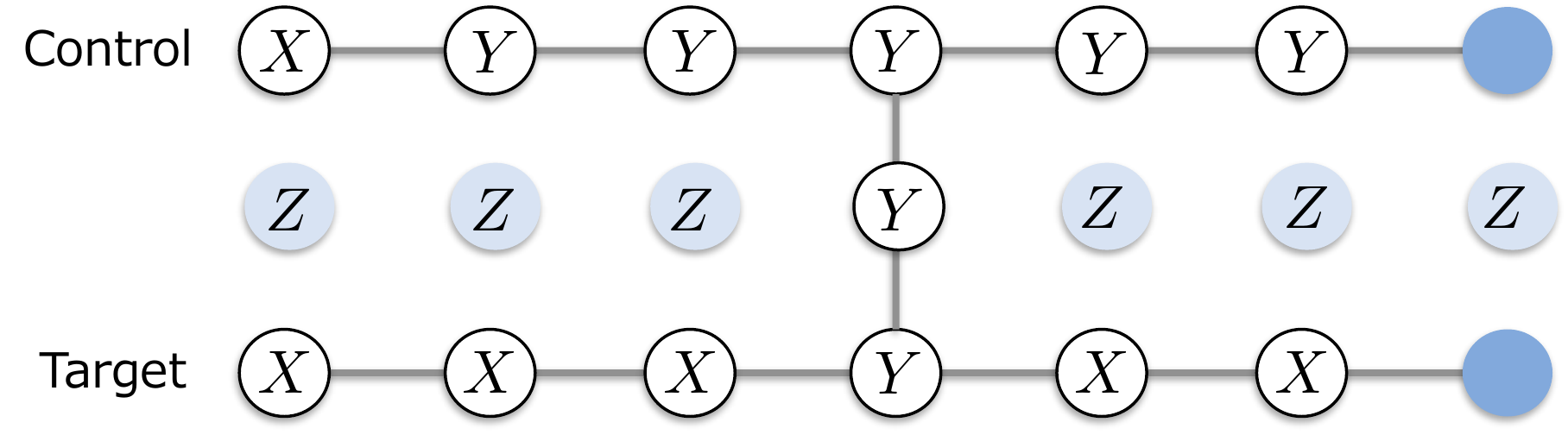}
\caption{Measurement-based CNOT operation}
\label{MBQCvsCM}
\end{figure}

\begin{figure}[!hbt]
\center
\includegraphics[keepaspectratio,scale=0.5]{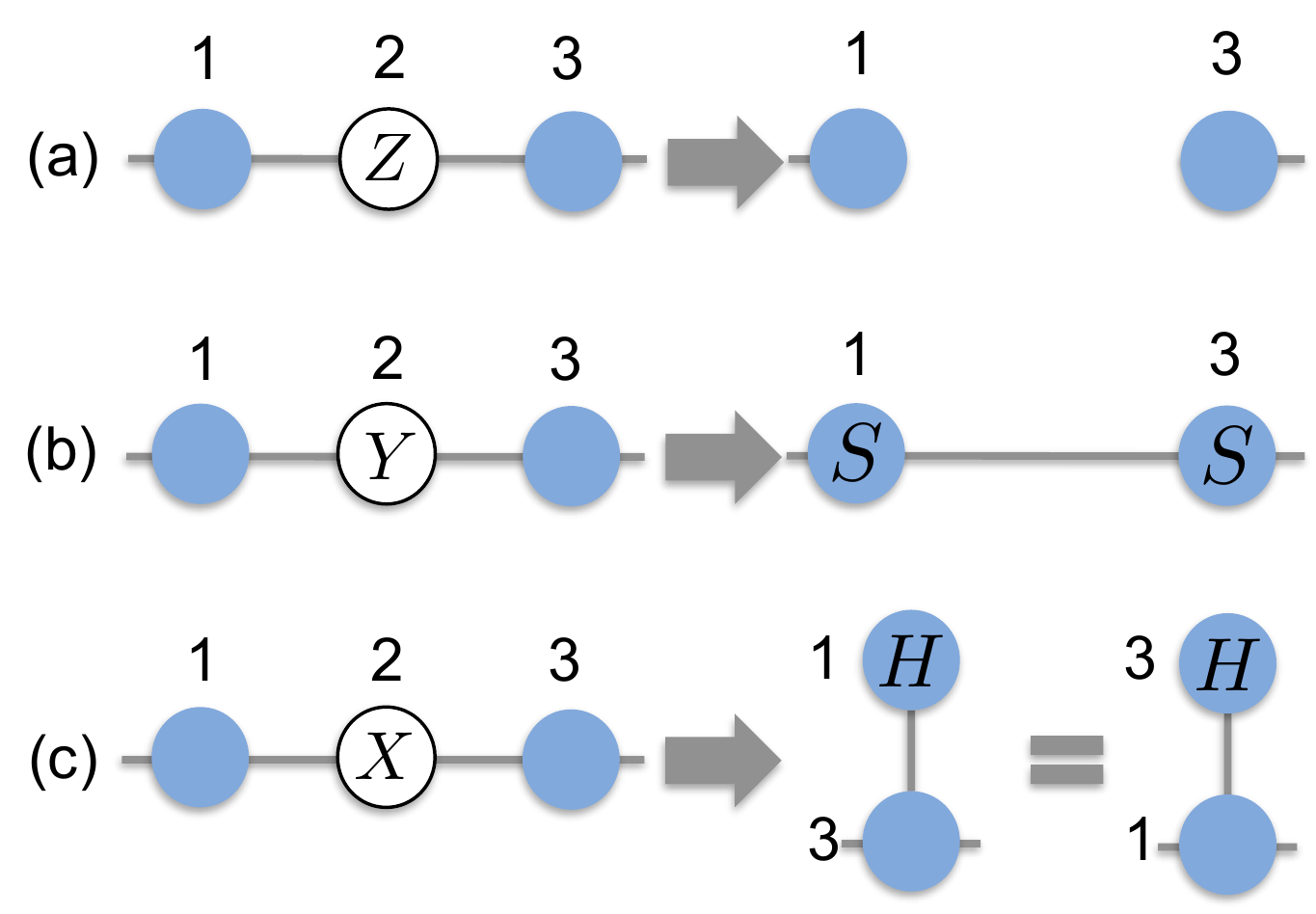}
\caption{Topological transition by measurements on 1D cluster states}
\label{chain_Xm}
\end{figure}

Performing Pauli basis measurements transforms the graph states.
Considering a 1D cluster state, the Z-basis measurement removes the measured qubit and disconnects the links, leaving two separate cluster states, or two independent qubits with a state $\ket{1}$ respectively (see Figure \ref{chain_Xm}(a)).
The Y-basis measurement also removes the measured qubit but directly connects the neighbors up to the phase operations as a byproduct (see Figure \ref{chain_Xm}(b)).
Unlike the other Pauli basis measurements, the X-basis measurement transforms the linear graph into a non-linear graph. One X gate as an byproduct, and an additional Hadamard gate needs to be applied to a neighboring qubit of the measured target. While the target qubit of the byproduct X gate does not affect the overall state but fixes the phase, the target qubit of the Hadamard gate directly affects the graph as in Figure \ref{chain_Xm}(c).

\begin{figure}[!hbt]
\center
\includegraphics[keepaspectratio,scale=0.35]{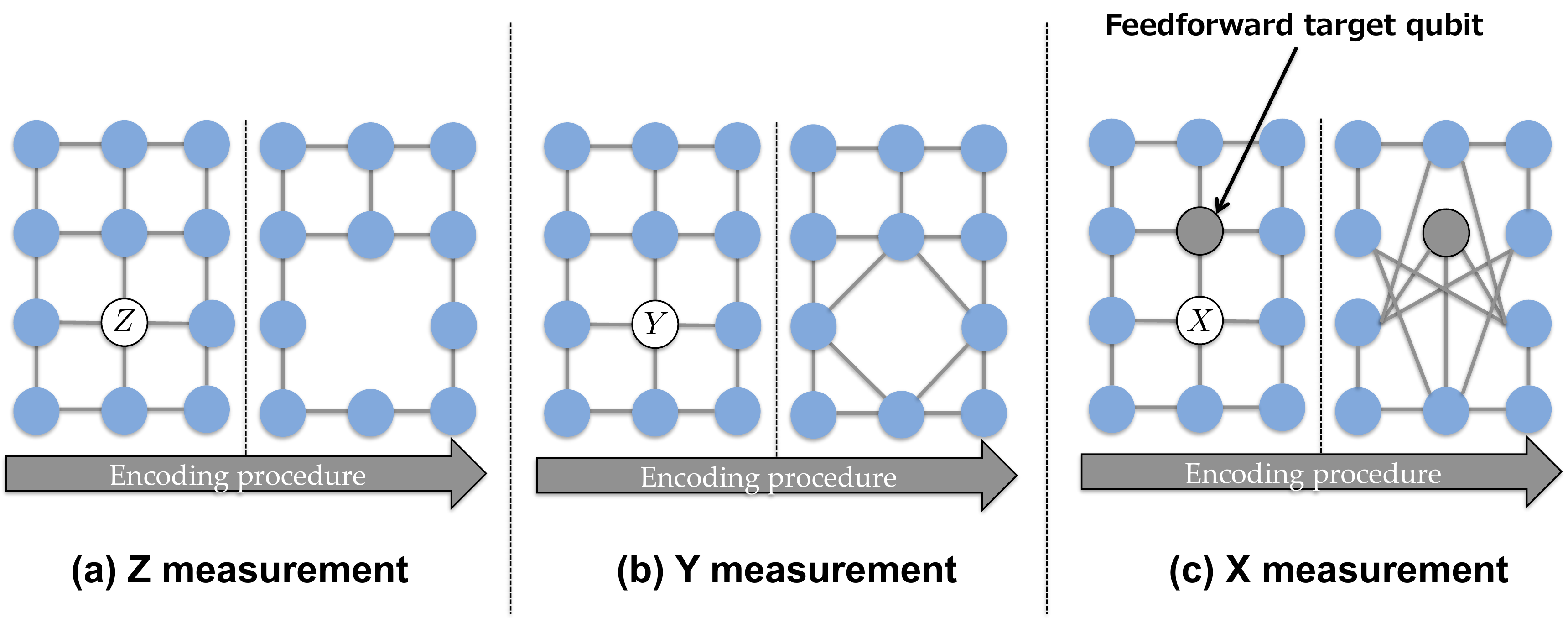}
\caption{Topological transition by measurements on 2D cluster states}
\label{mbqc_measurements}
\end{figure}

Similar to the 1D graph, performing measurements on a 2D cluster states affect the graph in a 2-dimensional manner. The Z-basis measurement performed on an arbitrary qubit will simply result in a new topology without the measured qubit and its links as in Figure \ref{mbqc_measurements}(a). The Y-basis measurement also removes the measured qubit but leaves additional complementary links between the neighbors of the measured qubit (Figure \ref{mbqc_measurements}(b)). The output topology after X-basis measurement differs according to the feedforward target, which is a neighbor qubit of the measurement target (filled with gray color in \ref{mbqc_measurements}(c)). After the measurement, new complementary links are formed between the neighbors of the feedforward targeted qubit and the neighbors of the measured qubit, between the mutual neighbors of the feedforward targeted and the measured qubit, and between the feedforward targeted qubit and the neighbors of the measured qubit. Similar to other measurements, the measured qubit is removed and its links will be disconnected from other qubits.

\chapter{Related Work}
\label{related_work}

Quantum networking in general, which includes the thesis proposal in Section 5, strongly depends on the ability of experimentalists to successfully demonstrate the capability of establishing quantum communication using entangled states and distributing those entangled qubits across a long distance.
Entangled states may be physically generated via photons directly or via stationary memories such as the NV diamond~\cite{PhysRevLett.97.087601}, trapped ion~\cite{Leibfried2003} and quantum dot~\cite{RevModPhys.79.1217,Elzerman2004}.
The experimental demonstration of quantum teleportation was accomplished in the late 1990s~\cite{Bouwmeester1997,Furusawa1998}.

\section{Violation of a Bell inequality}
One of the biggest questions that arise when dealing with quantum computing is, whether entanglement is real and is actually different from Newtonian physics based classical correlation.
The CHSH inequality, which is one example of the Bell inequality, allows us to mathematically prove that quantum mechanics cannot be explained by a local hidden variable theory.
In classical Newtonian physics, the correlations between outcomes of measurements on distant particles must satisfy an inequality $\mid S \mid < 2$ (S is the correlation between measurement outcomes).
Since quantum mechanics does not obey locality and realism, the same experiment on entangled particles violates the inequality and result in a different boundary $2\sqrt[]{2} \approx 2.82843$ (see subsection~\ref{CHSHsection} for details regarding CHSH inequality) - known as {\it Tsirelson's bound}.

In 2015, Poh et al. experimentally measured the correlation of maximally entangled photon polarization, and observed
$S = 2.82759 \pm 0.00051$, which is very close to Tsirelson's bound $ S - 2\sqrt[]{2}= 0.00084 \pm 0.00051$ \cite{PhysRevLett.115.180408} and a tremendous advance in precision.

In the same year, Hensen et al. succeeded in demonstrating a loophole-free Bell inequality violation using NV electron spins separated by 1.3km \cite{Hensen2015}.
Previously reported experimental results required extra assumptions which resulted in loopholes \cite{2014RvMP...86..419B, PhysRevD.35.3831, PhysRevA.47.R747, 2002PhRvA..66d2111B}. Using entangled electron spins of an estimated fidelity $ F= 0.92 \pm 0.03$, 245 trials of direct CHSH-Bell inequality test have been performed, and observed $S = 2.42 \pm 0.20$.

Many other Bell inequality experiment related papers are summarized well in Wikipedia \cite{Bell_test}.

\section{Long range entanglement distribution}

In order to establish a long distance quantum communication, it is necessary to distribute the entangled qubits beforehand. In real networking, not only the fidelity but also the distribution rate plays a significant role.

In 2017, Stockill et al. succeeded in coupling two quantum dots located 2m apart. The observed average Bell state fidelity for $\ket{\Psi^+}$ and $\ket{\Psi^-}$ is $F = 0.616 \pm 0.023$ with a high entanglement generation rate of 7.3kHz \cite{Stockill2017}.

In the same year, China achieved the longest ever entanglement distribution using their quantum satellite, Micius, which was designed to produce two separate photons with entangled polarizations as $\ket{\Psi^{+}}= \frac{1}{\sqrt{2}} (\ket{01} + \ket{10})$ \cite{Juan2017}. Pairs of photons are beamed down to earth under a pump power of ~30 mW, with a rate of 5.9 million entangled pairs per second with a fidelity of 0.907 $\pm$0.007, and measured at ground stations separated by 1203km. Overall the experiment succeeded in distributing 1.1 entanglements per second on average across over 1203km distance with a fidelity 0.869 $\pm$ 0.085.

\begin{figure}[!hbt]
\center
\includegraphics[keepaspectratio,scale=0.5]{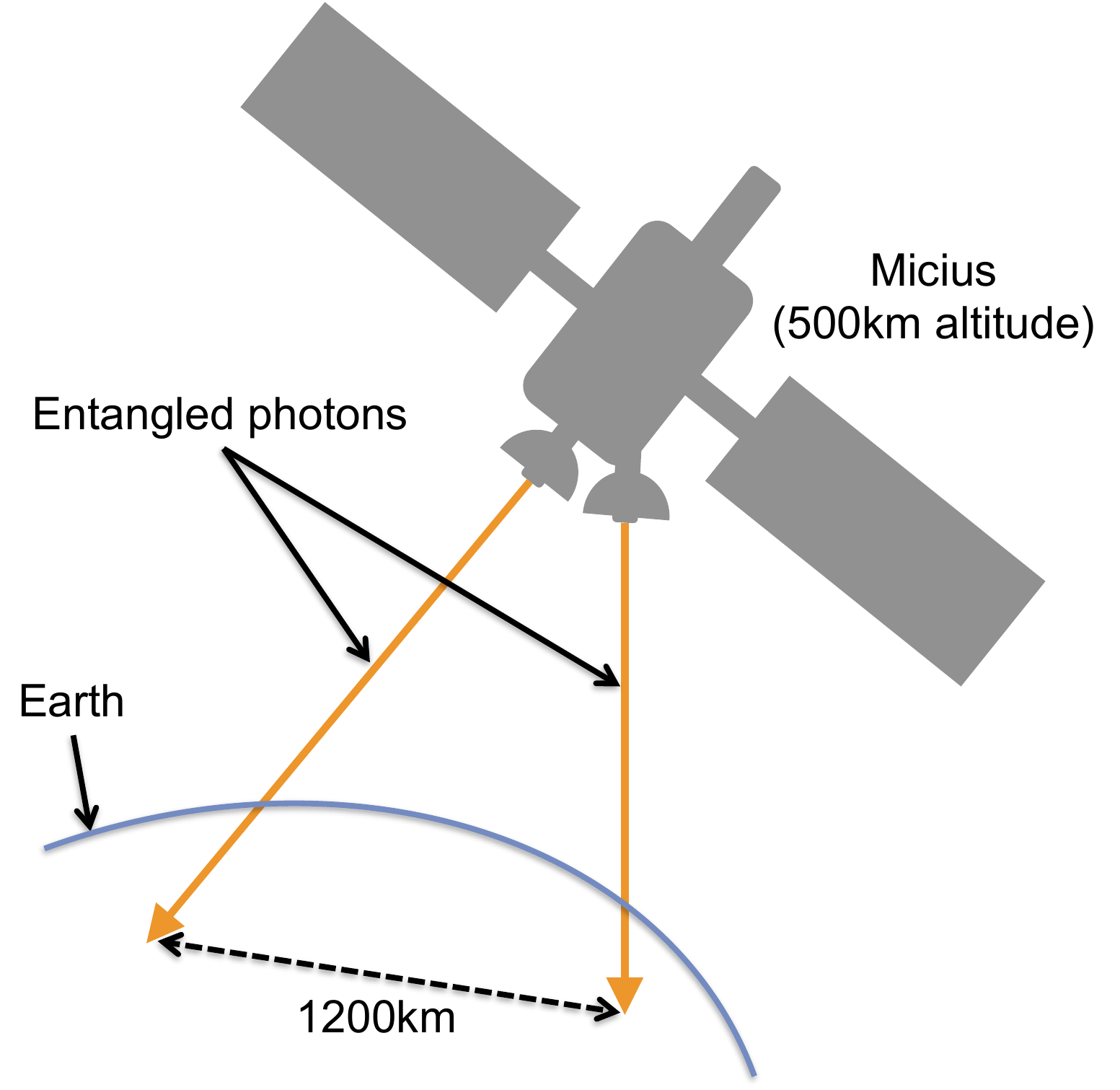}
\caption{Chinese satellite beaming down entangled photons to earth}
\label{fig01}
\end{figure}

Note that the above experiments have no quantum memories used.
Entangled qubits are immediately measured after reception.

\section{Physical link architecture}
\label{PLAsection}

Jones \emph{et al.} introduced three different architectures for a repeater link; MeetInTheMiddle, SenderReceiver and MidpointSource~\cite{2016NJPh...18h3015J}.

\subsection{MeetInTheMiddle}
\begin{figure}[!hbt]
\center
\includegraphics[keepaspectratio,scale=0.6]{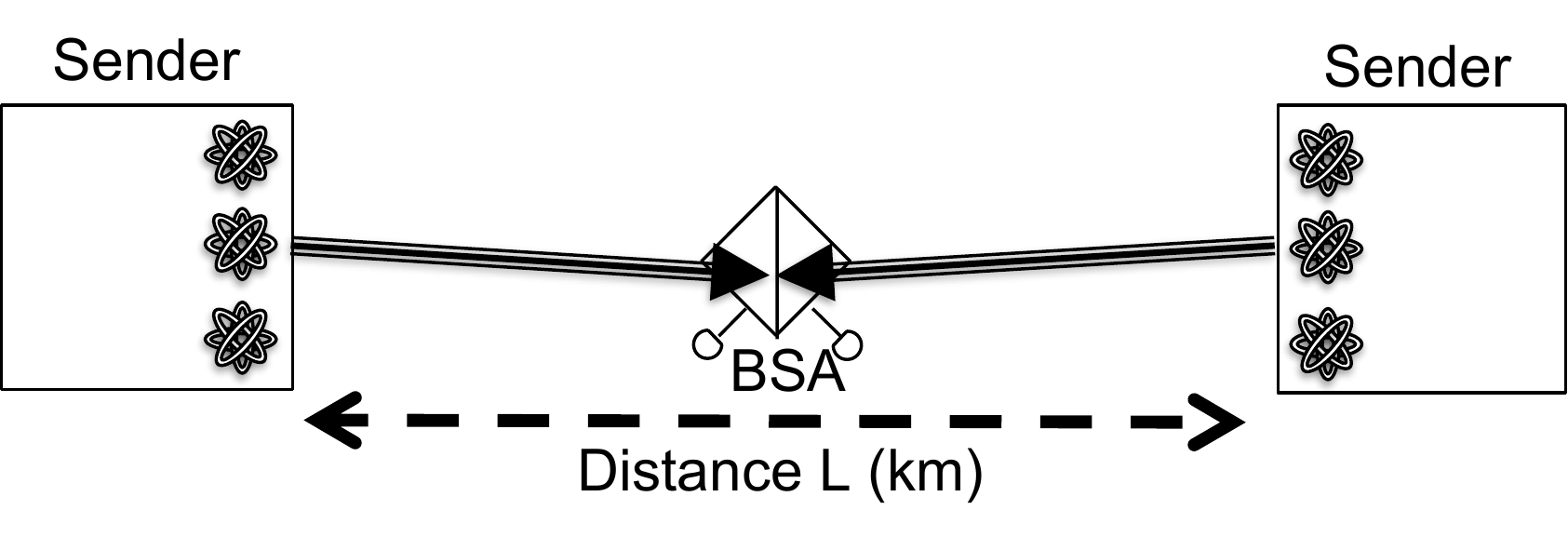}
\caption[The MeetInTheMiddle model.]{The MeetInTheMiddle model. This requires a stand-alone BSA node in between nodes.}
\label{MeetInTheMiddle}
\end{figure}

The pictorial model of the MeetInTheMiddle link is described in Figure~\ref{MeetInTheMiddle}.
As shown, the MeetInTheMiddle model includes two quantum repeater nodes with a stand-alone Bell State Analyzer (BSA) in between.
The positioning of the BSA may be limited, for example, due to geographical constraints.
Each node is connected via a quantum channel and a classical channel.
In this model, both quantum nodes must transmit photons through the quantum channel,
so that those photons reach at the BSA simultaneously for every attempt.
The photon emission timing and its burst rate, therefore, must be synchronized between the nodes.
For each attempt, the BSA must acknowledge the success/failure result to both nodes.

\subsection{SenderReceiver}
\begin{figure}[!hbt]
\center
\includegraphics[keepaspectratio,scale=0.6]{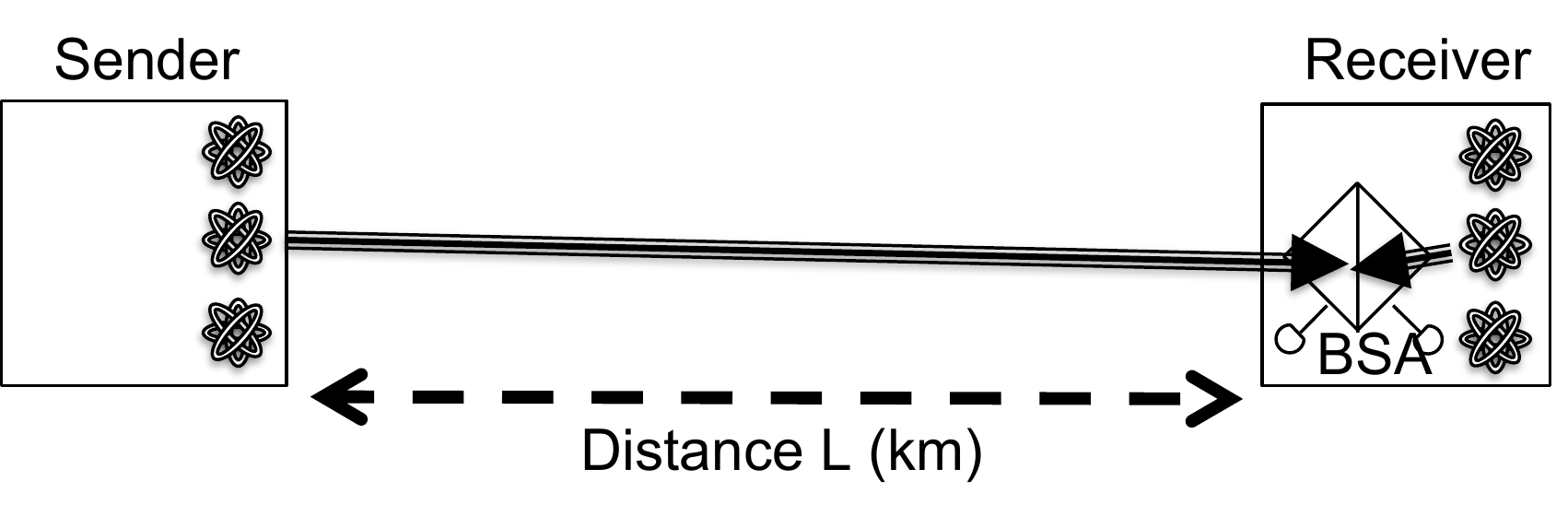}
\caption[The SenderReceiver model.]{The SenderReceiver model. The BSA is installed inside one endpoint.}
\label{SenderReceiver}
\end{figure}

The pictorial model of the SenderReceiver link is described in Figure~\ref{SenderReceiver}.
The SenderReceiver model is very similar to the MeetInTheMiddle model, but with an internally installed BSA module at one endpoint (Receiver Node).
The Sender node emits photons towards the Receiver node, and attempts to generate Bell pairs in a similar way as in the MeetInTheMiddle model.
However, the Receiver node is capable of resetting memory qubits in real time at every attempt, because the success/failure results can be immediately referred locally.
This is especially advantageous when the Receiver node's buffer size is smaller than that of the Sender node's.
The installation cost is also less expensive compared to the MeetInTheMiddle model because we have one node fewer, but classical latency increases by a factor of two over a link with the same length $L$.

\subsection{MidpointSource}
\begin{figure}[!hbt]
\center
\includegraphics[keepaspectratio,scale=0.6]{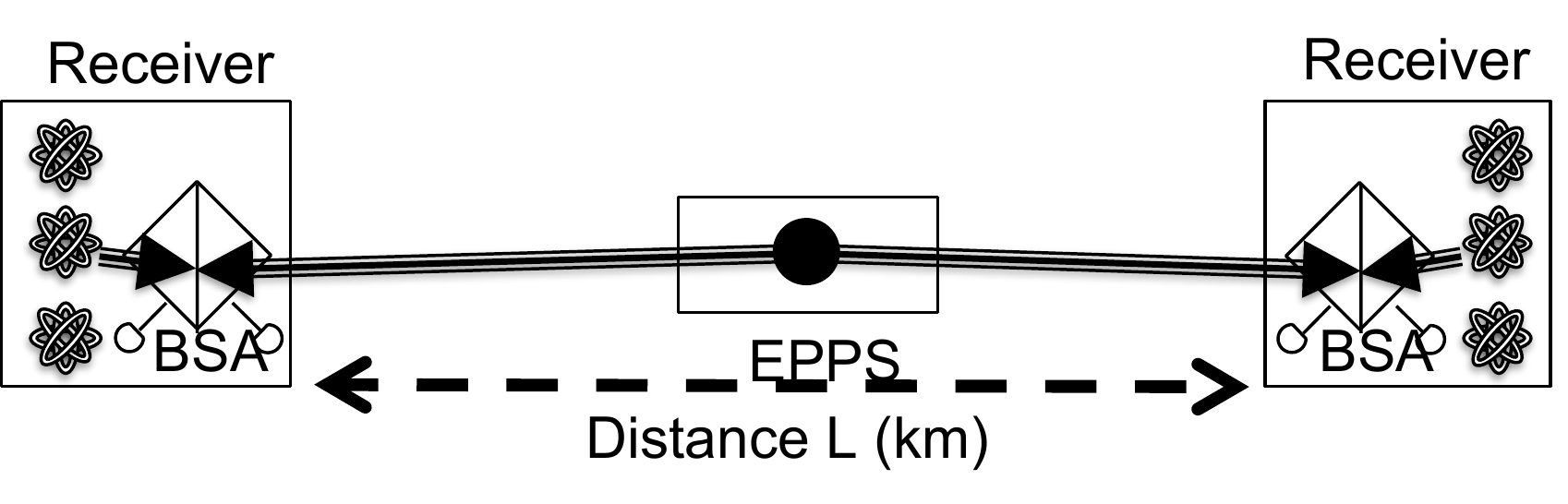}
\caption[The MidpointSource model.]{The MidpointSource model. The BSA is installed inside both nodes, with a stand-alone EPPS node in between.}
\label{MidpointSource}
\end{figure}

The pictorial model of the MidpointSource link is described in Figure~\ref{MidpointSource}.
This model requires one stand-alone Entangled Photon Pair Source (EPPS) node.
The EPPS is a device that is capable of generating entangled photon pairs, and forwards them to both nodes.
Each node, therefore, must transmit photons from their local memory,
so that its arrival time to the local BSA is synchronized to the arrival time of the photon from the EPPS.
In order to entangle memory qubits node-to-node, local Bell measurements of both nodes must succeed upon the same photon pair sent from the EPPS.

The trial rate of entangling qubits is higher than the other two link architectures, but the success probability is lower.

\section{Quantum data link protocol}

\begin{figure}[!hbt]
\center
\includegraphics[keepaspectratio,scale=0.37]{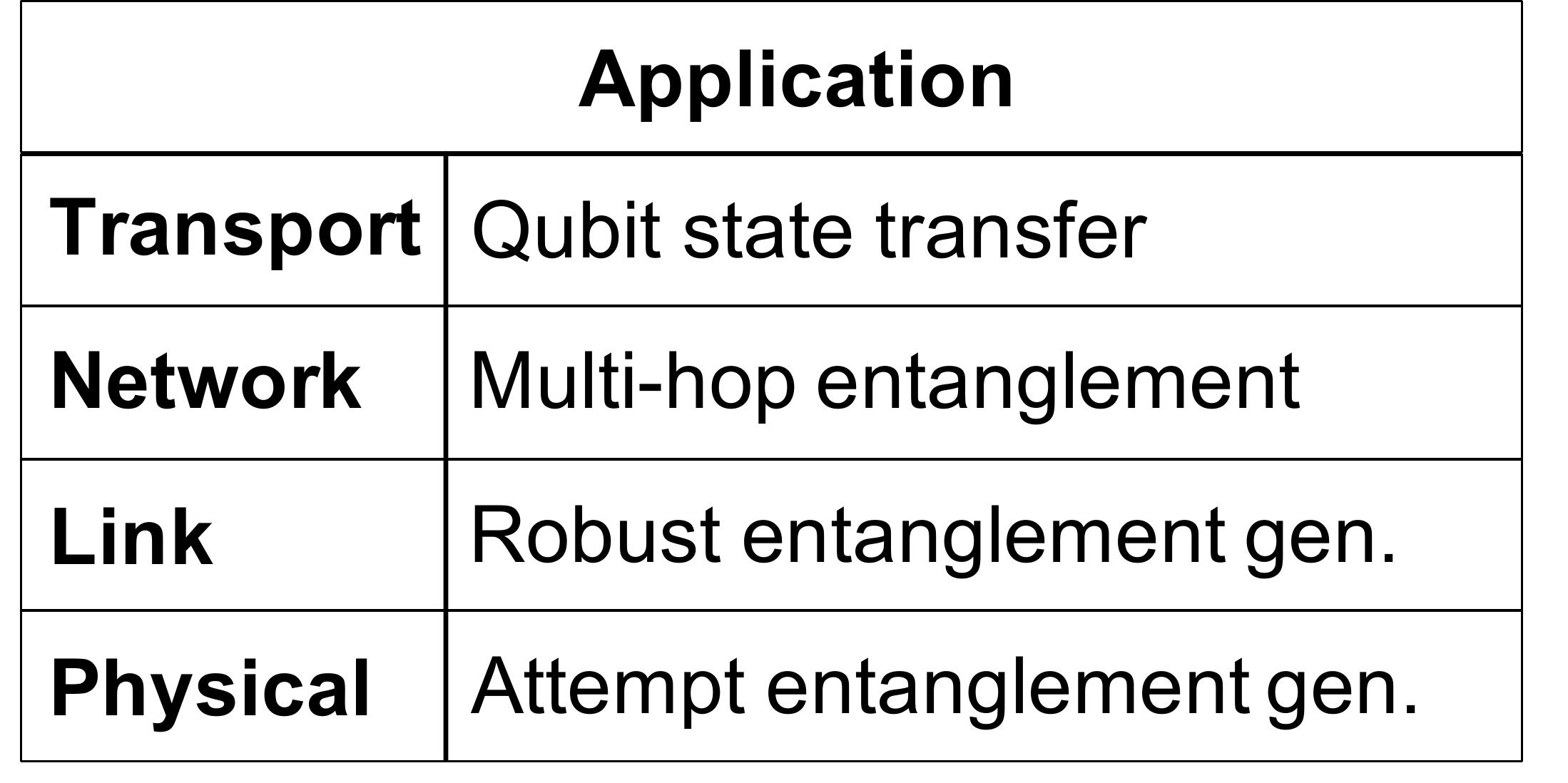}
\caption[Required functionality for each layer network stack]{Required functionality for each layer network stack as in~\cite{2019arXiv190309778D}.}
\label{QuantumLayer}
\end{figure}

\begin{figure}[!hbt]
\center
\includegraphics[keepaspectratio,scale=0.37]{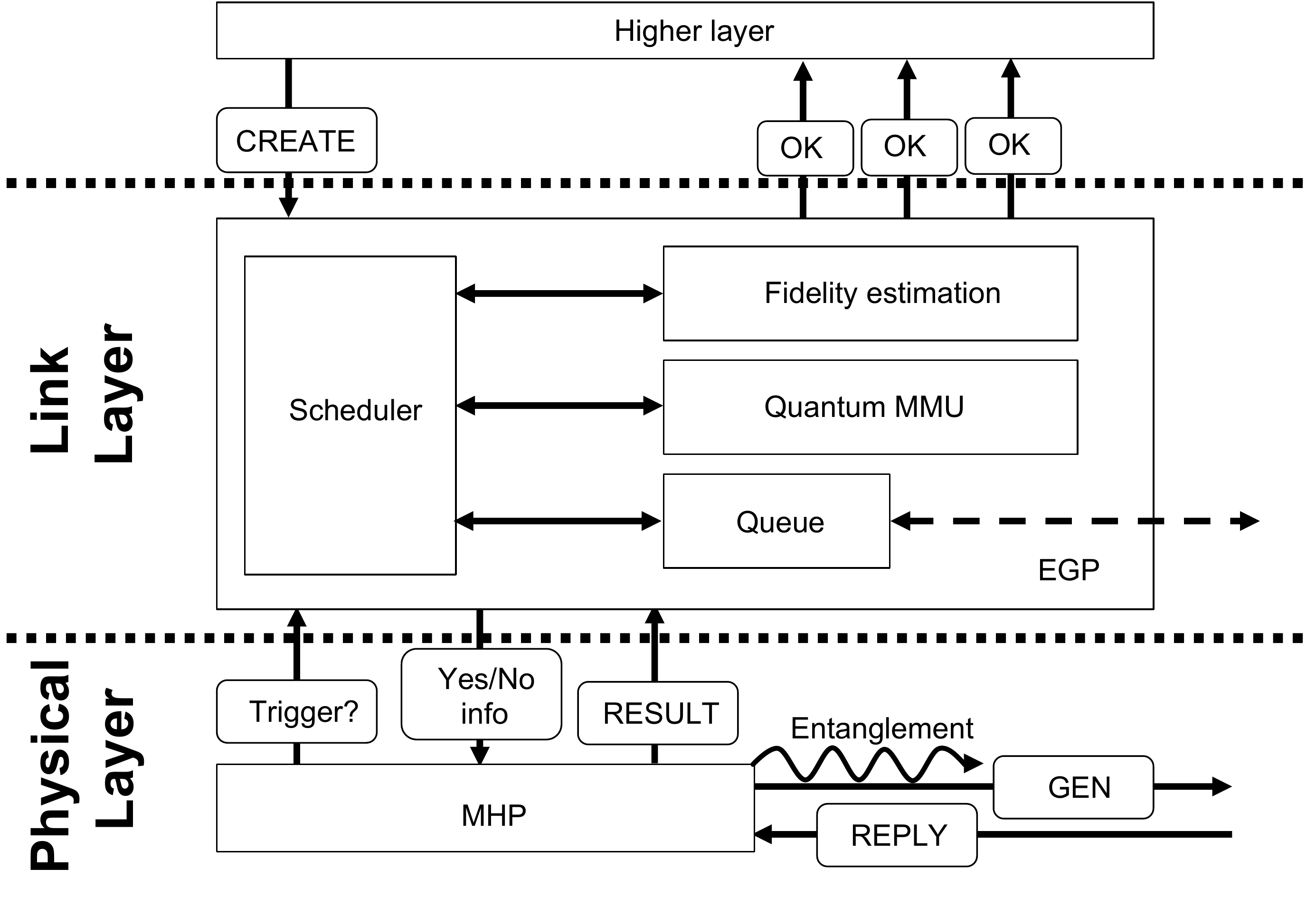}
\caption[Layered architecture for quantum communication]{Layered architecture for quantum communication as in~\cite{2019arXiv190309778D}.}
\label{AxelArchitecture}
\end{figure}

Dahlberg et al. introduced a quantum data link protocol, and analyzed its performance over a detailed network simulation composed of a few qubits, executed on a supercomputer~\cite{2019arXiv190309778D}.
The required functionality for each layer in the network stack is summarized in Fig.~\ref{QuantumLayer}, and the proposed architecture is shown in Fig.~\ref{AxelArchitecture}.

The higher layer requests quantum state creation of a resource with certain parameters by communicating with the link layer.
The Midpoint Heralding Protocol (MHP) periodically asks for triggering entanglement (via polling) to the link layer, and the link layer acknowledges the need with yes/no reply along with some other parameters.
There are two types of requests; the \emph{Create and Keep (K)} request or the \emph{Create and Measure (M)} request.
An example of included parameters in the \emph{K} request may be the ID, sequence of gate operations, the type of generation and etc.
The \emph{GEN} message is transmitted to the BSA node, and the generation attempt's success/failure \emph{REPLY} with its generation ID (which is used by the EGP later) is acknowledged to nodes.
This information will also be forwarded to the link layer.
The \emph{K} request specifies the measurement basis, and the qubit is measured along the basis immediately after the photon emission (even before receiving the REPLY message from the BSA node).
The measurement result is forwarded to the EGP accordingly.

The Queue in the Entanglement Generation Protocol (EGP) is used for coordinating the MHP triggering timing.
The Quantum Memory Management (QMM) tracks which physical qubit is used for which entanglement generation/storing.
The Fidelity estimation unit (FEU), based on known hardware capabilities,
estimates the minimal completion time for preparing an entangled resource with minimal fidelity $F_{\mbox{min}}$.
The minimal completion time is also used to reject requests from the higher layer based on the given timeout value.
The FEU also performs test rounds for improving the estimation.
The EGP scheduler manages which request in the queue should be served next.

\section{Quantum link bootstrapping}

\begin{figure}[!hbt]
\center
\includegraphics[keepaspectratio,scale=0.37]{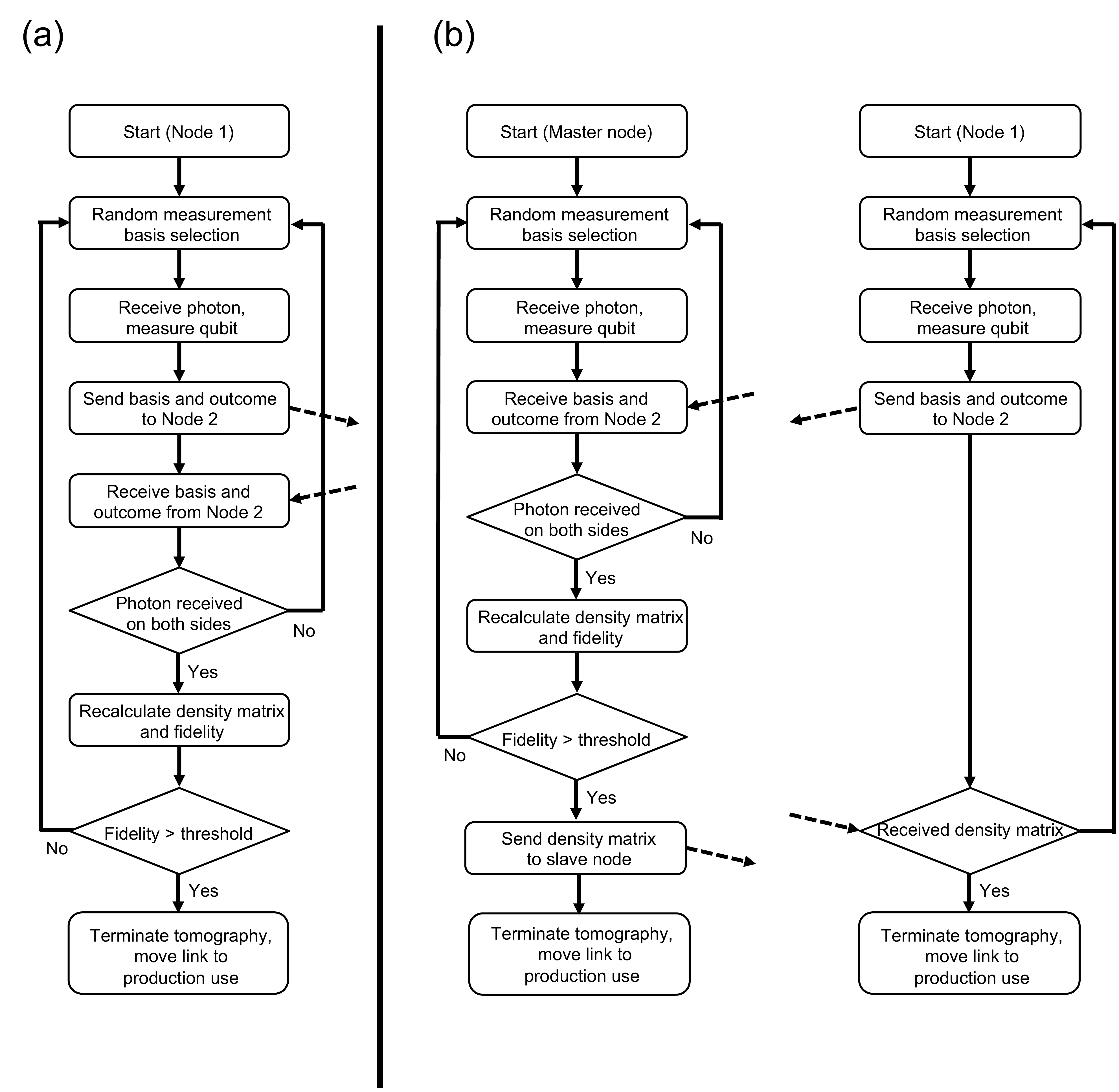}
\caption[Procedures of quantum link bootstrapping.]{Procedures of quantum link bootstrapping. Here, the fidelity is the reconstruction fidelity. (a) Distributed model. (b) Master-Slave model.}
\label{OkaTomography}
\end{figure}

Oka \emph{et al.} discussed procedures for distributed quantum link bootstrapping based on link-level tomography (see Fig.~\ref{OkaTomography}(a)),
and master-slave model based quantum link bootstrapping (see Fig.~\ref{OkaTomography}(b)), over a MidpointSource link~\cite{2016NJPh...18h3015J}.
The paper focuses on the accuracy of the reconstructed density matrix (reconstruction fidelity), based on the number of measured Bell pairs.
The simulation assumes infinite ideal memories.
As shown, both procedures assume the ability to raise the reconstruction fidelity up to a certain threshold.

\begin{figure}[!hbt]
\center
\includegraphics[keepaspectratio,scale=0.37]{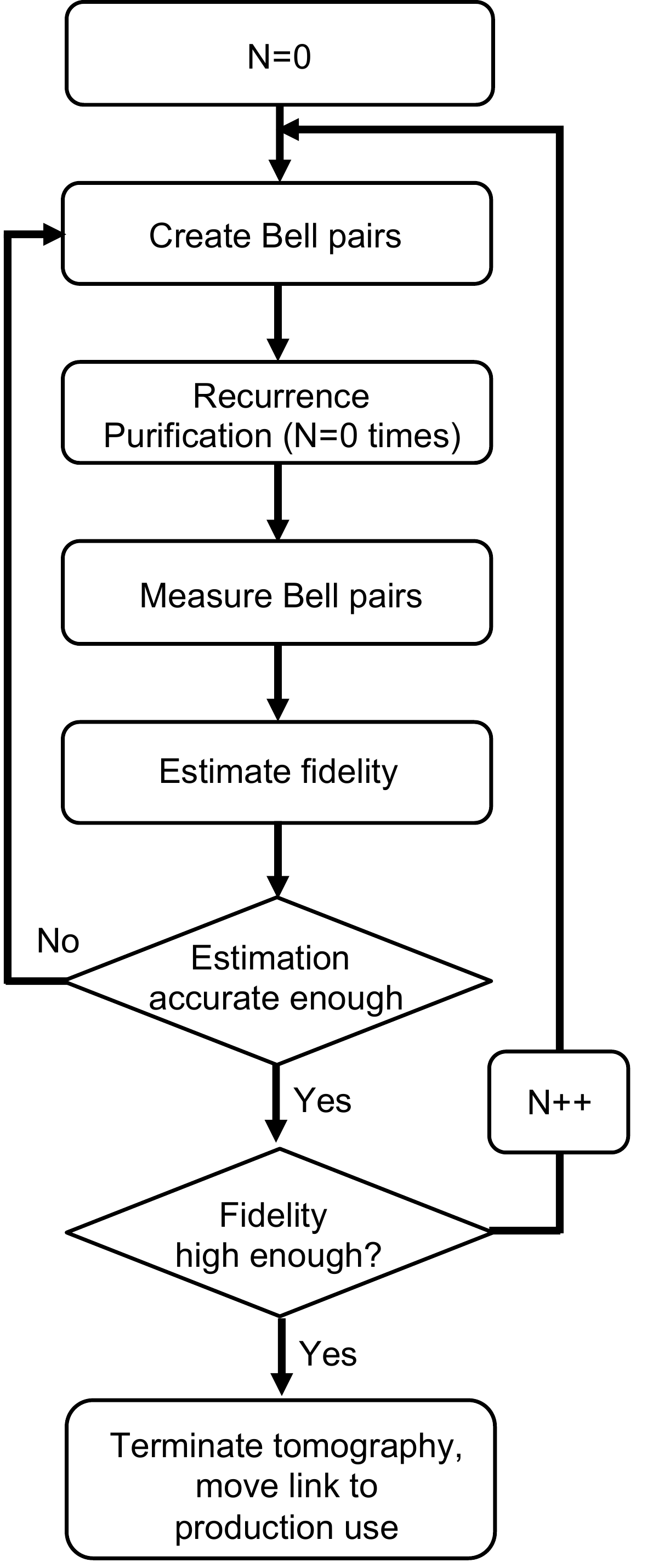}
\caption{Procedure of quantum link bootstrapping with purification.}
\label{OkaPurification}
\end{figure}

Oka's bachelor thesis~\cite{Oka:Thesis:2017} also considers recurrence purification for increasing the average resource fidelity, over a simulated noisy system with only Pauli errors.
The flowchart of the quantum link bootstrapping with recurrence purification is shown in Fig.~\ref{OkaPurification}.
The proposed protocol assumes the ability to raise the reconstruction fidelity and the resource fidelity up to a certain threshold.

\chapter{Protocol Design}
\label{protocol_design}

This chapter discusses the simulated protocols.
The simulation design is made based on the following assumptions.
\begin{itemize}
  \item Reliable, in-order timely delivery of classical messages.
  \item Paired classical and quantum channels installed between adjacent nodes.
  \item Perfectly synchronized clocks.
  \item Reliable nodes with no system failures.
  \item Nodes capable of determining the length of the quantum channels between neighboring nodes to sufficient accuracy.
\end{itemize}

\section{Quantum data link protocol}
\label{qdsec}
A quantum repeater requires shared entangled resources between neighboring nodes.
The role of the quantum data link protocol is to generate such resources continuously, yet autonomously.
This thesis will focus on designing a data link protocol for two closely related repeater link architectures,
the MeetInTheMiddle model and the SenderReceiver model (see section~\ref{PLAsection} for details regarding link architectures).

The core concept of the data link protocols introduced in~\cite{2016NJPh...18h3015J} will be kept untouched.
As in~\cite{2016NJPh...18h3015J}, this thesis does not elaborate how to synchronize clocks across a network,
but will start with the coordination of photon emissions between adjacent nodes.

\begin{figure}[H]
  \center
  \includegraphics[keepaspectratio,scale=0.5]{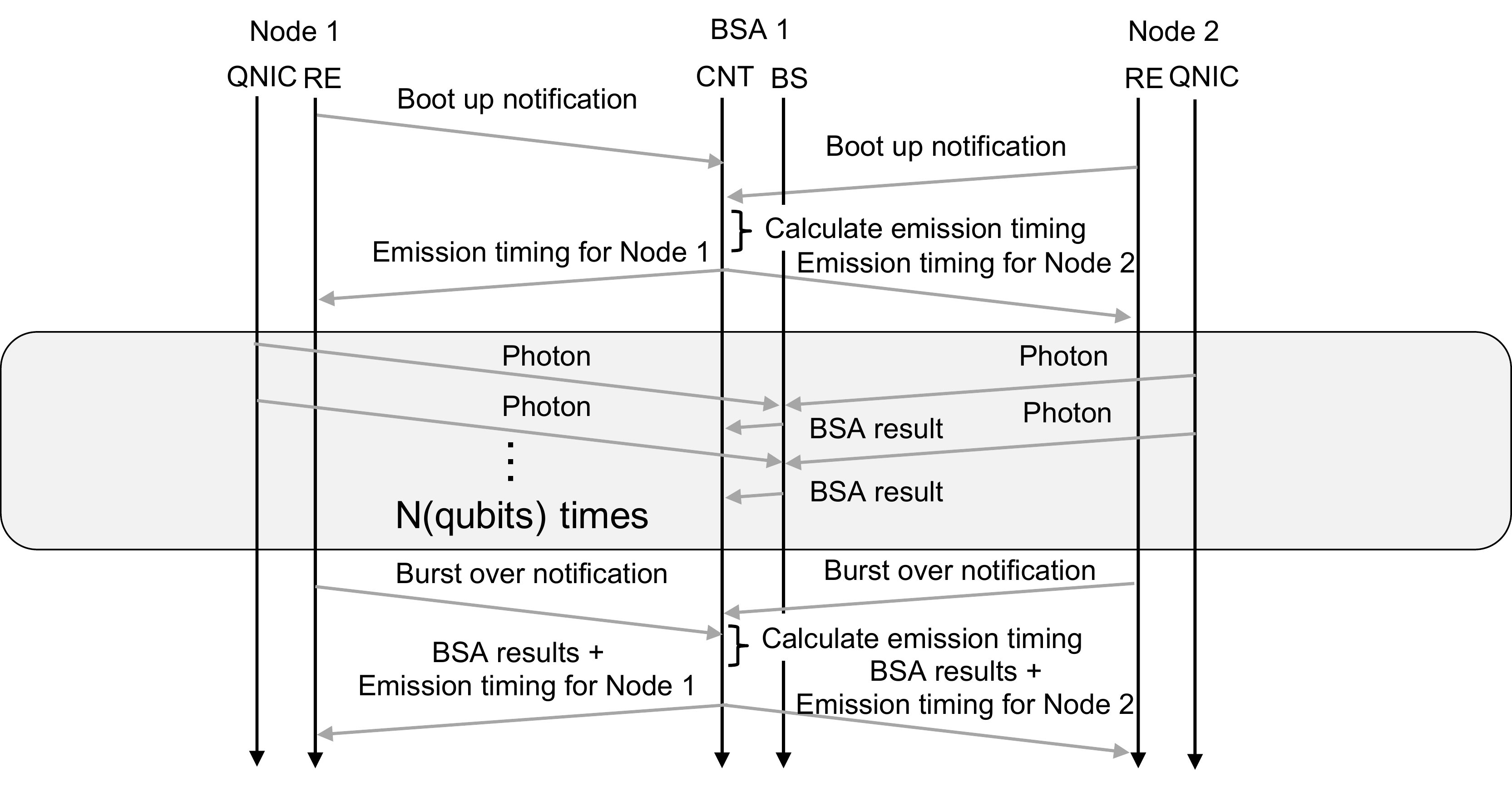}
  \caption[Flow chart of resource generation.]{Flow chart of resource generation. This process is looped to continuously generate new resources.
  QNIC is Quantum Network Interface Card, RE is Rule Engine, CNT is Controller and BS is Beamsplitter.
  For details regarding modules, see section \ref{QNarchitecture}.}
  \label{REflow}
\end{figure}

Supplemental components have been added to the protocol for the network simulation.
Coordinating the arrival time of photons at the BSA via messaging between the two nodes with memory is difficult.
Instead, the BSA node has been assigned to take responsibility in coordinating the entanglement generation.
To start, an activated quantum node classically transmits a {\it Boot Up Notification} to all neighboring BSA nodes.
For a SenderReceiver link, the sender transmits the message towards the receiver.
Once received, the BSA node calculates the photon emission timings and the corresponding burst rate for neighboring nodes,
based on the quantum channel lengths and its own single photon detector recovery time.
Such information will be classically forwarded to the neighboring quantum nodes to start the entanglement generation.
The quantum nodes also need to classically notify the BSA node of the end of the burst.
Reception triggers the BSA to recalculate the emission timing for the next round,
and returns the information together with a list of {\it success/failure for transmission \( i\)}, and node addresses specifying between which nodes the resources are shared.
We buffer the success/failure results, and send them as a single packet to prevent overflowing the classical channel.
Details regarding modules are also provided in section~\ref{QNarchitecture} and in section~\ref{BSAarchitecture}.
The rule engine also operates qubit emissions for the quantum data link protocol, which is an operation for generating entangled resources (see Figure.~\ref{REflow}).

\section{RuleSet-based communication protocol}
\label{RSPsection}

\begin{figure}[htbp]
  \center
  \includegraphics[keepaspectratio,scale=0.45]{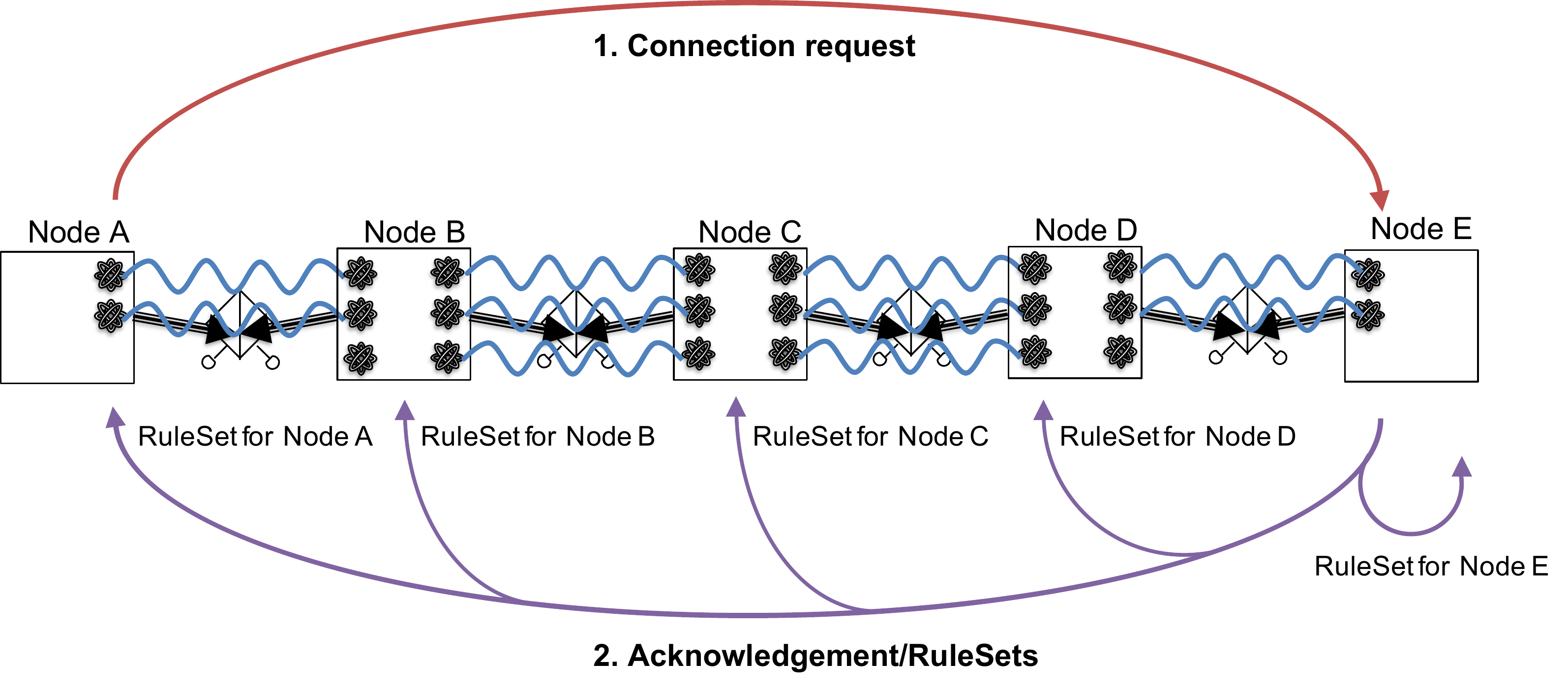}
  \caption[Node A generating and distributing RuleSets]{Node A generating and distributing RuleSets done at connection setup time. In this example, there are five nodes in total, which requires a total of five distinct RuleSets.}
  \label{RuleSetDistribution}
\end{figure}

Coordinated operation is a prerequisite functionality for quantum networking over both the link layer and the network layer~\cite{zimmermann1980osi}.
When nodes perform a particular operation using pre-shared entangled resources,
participants are assumed to perform the correct operation targeting the appropriate resources.
Such consistency should not be achieved by exchanging messages with each other, especially over long distances due to the latency incurred.
In this section, we introduce the concept of {\it RuleSet} for supporting quantum networking,
which allows us to synchronize operations over a network with minimal classical message transmission.
If a single connection follows a route involving $n$ nodes, the source node requests the destination node to generate $n$ RuleSets,
 and distributes them to all $n$ nodes respectively (see Fig.~\ref{RuleSetDistribution})~\cite{van-meter-qirg-quantum-connection-setup-00}.

A RuleSet is an object consisting of one or more {\it Rules}, each holding a {\it Condition} and an {\it Action}.
A Condition may have one or more {\it Clauses}, each of which is a conditional statement.
An Action holds a list of operations that a node needs to perform in order to accomplish a single task,
which may be for example, entanglement swapping, purification, measurement for tomography, etc.,
and is invoked if and only if each Clause in the corresponding Condition is fulfilled.
When the task involves at least one resource, the oldest available resource will be picked from the allocated set.
An Action may also generate a message to another node, lock and reinitialize resources.
In our current implementation,
a RuleSet also has a {\it Termination Condition}, which is a simple counter for determining when to discard the RuleSet and discontinue the connection.
This may also have multiple Clauses.
Each Rule, also holds more than one {\it partner address}, identifying where the paired entangled qubit (or the paired node for a particular operation) is physically located.
This is, of course, taken into consideration by the destination node generating the RuleSets for a particular connection.

\begin{figure}[htbp]
  \center
  \includegraphics[keepaspectratio,scale=0.45]{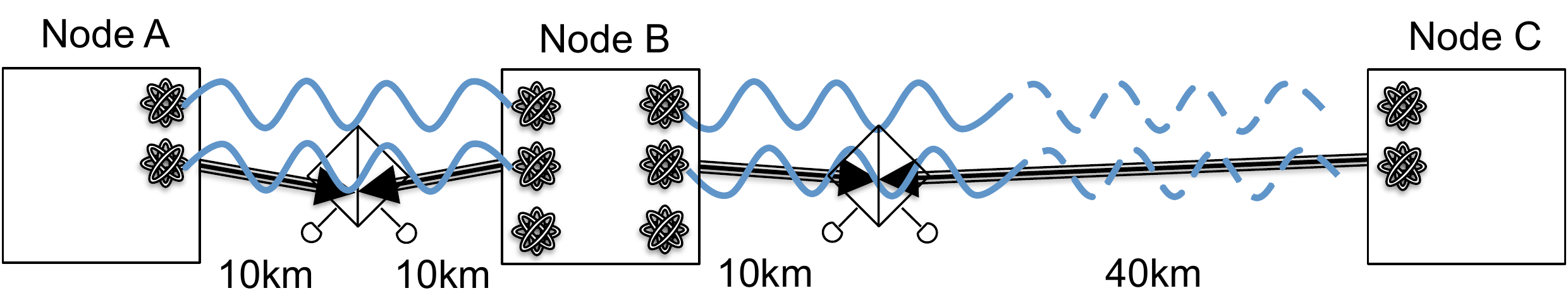}
  \caption[Bias in knowledge regarding the state of resources]{Bias in knowledge regarding the state of resources.  Node C receives the success/failure acknowledgement later than Node A and B.}
  \label{QI_inconsistency}
\end{figure}

RuleSets will be managed and executed by the RuleEngine,
which is a software module installed in all quantum nodes that is responsible for interpreting the RuleSet instructions, and executing them in real time.
The RuleEngine is event-driven, where the event may be a classical packet arrival, new resource allocation or RuleSet timeout.
If a single RuleSet consists of multiple Rules, the RuleEngine refers to each Rule based on top-down strategy, invoking Actions from top to bottom sequentially.
That is, if a node completes the Action in the first Rule on qubit A, if any, qubit A will be reassigned for the use of the second Rule in the same RuleSet.
Therefore, newly generated resources are first assigned to the first Rule in the set, when the generated resource is shared between the partner node of thee first Rule.
They must complete the first rule before being reassigned to the second Rule.
Timeout may be performed by tracking how well resources are upgraded to the upper Rules.
This RuleSet level resource allocation can be directly translated into multiplexing schemes, such as buffer space multiplexing~\cite{Aparicio2011}.
This top-down approach is also a simple solution to overcome the bias in knowledge regarding the resources across a network (see  Fig.~\ref{QI_inconsistency}).

\begin{figure}[H]
  \center
  \includegraphics[keepaspectratio,scale=0.55]{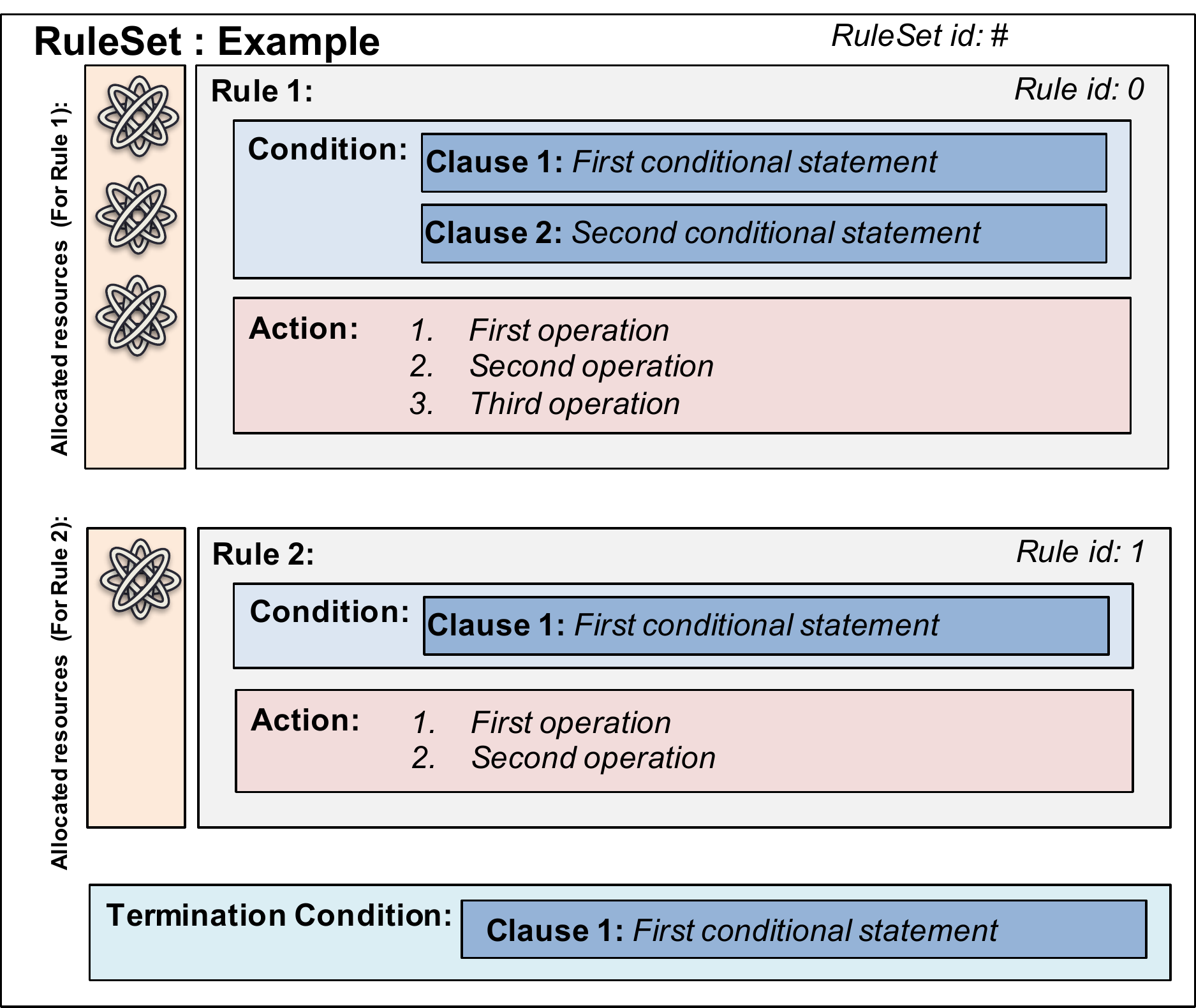}
  \caption[An example structure of a RuleSet]{An example structure of a RuleSet.
  The resource allocated in the second Rule (represented by the atom symbol) has been already passed through the first.}
  \label{RuleSetExample}
\end{figure}

All RuleSets for the same connection share the same RuleSet identifier ({\it RuleSetID}).
The RuleSet identifier is generated via a hash function using the time when the RuleSet was generated,
the IP address of the node which generated the RuleSet, and a random number as a seed to avoid global conflicts.
Rules are also indexed ({\it RuleID}) inside the RuleSet. Each Action holds a counter ({\it ActionIndex}), where the counter is incremented whenever the Action has been performed.
These identifiers are used, for example, when a node needs to share the measurement result with another node for link-level tomography.
Whenever an Action of a RuleSet requires a node to transmit a classical packet to another node, such as the measurement result,
it encloses the identifiers in the packet, so that the receiving node can uniquely identify and pair its own measurement result with the received one accordingly.
An example structure of a RuleSet composed of two Rules is shown in Fig.~\ref{RuleSetExample}.

\begin{figure}[H]
  \center
  \includegraphics[keepaspectratio,scale=0.45]{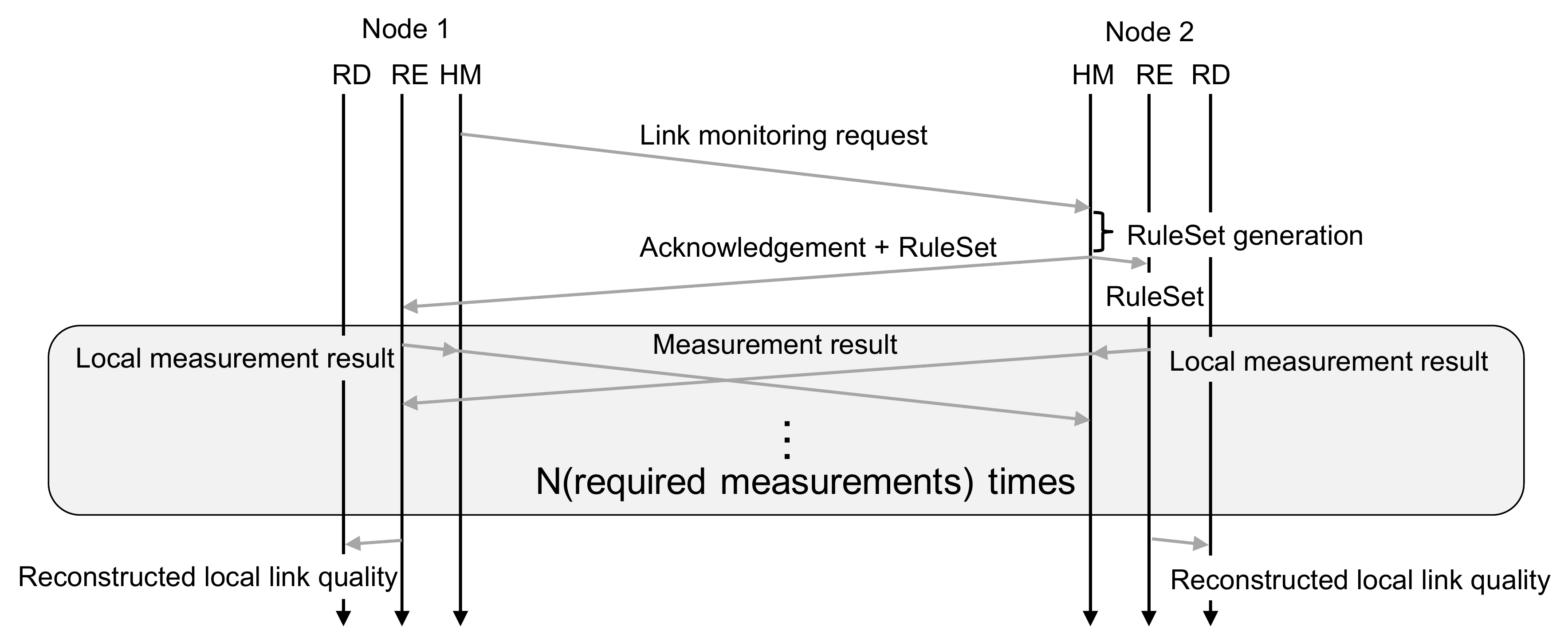}
  \caption[Flow chart of the hardware monitor working on link-level tomography.x]{Flow chart of the hardware monitor working on link-level tomography. RD is Routing Daemon, HM is Hardware Monitor and RE is Rule Engine. For details regarding modules, see section~\ref{QNarchitecture}.}
  \label{HMflow}
\end{figure}

When a RuleEngine executes an Action, the Action generally demands one or more entangled resources from its allocated list.
Resources are always picked from the oldest, so that operations are performed consistently between the partner node(s).
The flow chart of the RuleSet-based quantum link bootstrapping protocol is shown in Figugre.\ref{HMflow}.

\section{Pseudocode of Rules}
\label{EXAMPLERULESsection}
This subsection provides examples of Rules.

\subsection{Single selection-Single error Purification (Ss-Sp) Rule}

The Ss-Sp is the simplest method to detect either the X error or the Z error (see Fig.~\ref{SS-Xp} in section~\ref{purifications} for the circuit).

\subsubsection{Condition}
The Condition for the Ss-Sp Rule requires a single Clause, which is the \emph{ResourceConditionClause} shown in Algorithm.~\ref{ResourceClause}.
This Clause's role is to check the number of available resources, where $numRequired=2$ in this case.
A resource is \emph{locked} when it should not be disturbed -- e.g. waiting for purification result.

\begin{algorithm}[H]
  \algrenewcommand\algorithmicrequire{\textbf{Clause:}}
  \caption{ResourceConditionClause}
  \label{ResourceClause}
  \begin{algorithmic}[1]
    \Statex {\bf{This Clause checks if enough resources are available for the corresponding Action.}}
    \Statex {Input: resourceList $\gets$ List of allocated resources for the Rule. A resource may be locked when it is waiting for the classical packet to arrive, such as for purification.}
    \Statex {Output: enoughResources $\gets$ A boolean value.}
    \Statex {}
    \Procedure{ResourceConditionClause}{resourceList}
    \State {numRequired $\gets$ Number of required resources for the Action}
    \State {numFree $= 0$}
    \State {enoughResources $=$ false}
    \For {each resource in resourceList}
      \If {resource is not locked}
        \State {numFree$++$}
      \EndIf
      \If {numFree $>=$ numRequired}
        \State {enoughResources $=$ true}
        \State {break}
      \EndIf
    \EndFor
    \State {return enoughResources}
  \EndProcedure
  \end{algorithmic}
\end{algorithm}

\subsubsection{Action}
The Action for the Ss-Sp Rule simply picks two available resources, and performs an X purification.
The purified resource will be locked until the node gets the paired measurement result from the other node.

\begin{algorithm}[H]
  \caption{Ss-SpAction}
  \label{Ss-SpAction}
  \begin{algorithmic}[1]
  \Statex {\bf{This Action performs an X error purification based on single selection.
  The measurement result is stored, and then sent to the partner node as a message.}}
  \Statex{Input: resourceList $\gets$ List of allocated resources for the Rule.}
  \Statex {Output: msg $\gets$ A message for another node.}
  \Statex {}
  \Require {ResourceConditionClause == true}
  \Procedure{Ss-SpAction}{resourceList}
    \State {self\_addr $\gets$ Node address of this RuleSet owner.}
    \State {partner\_addr $\gets$ Node address of tomography partner.}
    \State {resource1 = Select a free resource from resourceList}
    \State {resource2 = Select another free resource from resourceList}
    \State {resource2.CNOT(resource1) /*Target.CNOT(Controlled)*/}
    \State {basis = Z}
    \State {outcome = resource2.Measure(basis)}
    \State {removeResourceFromList(resourceList,resource2)}
    \State {lockResourceInList(resourceList,resource1)}
    \State {Data.result = \{outcome, basis\}}
    \State {Data.ruleSetId = this.RuleSetID}
    \State {Data.ruleId = this.RuleID}
    \State {Data.actionIndex = this.ActionIndex}
    \State {save(Data)}
    \State {msg.destination = partner\_addr}
    \State {msg.source = self\_addr}
    \State {msg.data = Data}
    \State {this.ActionIndex$++$}
    \State {return msg}
  \EndProcedure
  \end{algorithmic}
\end{algorithm}

Each node will exchange its measurement outcome.
The locked resource will only be unlocked and reassigned to the next Rule when the measurement outcomes coincide.
Otherwise, purification failed and the locked resource is removed from the \emph{resourceList} and returned to the pool of unused resources for reinitialization and reuse.

\subsection{Double selection-Single error Purification (Ds-Sp) Rule}

The Ds-Sp is a purification method similar to Ss-Sp but with double selection (see Fig.~\ref{DS-Xp} in section~\ref{purifications} for the circuit)~\cite{fujii:PhysRevA.80.042308}.

\subsubsection{Condition}
The only Clause required for the Ds-Sp Rule is also the \emph{ResourceConditionClause} (see Algorithm.~\ref{ResourceClause} above).

\subsubsection{Action}
The Action for the Ds-Sp Rule picks three available resources, and performs an X purification based on double selection.
The purified resource will be locked until the node gets the paired measurement result from the other node.

\begin{algorithm}[H]
  \caption{Ds-SpAction}
  \label{Ds-SpAction}
  \begin{algorithmic}[1]
  \Statex {\bf{This Action performs an X error purification based on double selection.
  The measurement result is stored, and then sent to the partner node as a message.}}
  \Statex{Input: resourceList $\gets$ List of allocated resources for the Rule.}
  \Statex {Output: msg $\gets$ A message for another node.}
  \Statex {}
  \Require {ResourceConditionClause == true}
  \Procedure{Ds-SpAction}{resourceList}
    \State {self\_addr $\gets$ Node address of this RuleSet owner.}
    \State {partner\_addr $\gets$ Node address of tomography partner.}
    \State {resource1 = Select a free resource from resourceList}
    \State {resource2 = Select another free resource from resourceList}
    \State {resource3 = Select another free resource from resourceList}
    \State {resource2.CNOT(resource1) /*Target.CNOT(Controlled)*/}
    \State {resource2.CNOT(resource3) /*Target.CNOT(Controlled)*/}
    \State {resource3.H()}
    \State {basis = Z}
    \State {outcome1 = resource2.Measure(basis)}
    \State {outcome2 = resource3.Measure(basis)}
    \State {removeResourceFromList(resourceList,resource2)}
    \State {removeResourceFromList(resourceList,resource3)}
    \State {lockResourceInList(resourceList,resource1)}
    \State {Data.result = \{\{outcome1, basis\},\{outcome2, basis\}\}}
    \State {Data.ruleSetId = this.RuleSetID}
    \State {Data.ruleId = this.RuleID}
    \State {Data.actionIndex = this.ActionIndex}
    \State {save(Data)}
    \State {msg.destination = partner\_addr}
    \State {msg.source = self\_addr}
    \State {msg.data = Data}
    \State {this.ActionIndex$++$}
    \State {return msg}
  \EndProcedure
  \end{algorithmic}
\end{algorithm}

Each node will exchange its measurement outcome.
The locked resource will only be unlocked when the each of the paired measurement outcomes coincide.
Otherwise, purification failed and the locked resource is removed from the \emph{resourceList}.

\subsection{Full-state link-level tomography Rule}

\subsubsection{Condition}
The Condition for the tomography Rule requires two Clauses.
One of them is the \emph{ResourceConditionClause} (Algorithm.~\ref{ResourceClause}), and another is the \emph{MeasurementConditionClause} (Algorithm~\ref{MeasurementConditionClause}).
The \emph{MeasurementConditionClause} is a simple counter for checking the number of measured qubits.

\begin{algorithm}[H]
  \algrenewcommand\algorithmicrequire{\textbf{Clause:}}
  \caption{MeasurementConditionClause}
  \label{MeasurementConditionClause}
  \begin{algorithmic}[1]
    \Statex {\bf{This Clause checks if enough measurements has been performed.}}
    \Statex {Input: none.}
    \Statex {Output: enoughMeasurements $\gets$ A boolean value. True if enough measurements have been performed.}
    \Statex {}
    \Procedure{MeasurementConditionClause}{\ }
    \State {numRequired $\gets$ Number of required measurements for tomography}
    \State {numCurrent $= 0$}
    \State {enoughMeasurements $=$ false}
    \If{numCurrent $<$ numRequired}
      \State {numCurrent$++$}
      \State {return enoughMeasurements=false}
    \Else
      \State {return enoughMeasurements=true}
    \EndIf
    \State {return enoughMeasurements}
  \EndProcedure
  \end{algorithmic}
\end{algorithm}

\subsubsection{Action}

The Action for the tomography selects a qubit, and performs a measurement on it based on a randomly selected basis.

\begin{algorithm}[H]
  \caption{TomographyAction}
  \label{TomographyAction}
  \begin{algorithmic}[1]
  \Statex {\bf{This Action performs measurement on an entangled qubit.
  The measurement result and the selected basis is stored, and sent to the partner node as a message.}}
  \Statex{Input: resourceList $\gets$ List of allocated resources for the Rule.}
  \Statex {Output: msg $\gets$ A message for another node.}
  \Statex {}
  \Require {ResourceConditionClause == true \&\& MeasurementConditionClause == true}
  \Procedure{TomographyAction}{resourceList}
    \State {self\_addr $\gets$ Node address of this RuleSet owner.}
    \State {partner\_addr $\gets$ Node address of tomography partner.}
    \State {resource = Select a free resource from resourceList}
    \State {basis = RandomBasisSelect(\{X,Y,Z\})}
    \State {outcome = resource.Measure(basis)}
    \State {removeResourceFromList(resourceList,resource)}
    \State {Data.result = \{outcome, basis\}}
    \State {Data.ruleSetId = this.RuleSetID}
    \State {Data.ruleId = this.RuleID}
    \State {Data.actionIndex = this.ActionIndex}
    \State {save(Data)}
    \State {msg.destination = partner\_addr}
    \State {msg.source = self\_addr}
    \State {msg.data = Data}
    \State {this.ActionIndex$++$}
    \State {return msg}
  \EndProcedure
  \end{algorithmic}
\end{algorithm}

Selected measurement basis and its outcome will be stored locally, and shared to the partner node.

\chapter{Quantum network simulator}
\label{architecture}

The quantum network simulator implemented for this thesis is built on top of a publicly available discrete event network simulator named \emph{OMNeT++}~\cite{Varga:2008:OOS:1416222.1416290}.
OMNeT++ provides users an easy framework for defining network topologies and system area topologies for nodes.
Each software/hardware element is defined as a \emph{module} in OMNeT++, and their functionalities must be programmed by the user.
However, some of the basic functions, such as for running Dijkstra's algorithm (weighted or not), are provided as a standard function.
Different classical packets can also be defined very easily as OMNeT++ messages.
Simulations are runnable in CUI mode, or in GUI mode with animation.

Because OMNeT++ is a classical networking simulator, components for simulating the physics in quantum networking must be added.
Basic mathematics, such as for measuring a qubit with a specific density matrix, have been accomplished though the use of \emph{Eigen} package~\cite{eigenweb}.
Physical entanglement of a qubit module has been tracked using a pointer (pointing to another qubit module), and distributed states are directly accessed and updated on demand.

\begin{figure}[htbp]
  \center
  \includegraphics[keepaspectratio,scale=0.33]{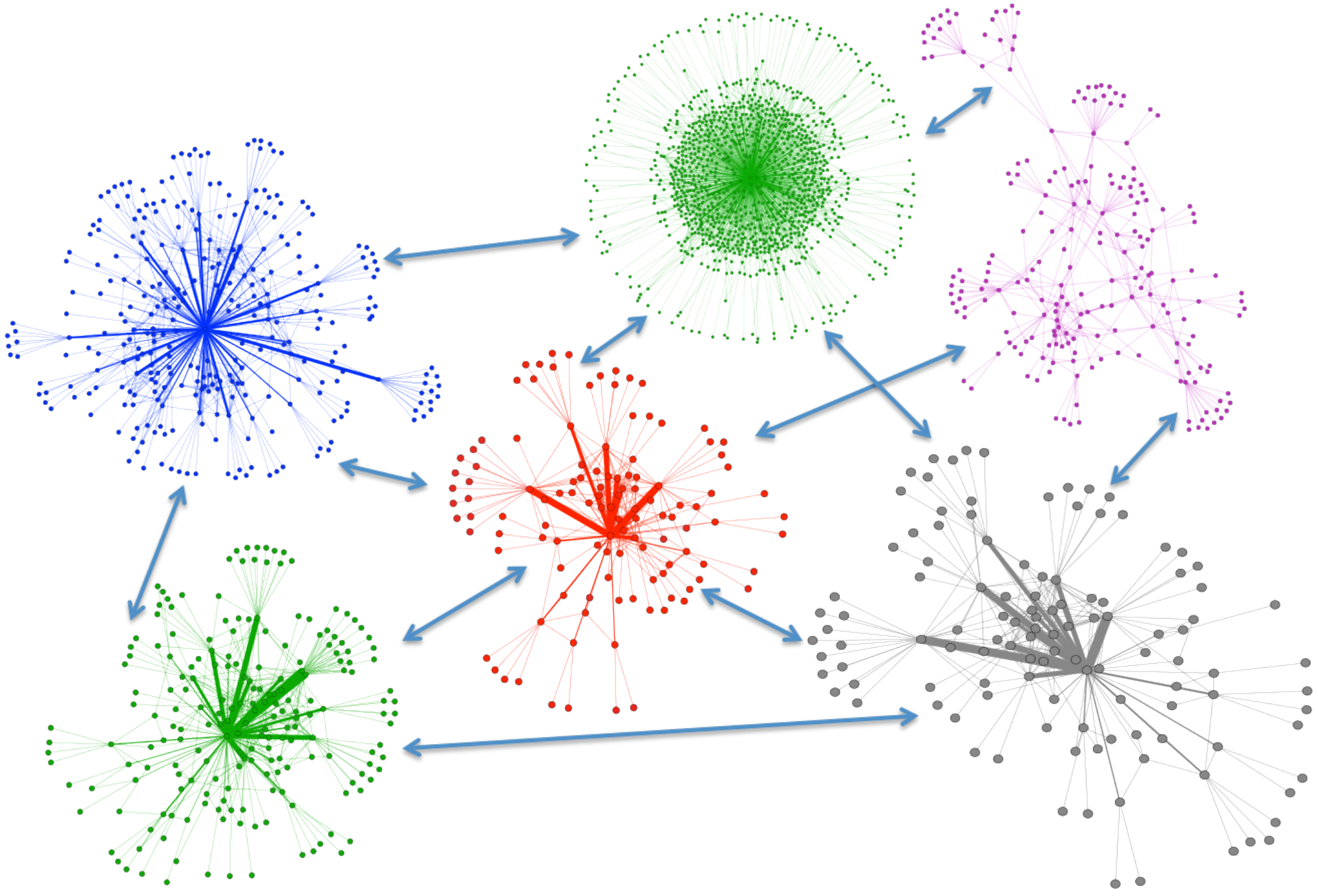}
  \caption{The Quantum Internet. Different networks are operated under different technologies. That includes the physical link architecture, data link protocol, routing protocol and any other necessary technology layered as in the OSI reference model~\cite{zimmermann1980osi}. Connecting those networks forms the Quantum Internet.}
  \label{QuantumInternet}
\end{figure}

The overall simulator size for this thesis is approximately 10,000 lines of code (see Appendix.~\ref{SIMappendix}) exclusive of OMNeT++ and Eigen.
The following subsections discuss the overall structure of the simulator and its flow.
The simulator is designed to evaluate various protocols over a single network, and over the quantum Internet.
The Quantum Internet will be a world-wide network interconnecting diverse quantum networks, both small scale and large scale~\cite{Kimble2008,VanMeter:2014:QN:2683776, Wehnereaam9288,irtf-qirg-principles-00}.
These independent, interconnected networks are utilizing different technologies and managed by different organizations (see Fig.~\ref{QuantumInternet}), known as Autonomous Systems in the classical Internet.
The role of such a network is similar to the classical Internet: to provide to users a quantum information service between arbitrary nodes.

\section{Design of a Quantum node architecture}
\label{QNarchitecture}

\begin{figure}[H]
  \center
  \includegraphics[keepaspectratio,scale=0.62]{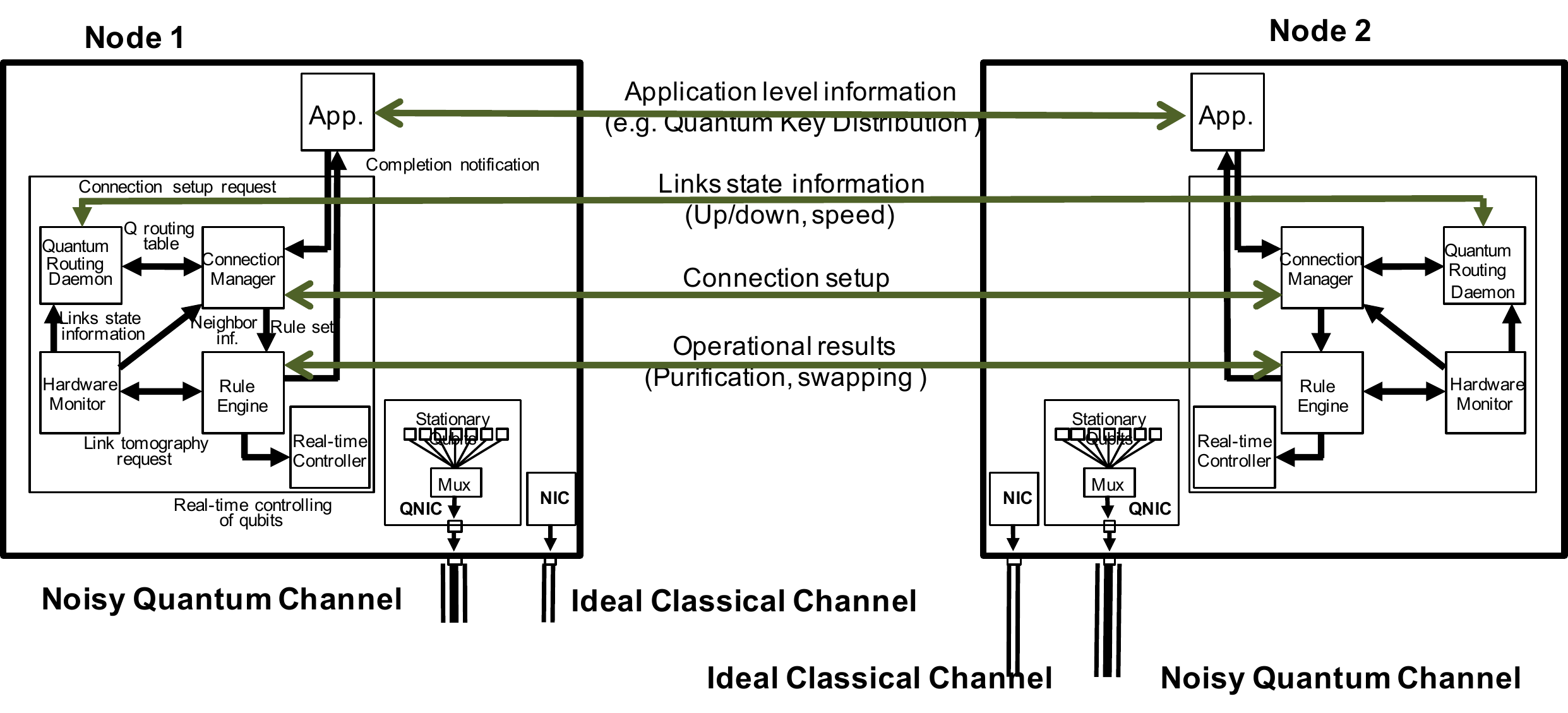}
  \caption{Quantum node architecture.}
  \label{QNode}
\end{figure}

Building a Quantum network simulator requires designing the hardware/software architecture of a quantum
node, which includes end-nodes, quantum repeaters and quantum routers.
The challenge is that designing the architecture requires understanding the required module
functionalities and protocol specifications across different
layers.
In our work, we define a node with a single quantum
network interface card (QNIC) as an \emph{end-node}. Similar
to the classical network interface card (NIC), the QNIC is
connected to another node's QNIC via a quantum channel.
The end-node regularly entangles its own qubits with its neighbor's.
An end-node also generates and transmits a classical connection setup request to another end-node through the classical channel.
A node with two QNICs is a \emph{quantum repeater}.
The role of a quantum repeater is to forward classical packets, to create base-level
entanglement over a link and to monitor and manage errors.
The third type of nodes are the \emph{quantum routers}, which have similar roles to quantum repeaters, but also takes responsibility for routing, for both classical and quantum information.
A border quantum router, which is a router connected to another border quantum router of a different network, may have minimum trust between
networks -- a router may have no knowledge of the internals of other autonomous networks. Backbone routers
also physically and logically link heterogenous networks.
Below is the architecture of the simplest quantum node, which is the end-node (see Fig.~\ref{QNode})
 -- a quantum repeater or a quantum router has the same architecture, except for the number of QNICs and its need to correctly route packets.
Note that the required hardware/software still varies depending on the design decisions.
The design is intended for enabling the RuleSet-based communication protocol.

\subsection{Network Interface Card (NIC)}
A NIC is connected to the classical channel, and is capable of forwarding classical packets out to the channel.

{\bf Assumption:}
All NICs are connected to ideal classical channels -- no error and infinite bandwidth and the same latency as corresponding QNICs.

\subsection{Quantum Network Interface Card (QNIC)}
A QNIC is connected to the quantum channel.
This module contains multiple quantum memories, each capable of emitting one entangled photon out to the channel through the multiplexer (MUX).

{\bf Assumption:}
The simulation assumes the ability to perform any multi-qubit operations between qubits in any local QNICs.
The error rate of a particular quantum gate can be set on a node-by-node basis but is fixed throughout the simulation.
Gate times are assumed to be negligible.

\subsection{Application module}
The application generates the connection setup request to an arbitrary end-node, and passes
the request to the connection manager. The connection setup request may include the required
number of Bell-pairs and their minimum fidelity for
the application. This module may directly be controlled by a user.

\subsection{Hardware Monitor}
The hardware monitor knows which local QNIC is connected to which neighboring node.
An example of the table is shown below in Tab.~\ref{table_itf}.

\begin{table}[H]
  \begin{center}
    \caption{Example of a table keeping the pair of the local QNIC id and its connected neighbor address. Neighboring nodes of Node 1 are assumed to be Node 2, 3 and 4}
    \label{table_itf}
    \begin{tabular}{|c||c|}
      \hline
      Next hop & QNIC id \\
      \hline
      \hline
      Node 2 & QNIC 0\\
      \hline
      Node 3 & QNIC 1\\
      \hline
      Node 4 & QNIC 2\\
      \hline
    \end{tabular}
  \end{center}
\end{table}

Its major role is to monitor and update the states of quantum links (up or down), by generating a request for checking the link quality based on tomography.
When received a request, the hardware monitor also generates a RuleSet for the link monitoring.
Based on the accumulated measurement results (basis and outcome), the hardware monitor reconstructs the density matrix based on a given algorithm (for details regarding the full-state reconstruction, see Sec.~\ref{TOMOGRAPHYsection}).
The acquired link information may be stored by the node as in Tab.~\ref{table_quality}.

\begin{table}[H]
  \begin{center}
    \caption{Example of a table keeping the link fidelity and the link throughout information.}
    \label{table_quality}
    \begin{tabular}{|c||c|c|}
      \hline
      QNIC id & Fidelity & Throughput (Bell pairs per second)\\
      \hline
      \hline
      QNIC 0 & 0.88 & 3312\\
      \hline
      QNIC 1 & 0.89 & 10011\\
      \hline
      QNIC 2 & 0.93 & 5504\\
      \hline
    \end{tabular}
  \end{center}
\end{table}

The estimated link quality is then provided to the local routing daemon.
The overall flow of the process has been provided in Figure~.\ref{HMflow} in section \ref{RSPsection}.

{\bf Assumption:}
The hardware monitor is assumed to be able to access to the exact length of the channels.

\subsection{Routing Daemon}
\begin{figure}[H]
  \center
  \includegraphics[keepaspectratio,scale=0.5]{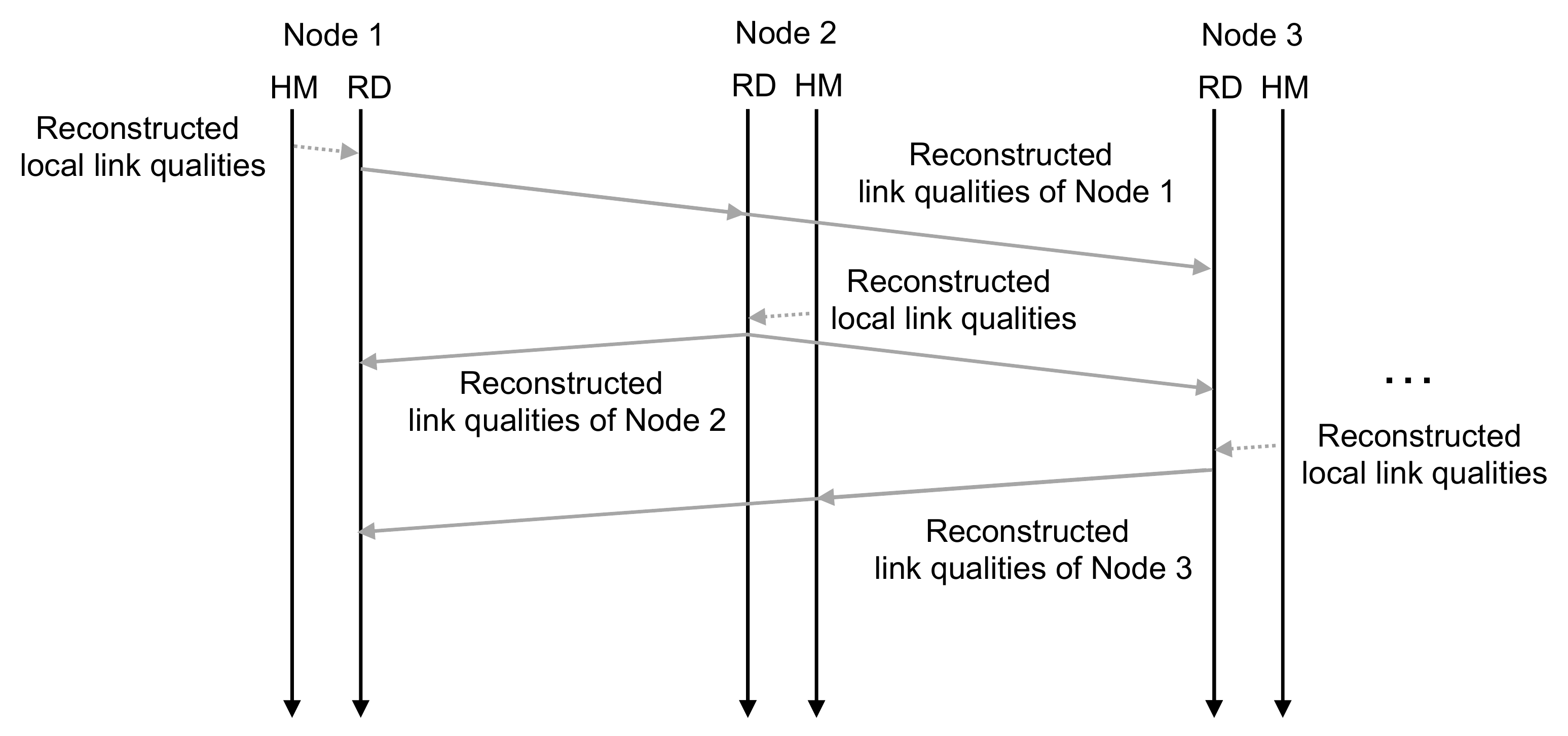}
  \caption[Message exchange of the routing daemon exchanging link qualities. ]{Message exchange of the routing daemon exchanging link qualities. RD is Routing Daemon and HM is Hardware Monitor. Link quality may include the achievable fidelity with its corresponding throughput.}
  \label{RDflow}
\end{figure}

The routing daemon communicates with the neighbors to exchange the link state information received from the hardware monitor.
The module accumulates all node/link information in a single network, and creates a quantum routing table based on a predefined algorithm (e.g. Dijkstra's algorithm~\cite{VanMeter2013}).
This daemon, like its classical counterpart, can theoretically support a variety of routing protocols, whether link state, distance vector, or other.
The BSA node is not visible in the routing system because the BSA can be considered as a part of the physical link.
The message exchange timing diagram with three nodes is provided in Fig.~\ref{RDflow}.

\begin{figure}[H]
  \center
  \includegraphics[keepaspectratio,scale=0.5]{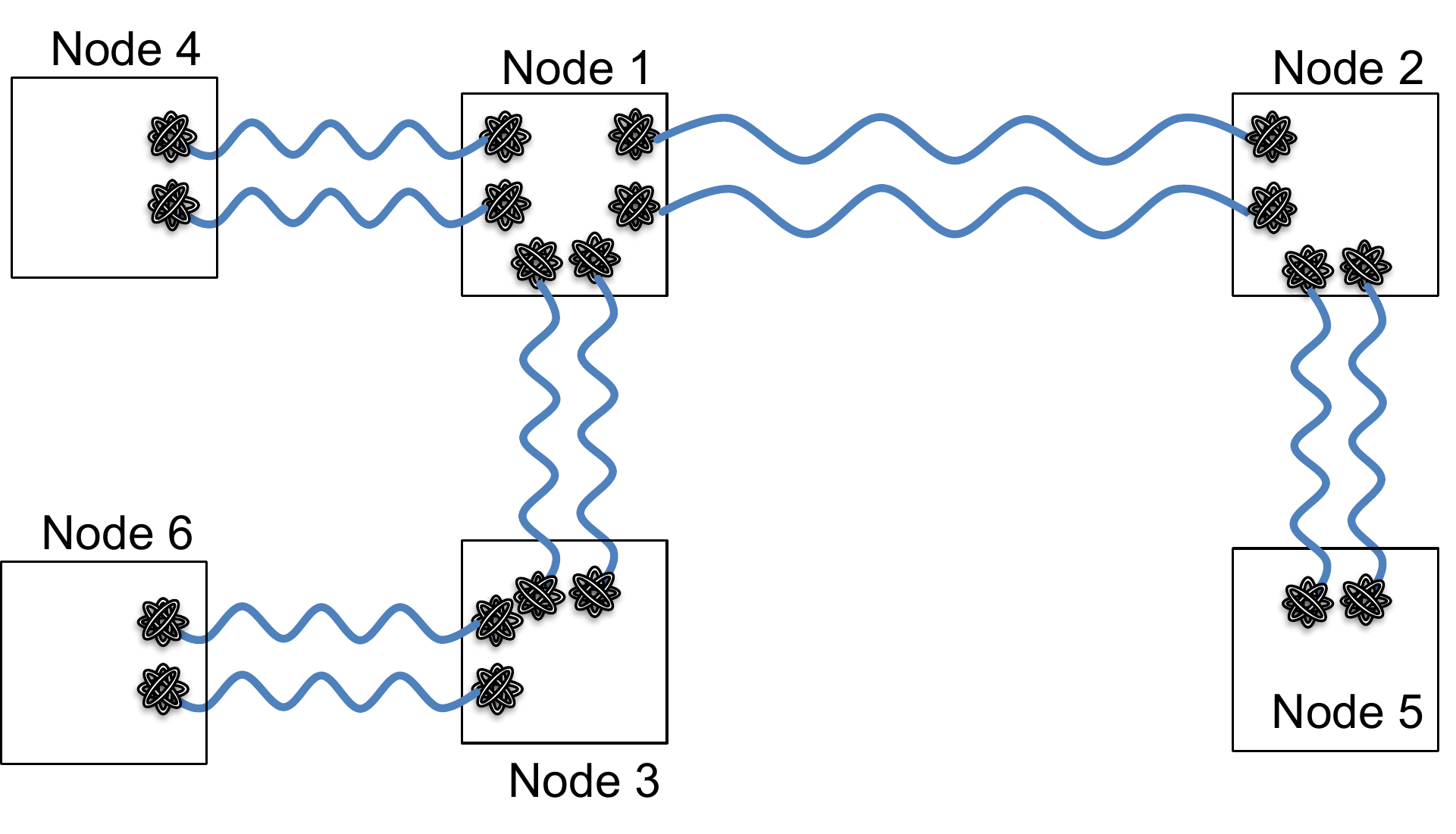}
  \caption[Example topology based on the routing table in Tab.\ref{table_routing}.]{Example topology based on the routing table in Tab.\ref{table_routing}. Here, there is only one path available to each node from Node 1.}
  \label{topoexample}
\end{figure}

A quantum routing table is distinct from the classical routing table, but the overall structure and use remains the same (see Tab.~\ref{table_routing}).
The corresponding example topolpogy is shown in Fig.\ref{topoexample}.

\begin{table}[H]
  \begin{center}
    \caption[Example quantum routing table of Node 1.]{Example quantum routing table of Node 1. A node only knows which neighbor to communicate with. Neighboring nodes of Node 1 are assumed to be Node 2, 3 and 4}
    \label{table_routing}
    \begin{tabular}{|c||c|}
      \hline
      Destination & Next hop \\
      \hline
      \hline
      Node 2 & Node 2\\
      \hline
      Node 3 & Node 3\\
      \hline
      Node 4 & Node 4\\
      \hline
      Node 5 & Node 2\\
      \hline
      Node 6 & Node 3\\
      \hline
    \end{tabular}
  \end{center}
\end{table}

\subsection{Connection Manager}
\begin{figure}[H]
  \center
  \includegraphics[keepaspectratio,scale=0.5]{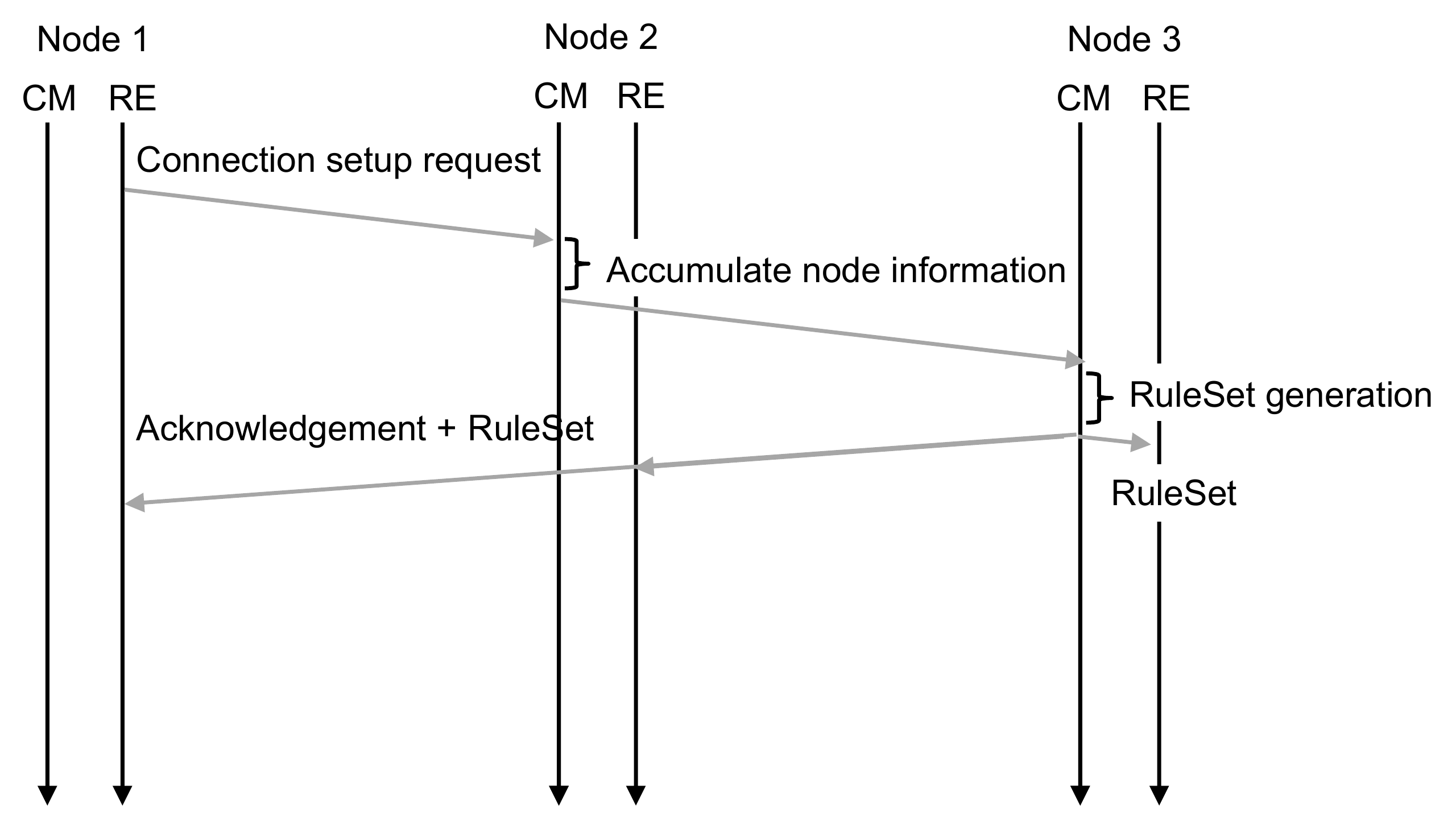}
  \caption[Flow chart of the connection setup including three nodes.]{Flow chart of the connection setup including three nodes. RE is Rule Engine and CM is the Connection Manager.}
  \label{CMflow}
\end{figure}

The connection manager forwards the connection
setup packet to the next quantum node by referring to the quantum routing Tab. The destination
node creates the rule set for a particular connection,
and distributes it back to all involving nodes (see Fig.~\ref{CMflow}).
A RuleSet consists of multiple Rules, each with an Action and a Condition.
The Action is a set of operations for accomplishing a single task, and the Condition tells when to execute it.
For details, see Sec.~\ref{RSPsection}.

\subsection{Rule Engine}
The rule engine is event-driven (e.g. classical packet arrival, new resource allocation or RuleSet timeout).
It is responsible for interpreting and executing operations in real time through the real-time controller based on the received RuleSet(s).
The rule engine also executes the quantum data link protocol for generating new resources (see Fig.~\ref{REflow} in section~\ref{qdsec}).
Those newly generated resources are assigned to the running RuleSets by the RuleEngine.

{\bf Assumption:}
Rule engines have perfectly synchronized clocks between nodes.
Therefore, for example, each node is capable of transmitting a photon exactly at the desired timing.

\subsection{Real-time Controller}
The real-time controller has direct access to all local QNICs and their stationary qubits.
Quantum operation, such as qubit measurement, is accomplished through this module \footnote{In the simulator, however, the Rule Engine directly executes the function of an Action.}.

\section{Design of a Bell State Analyzer (BSA) node/module architecture}
\label{BSAarchitecture}
\begin{figure}[H]
  \center
  \includegraphics[keepaspectratio,scale=0.45]{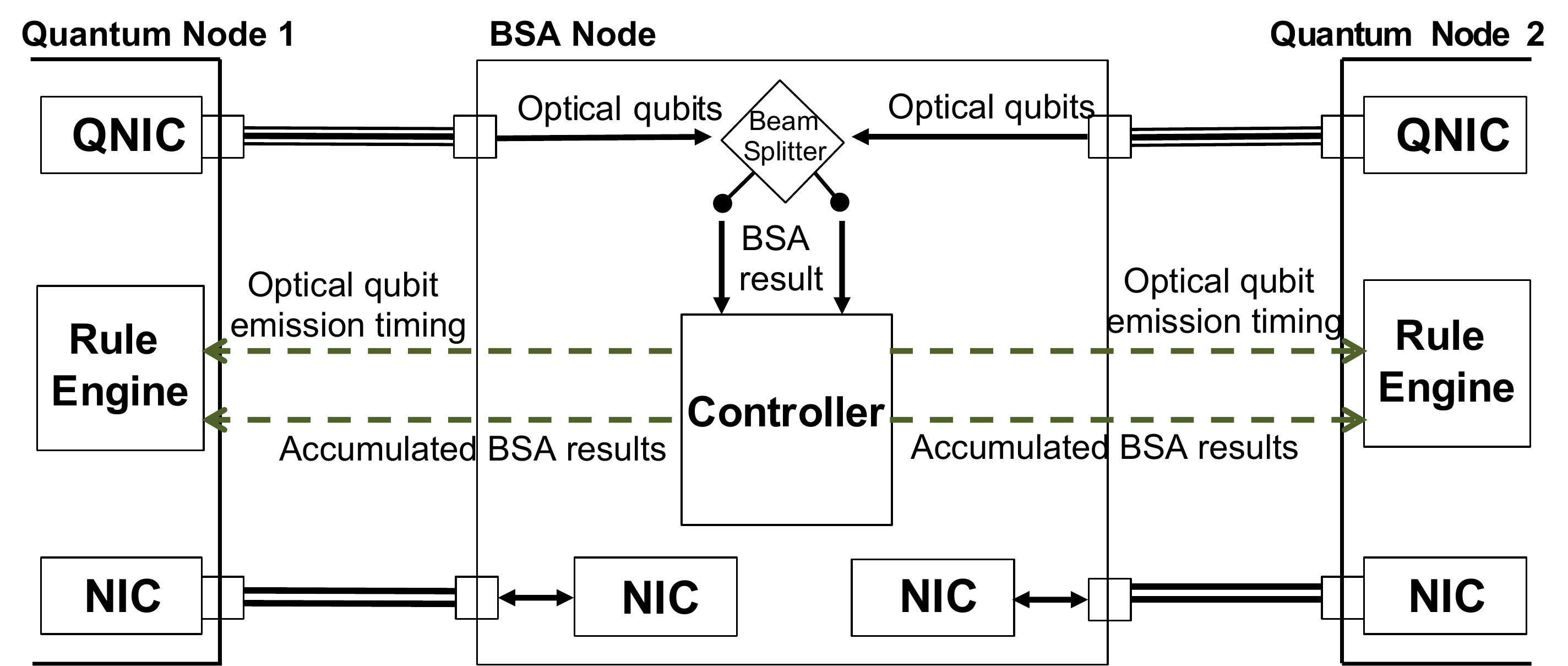}
  \caption[BSA node architecture.]{BSA node architecture. Classical communications, shown as dashed arrows, will be sent through the classical channel via NICs.}
  \label{BSAnode}
\end{figure}

Below is the architecture of a BSA node (see Fig.~\ref{BSAnode}).
The BSA is installed in between two nodes along the quantum channel for a MeetInTheMiddle link, or is directly installed at one endpoint's QNIC for a SenderReceiver link.
This node's main role is to coordinate the resource generations of neighboring nodes.

\subsection{Beamsplitter \& Photon detectors}

Optical qubits arrive at the beamsplitter from QNICs via quantum channels.
The beamsplitter with its detectors' role is to perform Bell measurements on the received qubits to stochastically generate entangled resources.
The two detectors' results will be forwarded to the controller module.
The first step is to establish that photons can be made indistinguishable from the two nodes.
This is confirmed using a Hong-Ou-Mandel test.
Then hardware parameters are adjusted to erase the which-path information while detecting photons at both detectors.

\subsection{Controller}
The controller calculates the emission timings for both nodes, and also forwards the accumulated success/failure results.

{\bf Assumption:}
The controller also assumes that each node has the ability to emit photons at the exact timing provided.

\section{Simulating errors}

The error evolution is modeled based on Markov-Chain Monte-Carlo simulation.

\subsection{Noisy memory errors}
When a qubit is waiting for some time $t$ since its initialization, we may observe an error on that qubit.
The simulator uses a row vector to describe the present state of a qubit, which is one of the seven distinguishable states -- no error, X error, Z error, Y error, excited, relaxed or completely mixed.
For example, a Bell pair with no error as an initial input state can be described as in Eq.~\ref{inputvector}.

\begin{eqnarray}
\label{inputvector}
\vec{\pi}(0) =\;\begin{blockarray}{cccccccc}
\scriptstyle Clean & \scriptstyle X{ }error & \scriptstyle Z{ }error & \scriptstyle Y{ }error & \scriptstyle Excited & \scriptstyle Relaxed & \scriptstyle Mixed \\
\begin{block}{(ccccccc)c}
  1 & 0 & 0 & 0 & 0 & 0 & 0 &  \\
\end{block}
\end{blockarray}
\end{eqnarray}

Accordingly, the infinitesimal generator for our continuous-time Markov-Chain~\cite{kleinrock1974qsv1} we use for the memory error simulation also consists of seven states,

\begin{eqnarray}
\label{infinitesimalgenerator}
\overline{Q} =\;\begin{blockarray}{cccccccc}
\scriptstyle Clean & \scriptstyle X{ }error & \scriptstyle Z{ }error & \scriptstyle Y{ }error & \scriptstyle Excited & \scriptstyle Relaxed & \scriptstyle Mixed \\
\begin{block}{(ccccccc)c}
  (1-\Sigma_{0}) & P_X & P_Z & P_Y & P_E & P_R & 0 &  \\
  P_X & (1-\Sigma_{1}) & P_Y & P_Z & P_E & P_R & 0 & \\
  P_Z & P_Y & (1-\Sigma_{2}) & P_X & P_E & P_R & 0 & \\
  P_Y & P_Z & P_X & (1-\Sigma_{3}) & P_E & P_R & 0 & \\
  0 & 0 & 0 & 0 & (1-\Sigma_{4}) & P_R & 0 & \\
  0 & 0 & 0 & 0 & P_E & (1-\Sigma_{5})  & 0 & \\
  0 & 0 & 0 & 0 & P_E & P_R & (1-\Sigma_{6}) & \\
\end{block}
\end{blockarray}.
\end{eqnarray}

Element $P_X$, $P_Y$ and $P_Z$ correspond to Pauli X, Y and Z error accordingly.
$P_E$ and $P_R$ are memory excitation and relaxation errors,
which are used to represent $T_1$ time and the Boltzmann thermal distribution.
$\Sigma_{i}$ is the sum of every other element in row $i$.
Mixed state occurs due to dark counts, or when the paired qubit gets excited/relaxed.
The probability distribution of a qubit output state after time $t$ can be found by:

\begin{eqnarray}
 \vec{\pi}(t)=\vec{\pi}(0)\overline{Q}^t.
\end{eqnarray}

Each element in the output vector $\vec{\pi}(t)$ represents the probability for the qubit being in the corresponding state.
The qubit state can be determined via a random selection based on the output probability distribution,
 and the density matrix is constructed on demand.
 This process is independently executed whenever a qubit is soon to be used by the simulator.

 \subsection{Noisy channels}
The simulation of a noisy channel is done similarly to the memory error simulation, except the error rate will be per unit distance.
Because each channel between nodes have a static length, we only have to exponentiate the matrix once for each channel at the beginning of the simulation.
The infinitesimal generator describing the error model per unit distance for the quantum channel is:

\begin{eqnarray}
\label{infinitesimalgenerator}
\overline{Q} =\;\begin{blockarray}{ccccc}
\scriptstyle Clean & \scriptstyle X{ }error & \scriptstyle Z{ }error & \scriptstyle Y{ }error & \scriptstyle Lost \\
\begin{block}{(ccccc)}
  (1-\Sigma_{0}) & P_X & P_Z & P_Y & P_L  \\
  P_X & (1-\Sigma_{1}) & P_Y & P_Z & P_L  \\
  P_Z & P_Y & (1-\Sigma_{2}) & P_X & P_L  \\
  P_Y & P_Z & P_X & (1-\Sigma_{3}) & P_L  \\
  0 & 0 & 0 & 0 & 1 \\
\end{block}
\end{blockarray}.
\end{eqnarray}

Element $P_X$, $P_Y$, $P_Z$ and $P_L$ correspond to Pauli X, Y, Z, and photon loss error accordingly.
$\Sigma_{i}$ is the sum of every other element in row $i$.
Given the input state $\vec{\pi}(0)$, the probability distribution of a qubit output state after traveling a channel with distance of $d$ can be found by:

\begin{eqnarray}
 \vec{\pi}(d)=\vec{\pi}(0)\overline{Q}^d.
\end{eqnarray}

\subsection{Bell State Analysis (BSA) error}

\begin{figure}[H]
  \center
  \includegraphics[keepaspectratio,scale=0.7]{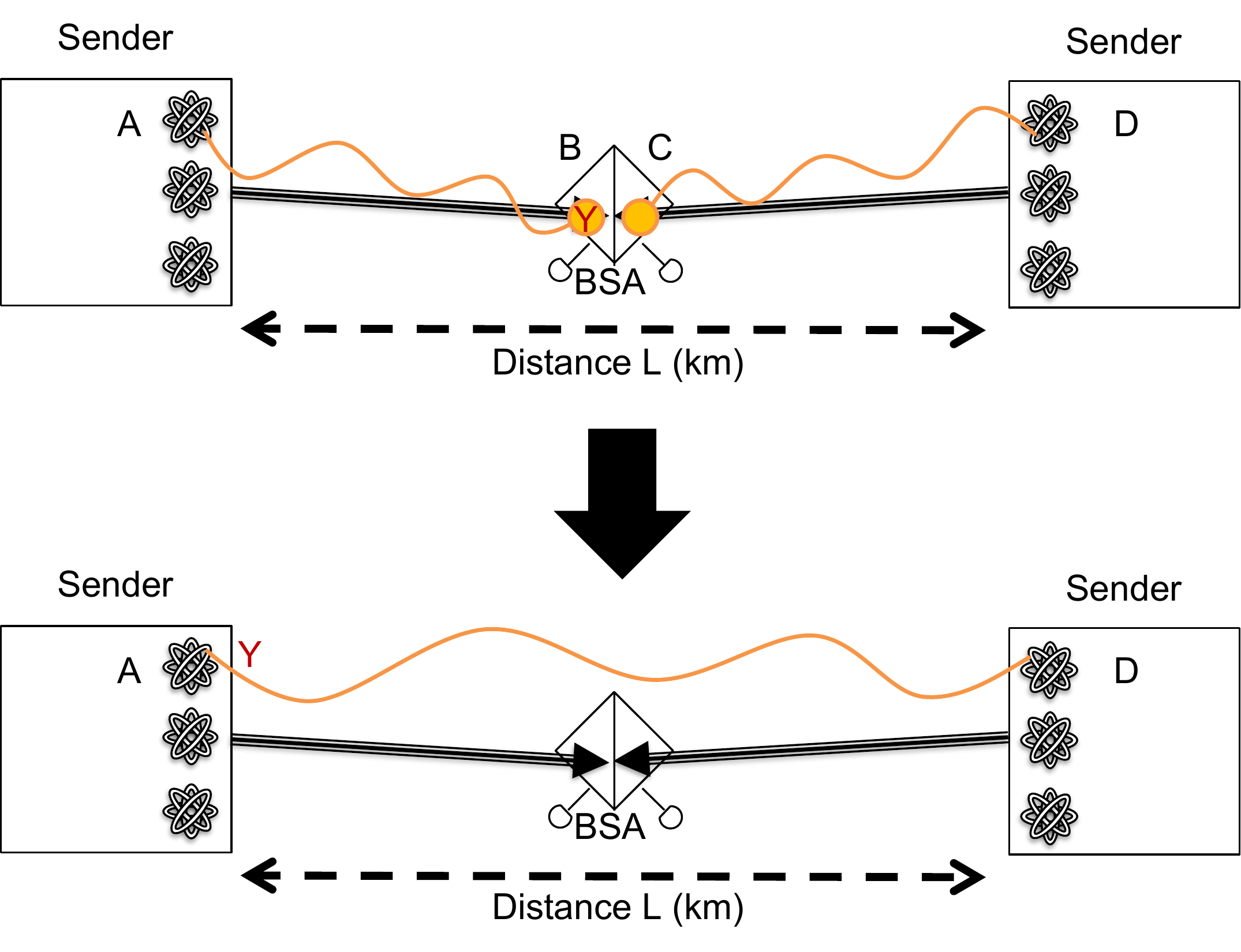}
  \caption{Bell measurement with a Y error on the received photon.}
  \label{BSAerror}
\end{figure}

There are cases where the BSA succeeds in entangling the memory qubits with errors on the consumed photons.
The Fig.~\ref{BSAerror} is a case where one of the received photon included a Y error.
In this case, if the BSA succeeds in entangling the memory qubits, the Y error will be reapplied as an error on the source memory qubit.
Notice that there is no physical difference between $Y_A\ket{\Psi^+_{A,B}}$ and $Y_B\ket{\Psi^+_{A,B}}$.

A photon may also be lost while traveling through the channel.
Most of the cases, this error will be detected because less than 2 detector will click.
However, depending on the detector darkcount rate, the system may mis-recognize qubits as entangled.
In this case, both memory qubits will have either $\ket{0}$ or $\ket{1}$ classically.

\begin{figure}[H]
  \center
  \includegraphics[keepaspectratio,scale=0.7]{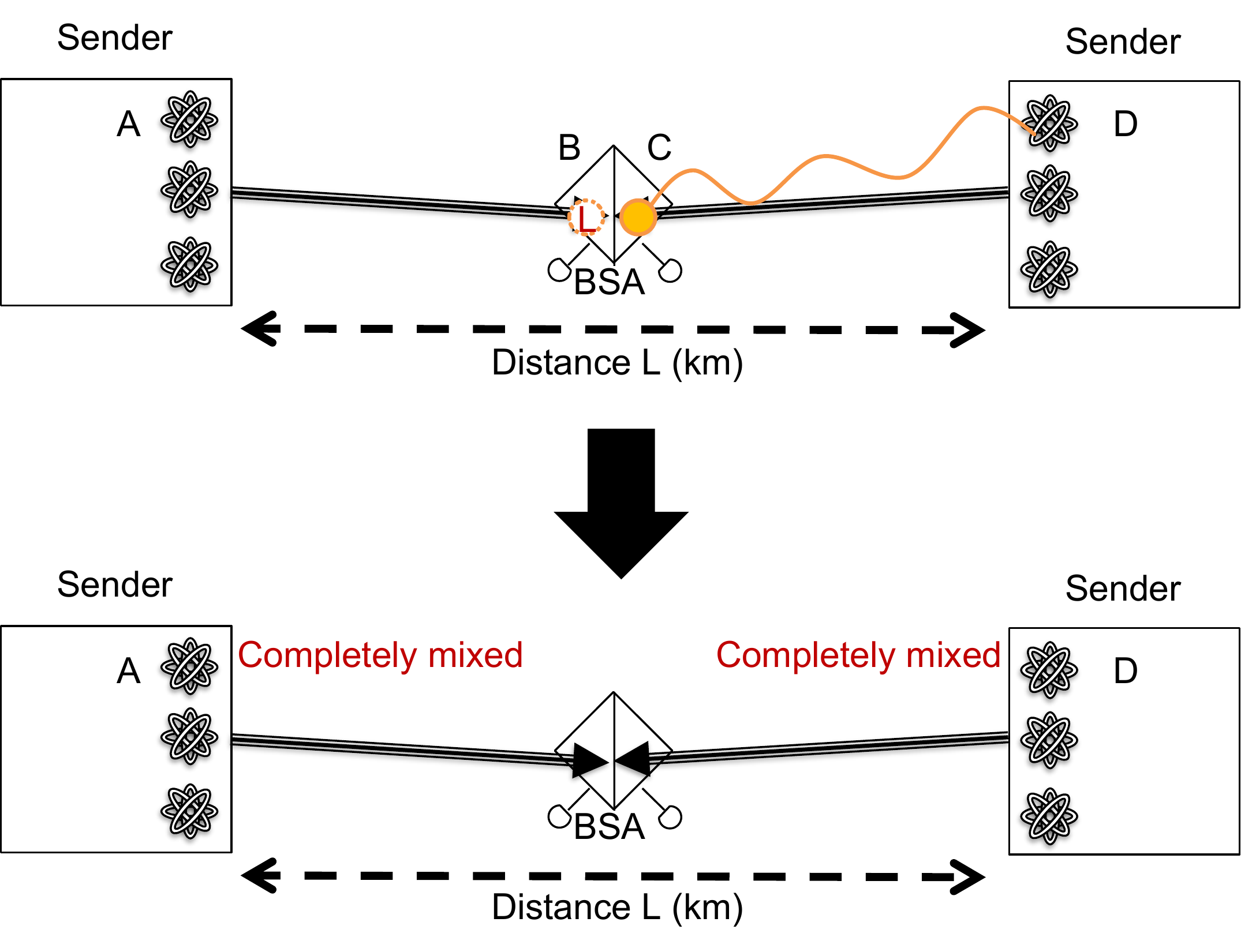}
  \caption{Bell measurement with a lost error on the received photon.}
  \label{BSAerror2}
\end{figure}

 \subsection{Noisy quantum gates}
Quantum gates are modeled to stochastically generate one of the Pauli errors (X, Y or Z error) per each attempt based on a user-defined static probability distribution.
To simulate the imperfection of a quantum circuit, the simulator classically tracks the error propagation throughout the circuit based on a standard Monte-Carlo simulation.
Notice that errors on the input state will of course propagate through the circuit, but each gate may also generate a new error.
For details regarding error propagations, see Sec.~\ref{ERRORsection}.

\chapter{Evaluation}
\label{evaluation}

In this chapter, we study the capability of a first generation quantum repeater link, with a limited amount of resources,
 and under noisy conditions generated by various hardware imperfections.
Simulated error sources include quantum memories, quantum channels, quantum gates and single photon detector darkcounts.
Our Markov-Chain Monte-Carlo simulation uses similar parameters as in~\cite{2018arXiv180900364R} but with larger memory buffers (see Table~\ref{err} for details).
We use the memoryless Markov-Chain in the simulation to dynamically model errors on memory qubits independently, based on a given lifetime $T_1$~\cite{van-meter19:errm}.
Other error sources, such as the gate error, are simulated via a random selection with static probabilities.
In order to reduce the computation time, we only propagate Pauli errors through circuits as in~\cite{PhysRevA.86.032331, PhysRevA.97.062328}.
The simulated purification circuit, therefore, is incapable of stochastically detecting excited/relaxed/completely mixed errors.
Hence, our simulation generates a pessimistic output fidelity, and an optimistic output resource generation rate.
To concretely identify the real impact of imperfect quantum systems, we assume ideal classical communication channels.
Because classical communication latencies cover the majority of the simulated time, we assume negligible gate times.
Furthermore, multi-qubit operations can also be performed between arbitrary qubits across QNICs within the same node.
For each data point in the figures, a total of 25 simulations has been performed to quantify the average behavior.

  \begin{table}[H]
    \caption{Default parameters used for all simulations (unless explicitly mentioned).}
    \label{err}
    \centering
    \begin{tabular}{|l|c|}
    \hline
    Key & Value \\
    \hline\hline
    Fiber refractive index & $1.44$~\cite{fiber} \\
   Fiber Pauli error rate (per km) & $0.03$ \tablefootnote{\label{pauli}Including X, Y and Z error.}\\
   Fiber photon loss rate (per km) & 0.04501 \tablefootnote{\label{db}Equivalent to 0.2dB/km.}~\cite{fiber}\\
   Memory Pauli error rate (per sec) & 1/3 \textsuperscript{\ref{pauli}}~\cite{Maurer1283}\\
   Memory lifetime & 50ms \tablefootnote{\label{memory}Memory lifetime can be, for example, up to the order to seconds~\cite{Maurer1283} depending on the system. We use an artificial value of 50ms for the simulation. The ratio of excitation and relaxation probability is set to 100:1.}\\
   Emission probability into the zero phonon line & 0.46 \tablefootnote{\label{emission}Photon emission probability (memory-to-fiber) is the product of the emission probability into the zero phonon line and the collection efficiency.}~\cite{PhysRevX.7.031040} \\
   Photon collection efficiency & 0.49 \textsuperscript{\ref{emission}}~\cite{HensenNature, doi:10.1063.1.5001144}\\
   Photon detector efficiency & 0.8~\cite{doi:10.1063.1.5001144} \\
   Photon detector darkcount rate (per sec) & 10~\cite{doi:10.1063.1.5001144} \\
   Photon detector recovery time  & 1ns~\cite{SNSPDs} \\
   Single-qubit gate error rate  & 0.0005 \textsuperscript{\ref{pauli}}~\cite{PhysRevX.6.021040} \\
   Multi-qubit gate error rate & 0.02 \textsuperscript{\ref{pauli}}~\cite{PCNature} \\
   Measurement error rate & 0.05 \textsuperscript{\ref{pauli}}~\cite{Kalb928} \\
   Number of memory qubits (per QNIC) & 100\\
    \hline
    \end{tabular}
  \end{table}

\section{Performing simple tomography with different numbers of measurements}
Performing quantum state tomography with more measurements typically results in a greater accuracy of the reconstructed density matrix.
Although the required accuracy may depend on the demanded precision of a particular application,
it is unlikely to perform an extremely large set of measurements solely for the reconstruction for purposes such as real time quantum channel monitoring.
We first investigate how the number of measurements for the tomography, $N_{M}$,
 impacts the accuracy of the \emph{reconstructed fidelity}, $F_{r}$, between adjacent nodes.
Here, $\mathrm{F_{r}} = \mathrm{Tr}[\rho_{r}\rho_{i}]$,
where $\rho_{i}$ is the ideal density matrix of a Bell pair $\ket{\Phi^{+}}$
and $\rho_{r}$ is the density matrix reconstructed through the RuleSet-based quantum link-level tomography.
This particular RuleSet consists of a single Rule composed of two Clauses and an Action.
The first Clause, {\it MeasurementConditionClause}, tracks how many measurements have been performed.
The second Clause, {\it ResourceConditionClause}, checks the availability of resources (see Algorithm \ref{ResourceClause} in section~\ref{EXAMPLERULESsection}).
The Action performs the measurement for the link-level tomography (see Algorithm \ref{TomographyAction} in section~\ref{EXAMPLERULESsection}).
Because each node picks a measurement basis at random from \{X,Y,Z\}, there are 9 detector settings in total used for the process.
In this case, the Termination Condition also tracks the number of performed measurements, and stops the execution when the requirement is fulfilled.

\begin{figure}[H]
  \center
  \includegraphics[keepaspectratio,scale=0.4]{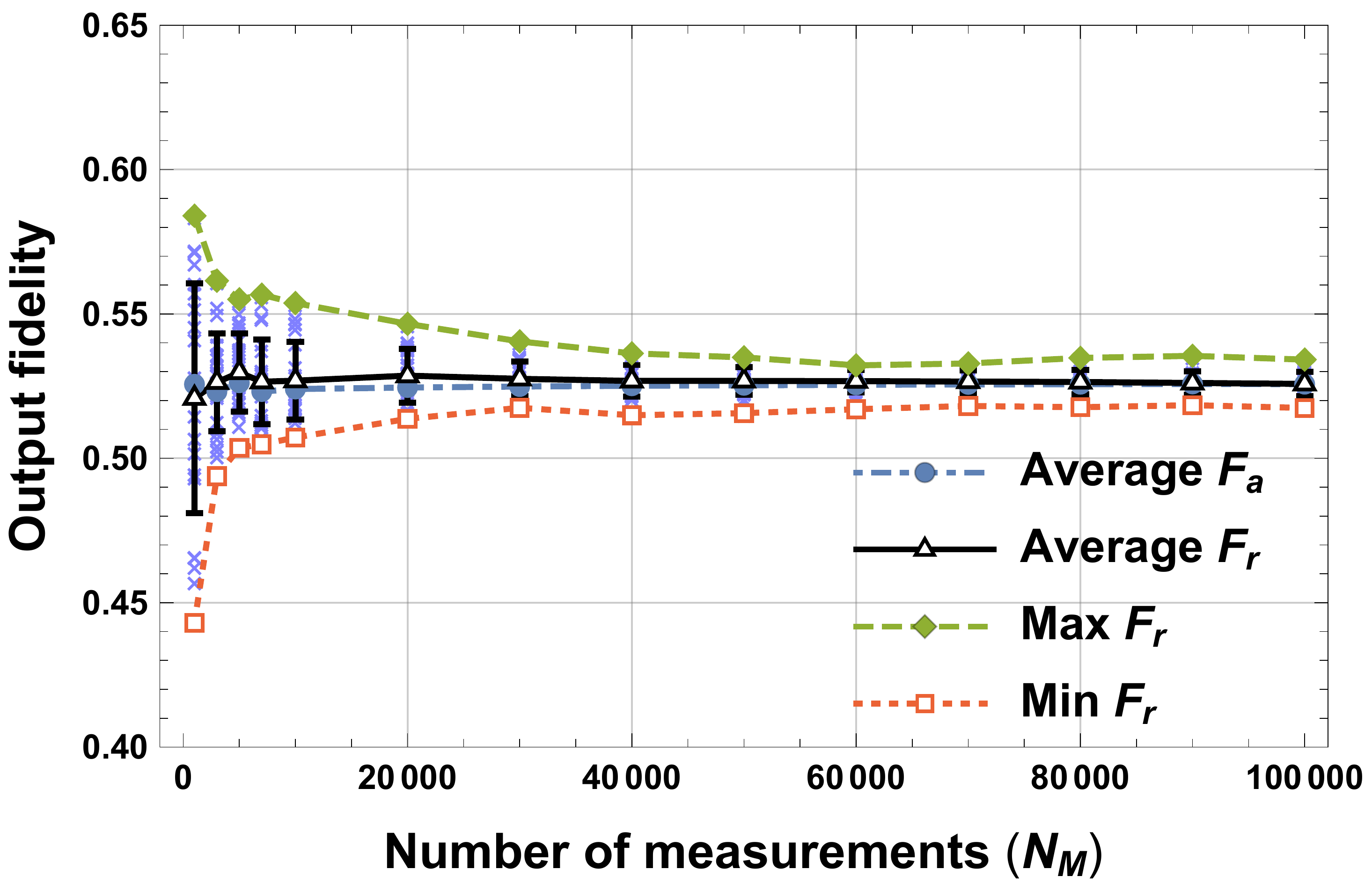}
  \caption[Impact of the number of measurements on the reconstructed fidelity error over the MeetInTheMiddle link]{Impact of the number of measurements on the reconstructed fidelity error over the MeetInTheMiddle link. $F_{r}$ is the reconstructed fidelity and $F_{a}$ is the actual fidelity.}
  \label{MeasurementCount}
\end{figure}

The simulated result of a repeater network composed of two nodes based on a MeetInTheMiddle link,
each 10km from the BSA node ($L=20km$), is shown in Fig.~\ref{MeasurementCount}.
Error bars represent one standard deviation uncertainty ($\sigma$) from the average $F_{r}$.
$\mbox{Max } F_{r}$ and $\mbox{Min } F_{r}$ are the highest and lowest fidelity observed within the 25 trials.
Purple data points marked with cross between the minimum and the maximum are individual simulation results.

As in Fig.~\ref{MeasurementCount}, the reconstructed fidelity has poor accuracy with only 1,000 measurements, and slightly underestimates the value simply due to the lack of samples -- $\sigma \approx 0.040$.
In order to achieve $\sigma < 0.015$, at least 7,000 measurement shots are required.
A total of 20,000 measurements is capable of reconstructing the fidelity with a standard deviation less than $1\%$,
and roughly converges to $\sigma \approx 0.004$ when $N_{M}>60,000$.

\begin{figure}[H]
  \center
  \includegraphics[keepaspectratio,scale=0.4]{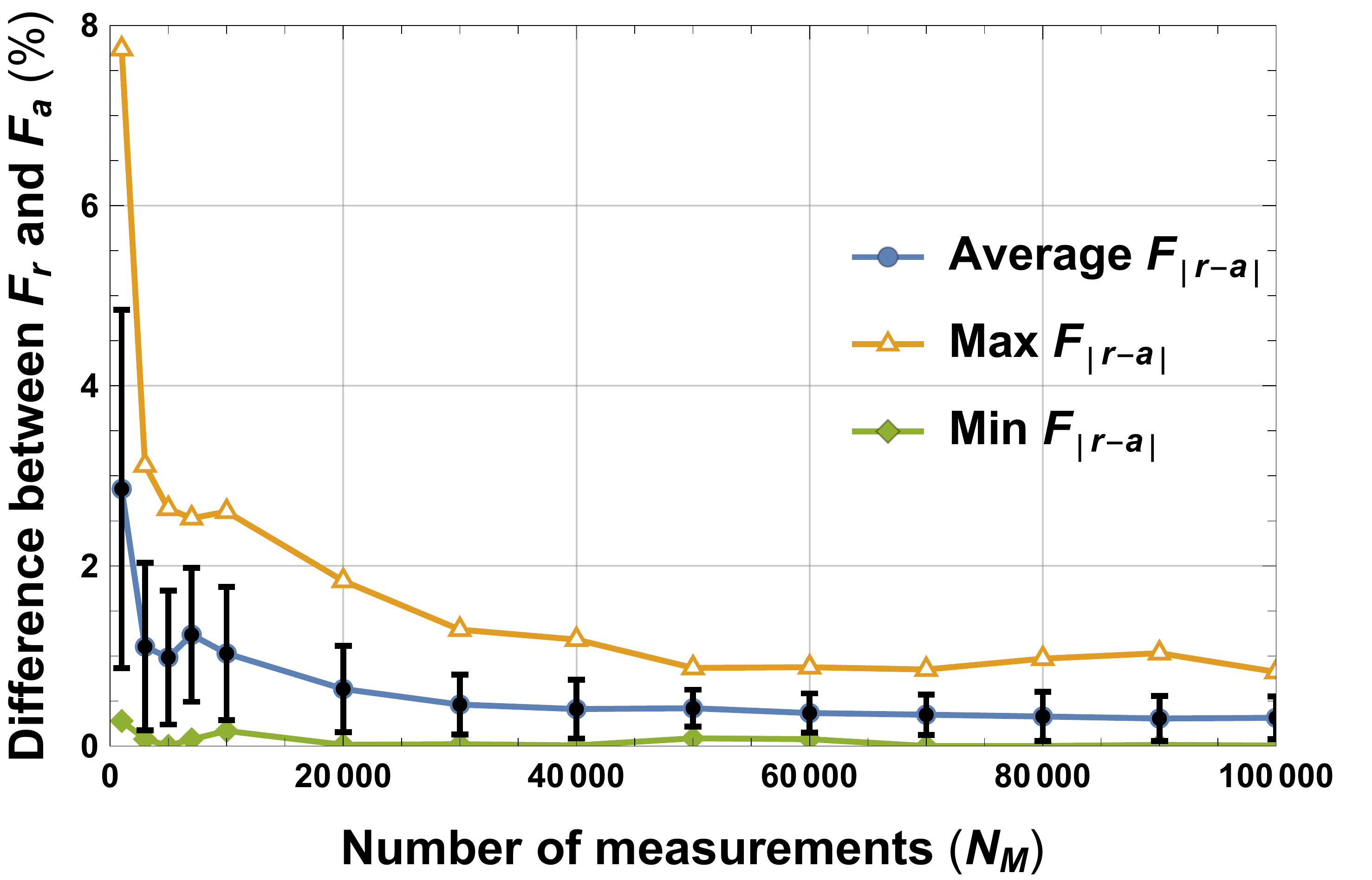}
  \caption{Change in the accuracy of the reconstructed fidelity relative to the actual fidelity.}
  \label{DifferenceF}
\end{figure}

$\mathrm{F_{a}}$ is the actual fidelity, accessible to us because this is a simulation, but of course hidden in the real world.
The change in the accuracy of the reconstructed fidelity compared to the actual fidelity is shown in Fig.~\ref{DifferenceF}.
The average difference between the reconstructed fidelity and the actual fidelity for a given $N_{M}$ is

\begin{eqnarray}
  \label{DiffF}
  \overline{F}_{\mid r-a \mid} = \frac{1}{N} \sum_{i=1}^{N} \mid F_{r(i)} - F_{a(i)} \mid,
\end{eqnarray}

where $N = 25$ is the total number of simulation trials.

The fidelity reconstruction process using only 1,000 measurement outcomes will result in about $3\%$ error from the actual value on average.
The accuracy improvement converges to $\Delta F_{\mid r-a \mid} > 0.3\%$ when $N_{M} \geq 40,000$,
in which case not much benefit can be gained from solely increasing the number of measurements.

The MeetInTheMiddle link and the SenderReceiver link have similar behavior regarding the fidelity,
with sufficiently long memory lifetime relative to the idle time caused by classical latencies.
Nevertheless, even over a single hop, the throughput strongly depends on the physical link architecture and the adopted data link protocol.
Accumulating more measurements will increase the tomography time linearly, for both links,
as the operation here consists only of measuring entangled qubits sequentially.
The entanglement generation rate over the MeetInTheMiddle link is twice of what the SenderReceiver link achieves,
simply because the BSA in the MeetInTheMiddle sits half-way through the link.
With the given parameter settings, architecture and protocols,
the MeetInTheMiddle link's Bell pair generation rate is around 6741/s (3368/s for the SenderReceiver link),
 when qubits are immediately measured when the system recognizes that they are entangled.

\section{Performing tomography with single shot purification over different distances}
\label{purifications}
Lengthening the channel between nodes increases the chance for optical qubits and memory qubits to experience errors.
One way to tolerate such errors is to perform quantum purification.

\begin{figure}[H]
  \center
  \includegraphics[keepaspectratio,scale=0.55]{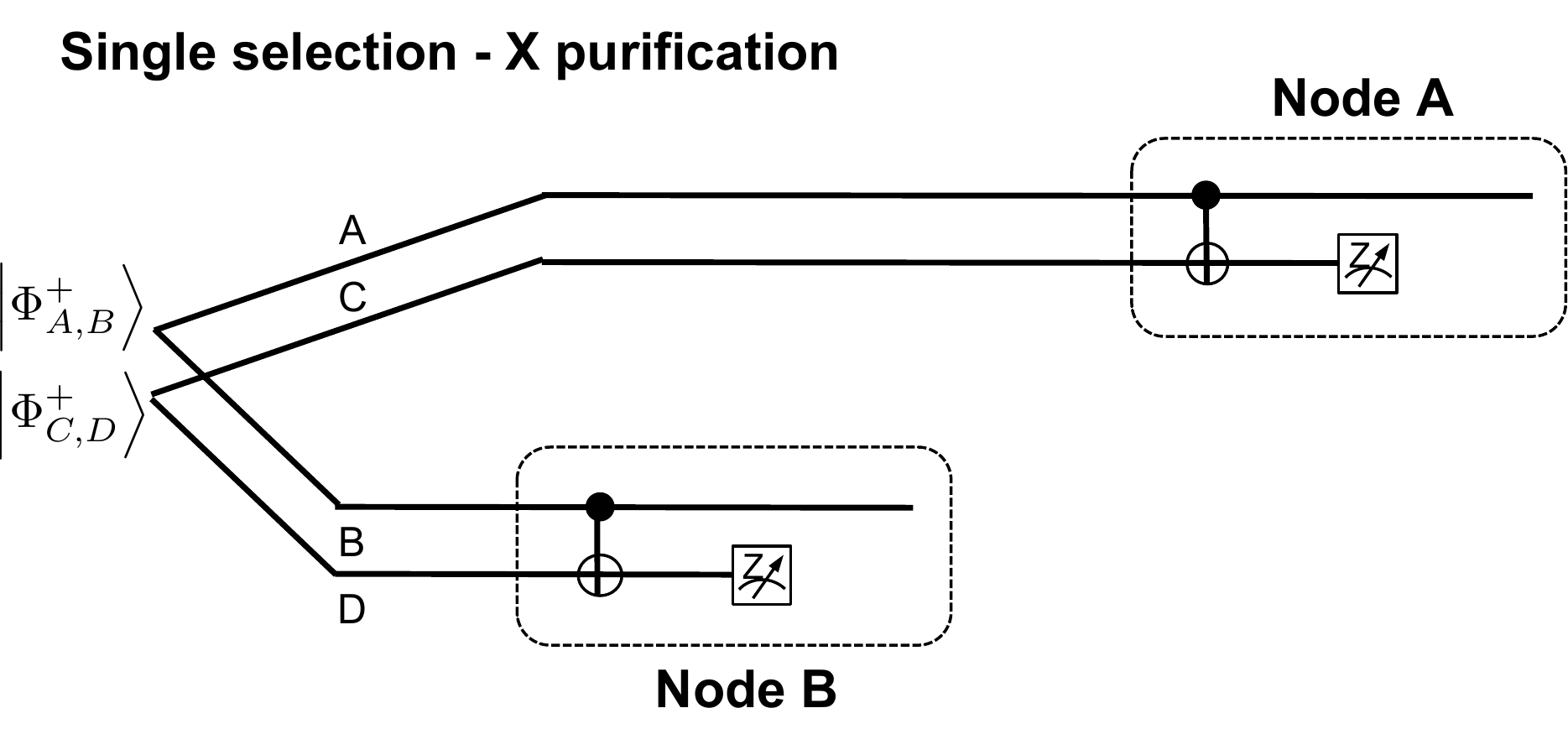}
  \caption[Single selection - Single error (X) purification (Ss-Sp)]{Single selection - Single error (X) purification (Ss-Sp). Consumes a single Bell pair $ \ket{\Psi^+_{C,D}}$ to detect the X error on $ \ket{\Psi^+_{A,B}}$.}
  \label{SS-Xp}
\end{figure}

The simplest purification scheme is the Single selection -  Single error purification (Ss-Sp)~\cite{Briegel1998}.
As in Fig.~\ref{SS-Xp}, the Ss-Sp requires two Bell pairs along the channel to be purified -- one Bell pair is consumed to detect the presence of an X error (or Z error) on the other Bell pair.
The resource is only purified successfully when the measurement outcomes of the consumed qubits (qubit C and qubit D in Fig.~\ref{SS-Xp}) coincide.
Otherwise, both nodes discard their resources.

\begin{figure}[H]
  \center
  \includegraphics[keepaspectratio,scale=0.55]{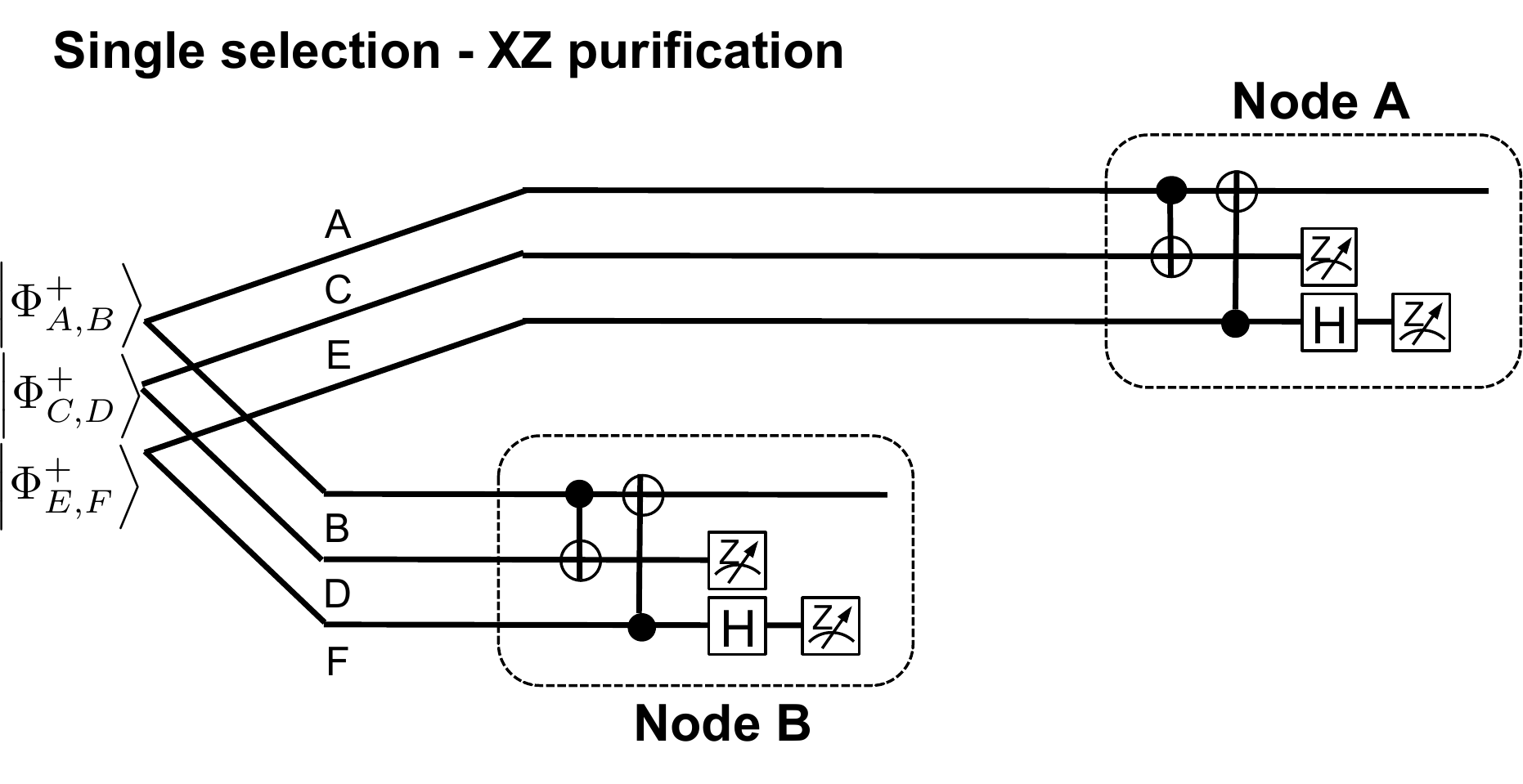}
  \caption[Single selection - Double error (XZ) purification (Ss-Dp).]{Single selection - Double error (XZ) purification (Ss-Dp). Consumes two Bell pairs, $ \ket{\Psi^+_{C,D}}$ and $\ket{\Psi^+_{E,F}}$ to detect X error and Z error accordingly.}
  \label{SS-XZp}
\end{figure}

By extending Ss-Sp, we can also try to purify both errors (see Fig.~\ref{SS-XZp}).
The Single selection - Double error purification (Ss-Dp) is similar to Ss-Sp,
but requires another Bell pair to detect the Z error on the resource.
However, because we perform the Z error purification after the X,
the X error can propagate to the purified resource from the consumed Bell pair.
In Fig.~\ref{SS-Xp}, the X error from qubit E and qubit F will propagate to qubit A and qubit B through the CNOT gate accordingly.

\begin{figure}[H]
  \center
  \includegraphics[keepaspectratio,scale=0.55]{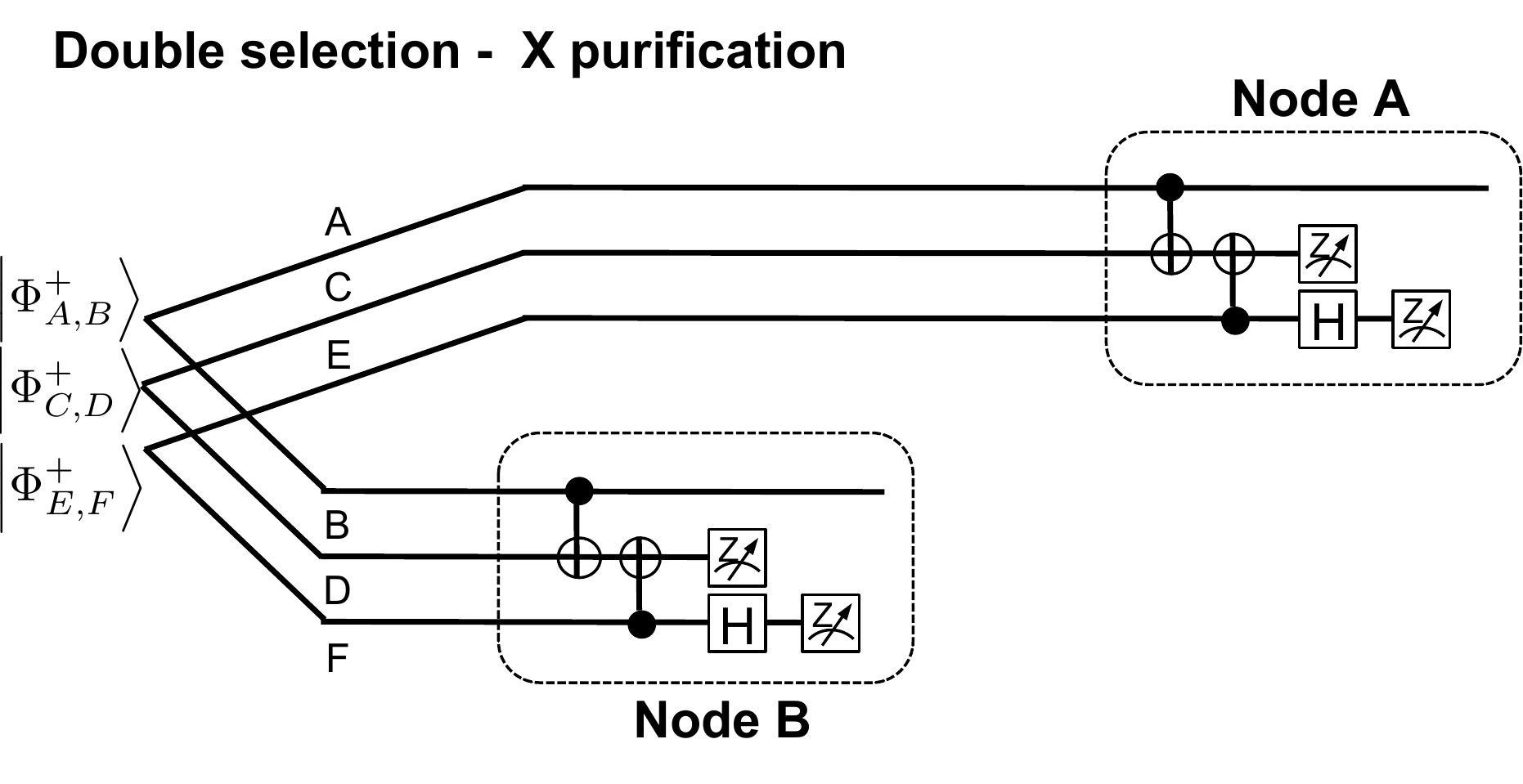}
  \caption[Double selection - Single error (X) purification (Ds-Sp)]{Double selection - Single error (X) purification (Ds-Sp).
  Consumes a single Bell pair $ \ket{\Psi^+_{C,D}}$ to detect X error on $ \ket{\Psi^+_{A,B}}$, and another Bell pair $ \ket{\Psi^+_{E,F}}$ to avoid the Z error propagation from $ \ket{\Psi^+_{C,D}}$ to $ \ket{\Psi^+_{A,B}}$.
  }
  \label{DS-Xp}
\end{figure}

The double selection method introduced by Fujii and Yamamoto in 2009~\cite{fujii:PhysRevA.80.042308} is a purification method utilizing an additional resource to improve the accuracy of post-selection by applying double verification.
The Double selection - Single error purification (Ds-Sp) does the same job as Ss-Sp,
but with an additional resource to detect the presence of a Z error on the consumed resource used for the X error purification (see Fig.~\ref{DS-Xp}).
This minimizes the error propagation to the purified resource.
Notice that any Z error that is originally present on the purified resource will not be detected.

\begin{figure}[H]
  \center
  \includegraphics[keepaspectratio,scale=0.55]{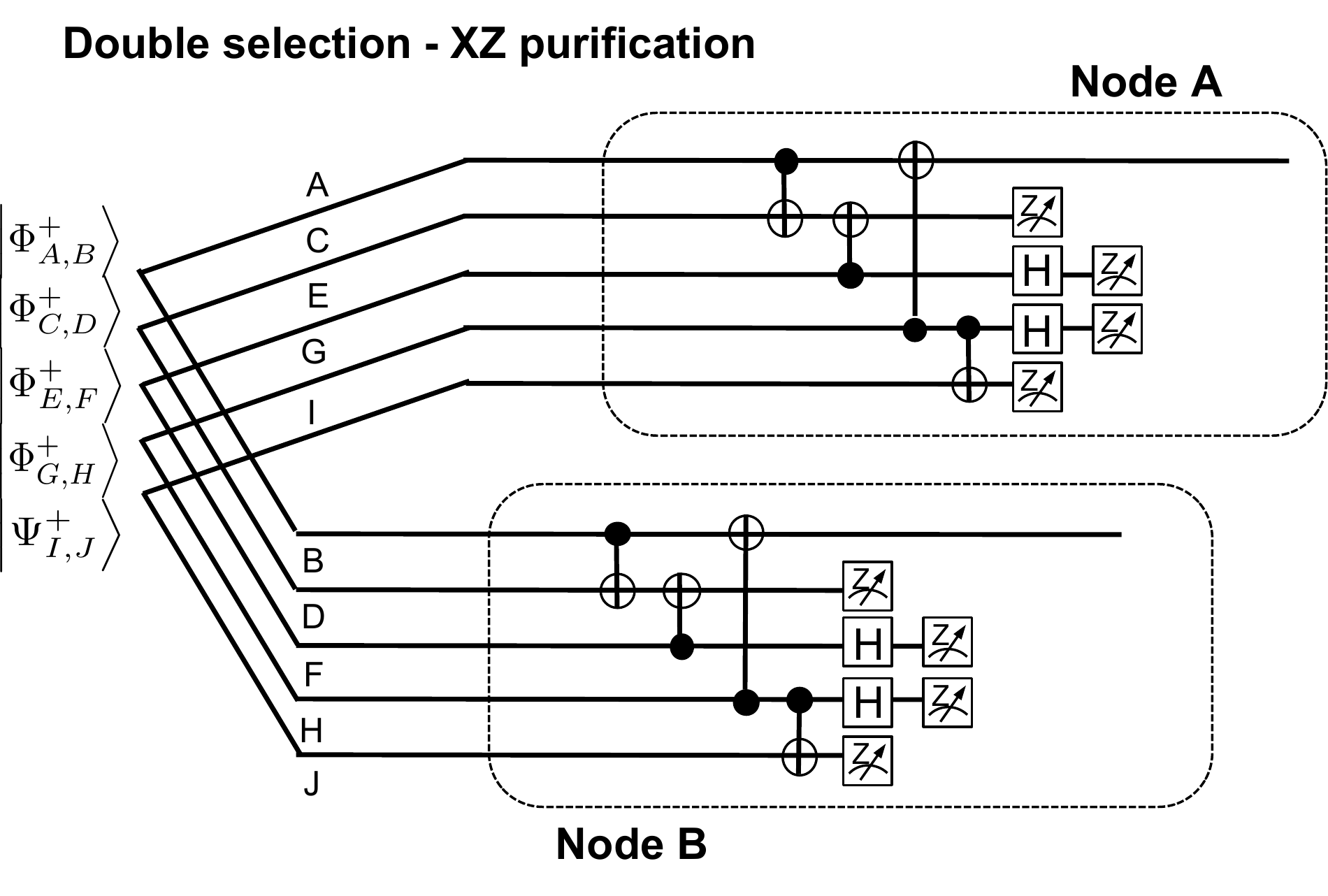}
  \caption[Double selection - Double error (XZ) purification (Ds-Dp).]{Double selection - Double error (XZ) purification (Ds-Dp).
  Consumes two Bell pairs, $ \ket{\Psi^+_{C,D}}$ and $ \ket{\Psi^+_{G,H}}$, to detect X error and Z error on $ \ket{\Psi^+_{A,B}}$. Each consumed Bell pair will also be verified through purification using $ \ket{\Psi^+_{E,F}}$ and $ \ket{\Psi^+_{I,J}}$ accordingly.}
  \label{DS-XZp}
\end{figure}

The double selection scheme can also be applied to the double error purification circuit (Ds-Dp).
The concept of this circuit is similar to Ds-Sp,
except we perform the double verification process on both X and Z purification to the resource we keep.
Thus, we can also minimize the error propagations that happen during the purification.
For the circuit diagram, see Fig.~\ref{DS-XZp}.

\begin{figure}[H]
  \center
  \includegraphics[keepaspectratio,scale=0.4]{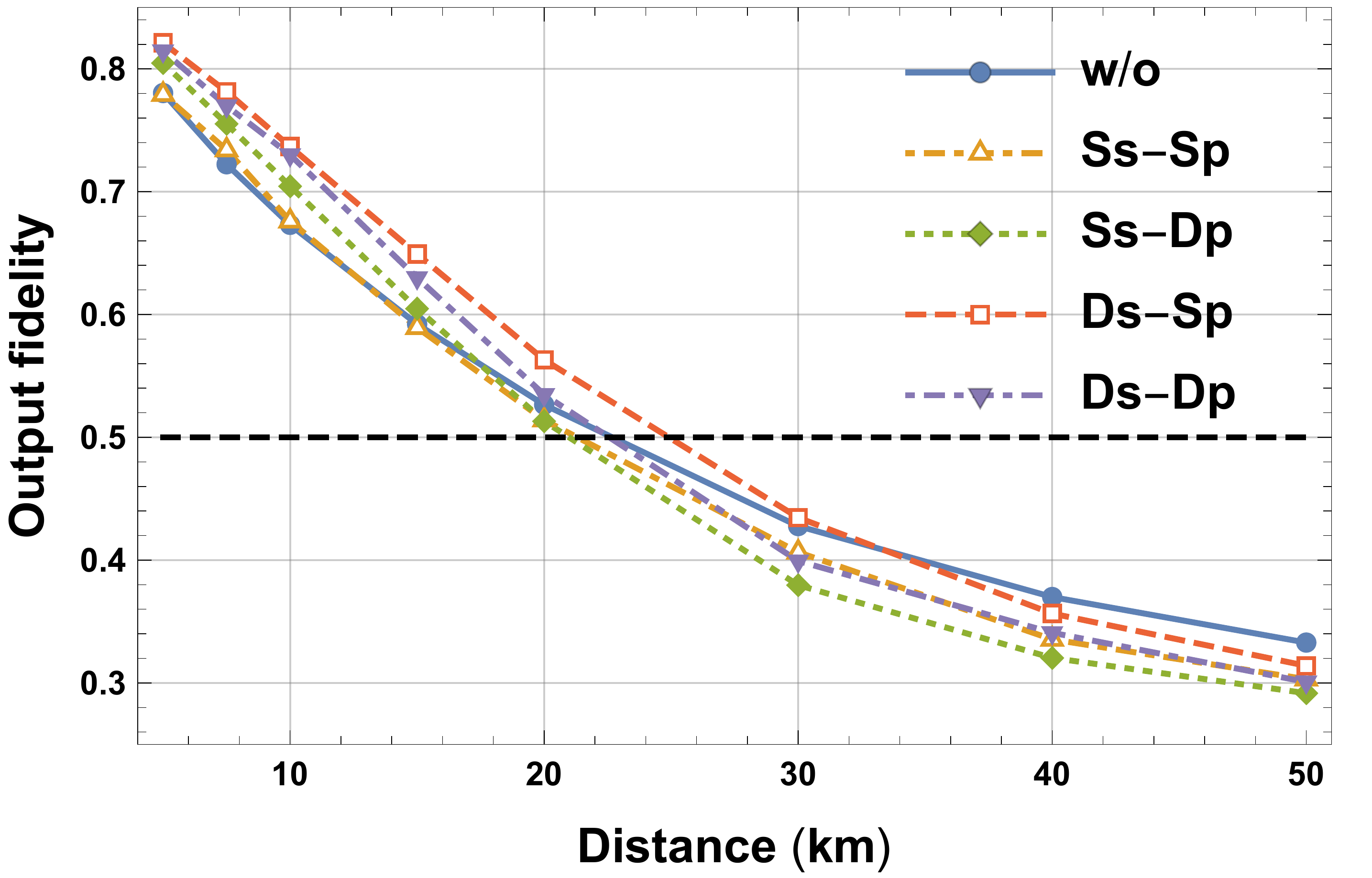}
  \caption[Impact of the MeetInTheMiddle channel length on the reconstructed fidelity with 7000 measurement outcomes w/o, with purification.]{Impact of the MeetInTheMiddle channel length on the reconstructed fidelity with 7000 measurement outcomes w/o, with purification.
  The steep drop in the fidelity is due to the high error rate of 1\% each X, Y and Z errors per km in the quantum channel chosen for this simulation.
  }
  \label{FidelityDistance}
\end{figure}

\begin{figure}[H]
  \center
  \includegraphics[keepaspectratio,scale=0.4]{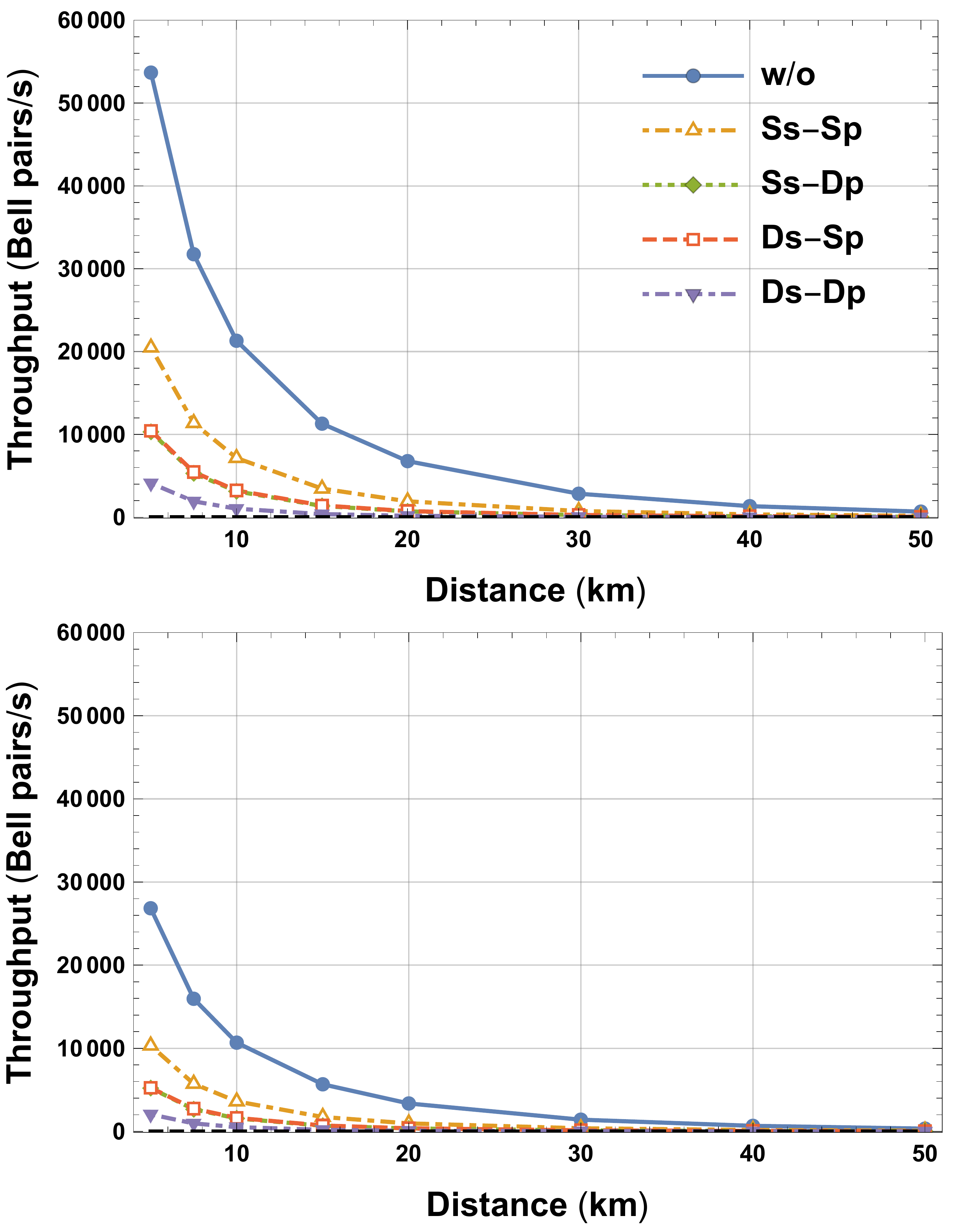}
  \caption[Impact of the channel distance on the channel throughput with and without performing purification.]{
  Impact of the channel distance on the channel throughput with and without performing purification.
  The throughput has been calculated as $N_{M}/T$, where $T$ is the tomography process time to complete $N_{M}$ measurements.
  (a) Throughput over the MeetInTheMiddle link.
  (b)  Throughput over the SenderReceiver link.
  }
  \label{ThroughputDistance}
\end{figure}

In this subsection, we describe simulated link-level tomography with 7000 measurements,
with and without Rules that perform one of the purification methods above beforehand, over different distances.
The simulated results of the fidelity reconstruction is shown in Fig.~\ref{FidelityDistance}, and Fig.~\ref{ThroughputDistance}.
The fidelity for SenderReceiver link is slightly worse than that of the MeetInTheMiddle link,
but the overall behavior is approximately the same.

As shown by the results in Fig.~\ref{FidelityDistance},
performing Ss-Sp once slightly improves the fidelity compared to the case without, only over relatively shorter distances,
but with a significant penalty on the throughput for both link architectures (see Fig.~\ref{ThroughputDistance}).
With a total distance $L=10km$, Ss-Sp suppresses about 5.6\% of the X errors and 5.4\% of Y errors,
but the Z error rate increases approximately 8.8\% due to the error propagation through the purification circuit.
Overall, the fidelity improves roughly by 3\%.
While the fidelity did not change dramatically,
the error distribution developed significantly different after performing the operation,
which we can still take advantage of in the recurrence purification protocol (see Section \ref{RpSec}).
By implementing double selection to the same circuit (Ds-Sp), the output fidelity improved significantly
-- about 6.4\% increase from the case without performing any purification.
The second verification decreased the Z error rate by about 4.6\% compared to the case of Ss-Sp with $L=10km$.
The Ss-Dp protocol purifies the X error first, and then the Z error afterwards.
In this case, over $L=10km$ distance,
the Z error rate decreases by 3.3\% and the Y error rate by 5.7\% -- the X error rate increases by 4.3\% due to propagation.
Simply applying double selection to this (Ds-Dp) improved the fidelity by approximately 2.6\% from Ss-Dp.
Both purifications based on double selection are capable of purifying resources under longer distances, which is roughly 20km.

With the given hardware qualities in Table~\ref{err},
Ds-Sp obtains the highest output fidelity.
The output fidelity for Ds-Dp is lower than that of Ds-Sp,
mainly because of larger $KQ$~\cite{2003PhRvA68d2322S} with noisy gates, and the increase waiting time for resources, especially over long distances,
 where $KQ$ is the product of the circuit depth ($K$) and the number of qubits ($Q$).
Under this scenario, because Ds-Sp only requires three resources at once, the throughput is also higher than that of Ds-Dp.

\section{Bootstrapping using Ruleset-based recurrence purification and tomography}
\label{RpSec}
Network operations, such as routing, often require knowledge regarding the links -- e.g. fidelity, throughput, etc.
Therefore, the channel characteristics to be shared need to be acquired beforehand via quantum link bootstrapping and characterization.
In quantum networking, we commonly focus on optimizing the link fidelity for realizing a robust connection,
where the goal may be achieved by performing nested purifications.
The recurrence protocol~\cite{2007RPPh70.1381D},
performs quantum purification on top of pre-purified resources to effectively improve the fidelity
-- the required number of resources for the operation, on the other hand, increases exponentially in the number of purification rounds $N_p$,
which alone linearly affects the total idle time due to classical communication latencies.

\begin{figure}[!htbp]
  \center
  \includegraphics[keepaspectratio,scale=0.55]{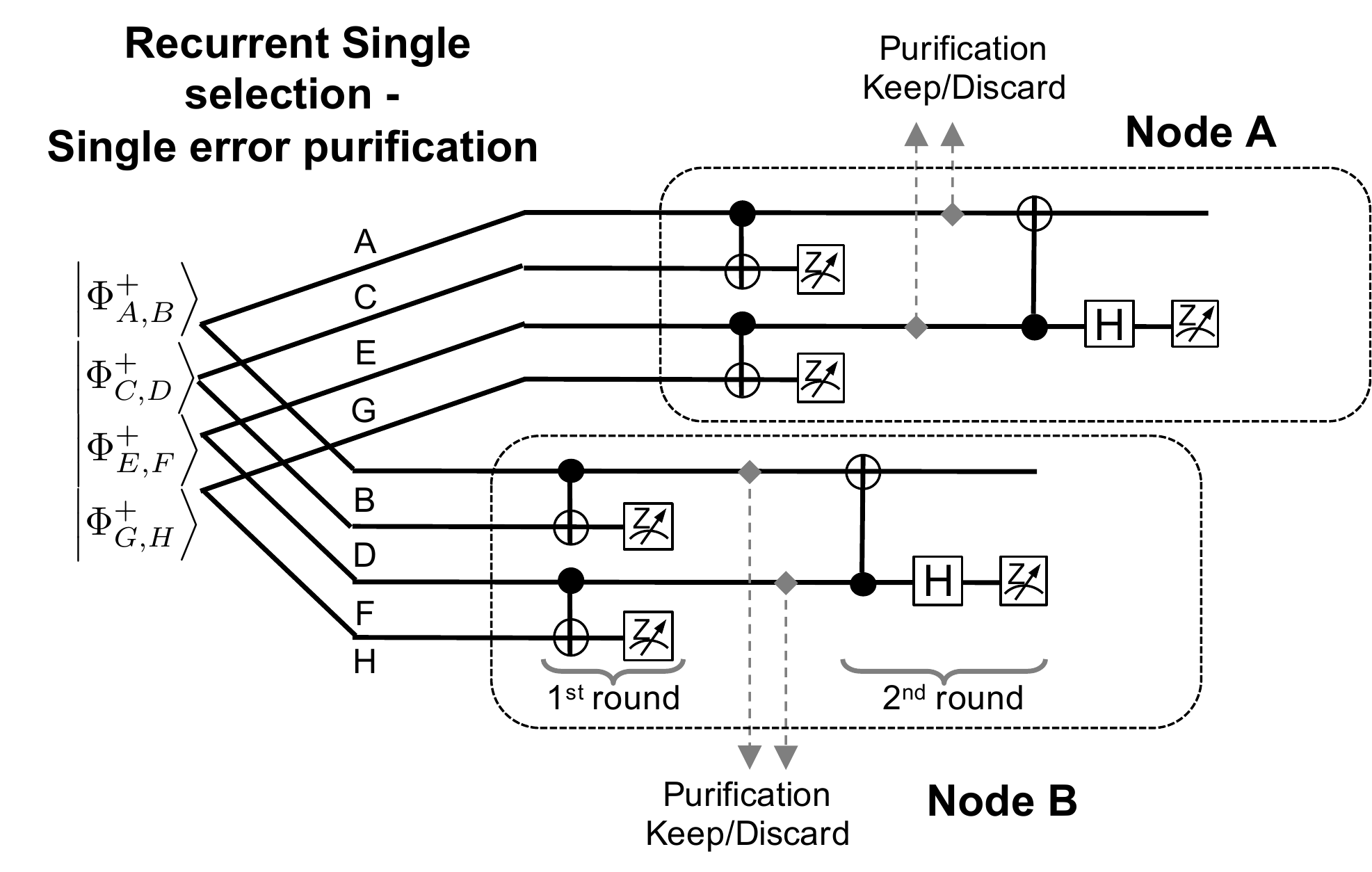}
  \caption[Recurrence purification protocol based on single selection X purification and Z purification.]{Recurrence purification protocol based on single selection X purification and Z purification.
  Each round of purification is a Rule in a Ruleset.
  }
  \label{Rp}
\end{figure}

In this subsection, we describe our simulation of quantum link bootstrapping to quantify the achievable link fidelity and its corresponding throughput,
using RuleSet-supported tomography with recurrence purification based on one of the four circuits shown in subsection \ref{purifications};
recurrent single selection single error purification (RSs-Sp),
recurrent single selection double error purification (RSs-Dp),
recurrent double selection single error purification (RDs-Sp) and recurrent double selection double error purification (RDs-Dp).
Each round of purification is a separate Rule in the same RuleSet.
In the given example of a two round RSs-Sp circuit shown in Fig.~\ref{Rp},
the first Rule performs X purifications to resources.
In the second Rule, we use those purified resources to perform another Z error purification.
Each round, we alternate the X purification and the Z purification to suppress the error propagation.
For double error purifications, we alternate the XZ purification and the ZX purification.
With probabilistic generation of Bell pairs and scarce memory qubits,
operations can stall in the middle, in which case the Rule waits for new resources to arrive.
Using RuleSets, the whole task is completed autonomously over a network.
The simulation continues reconstructing the density matrix every round,
and terminates the process when the same operation results in a lower reconstructed fidelity, or due to timeout.
For practical reasons, each RuleSet timeout is set to two minutes.

We assume perfect photon emissions from memories to fiber,
where in the real world, similar effects may be observed by increasing the number of memory qubits.
CNOT gates are also assumed to be ideal to solidify the real merit of recurrence protocols.
We first focus on a repeater network with a relatively short channel, where the total length is set to 10km.
Secondly, we perform similar simulations over a longer distance, $L=20km$.
For the MeetInTheMiddle link, each node is the same distance away from the BSA node.
The flowchart of the protocol is provided below in Fig.~\ref{FlowChartQLB}.
The stored fidelity and throughput by the selected recurrence purification method may be referred by other operations.
\begin{figure}[H]
  \center
  \includegraphics[keepaspectratio,scale=0.4]{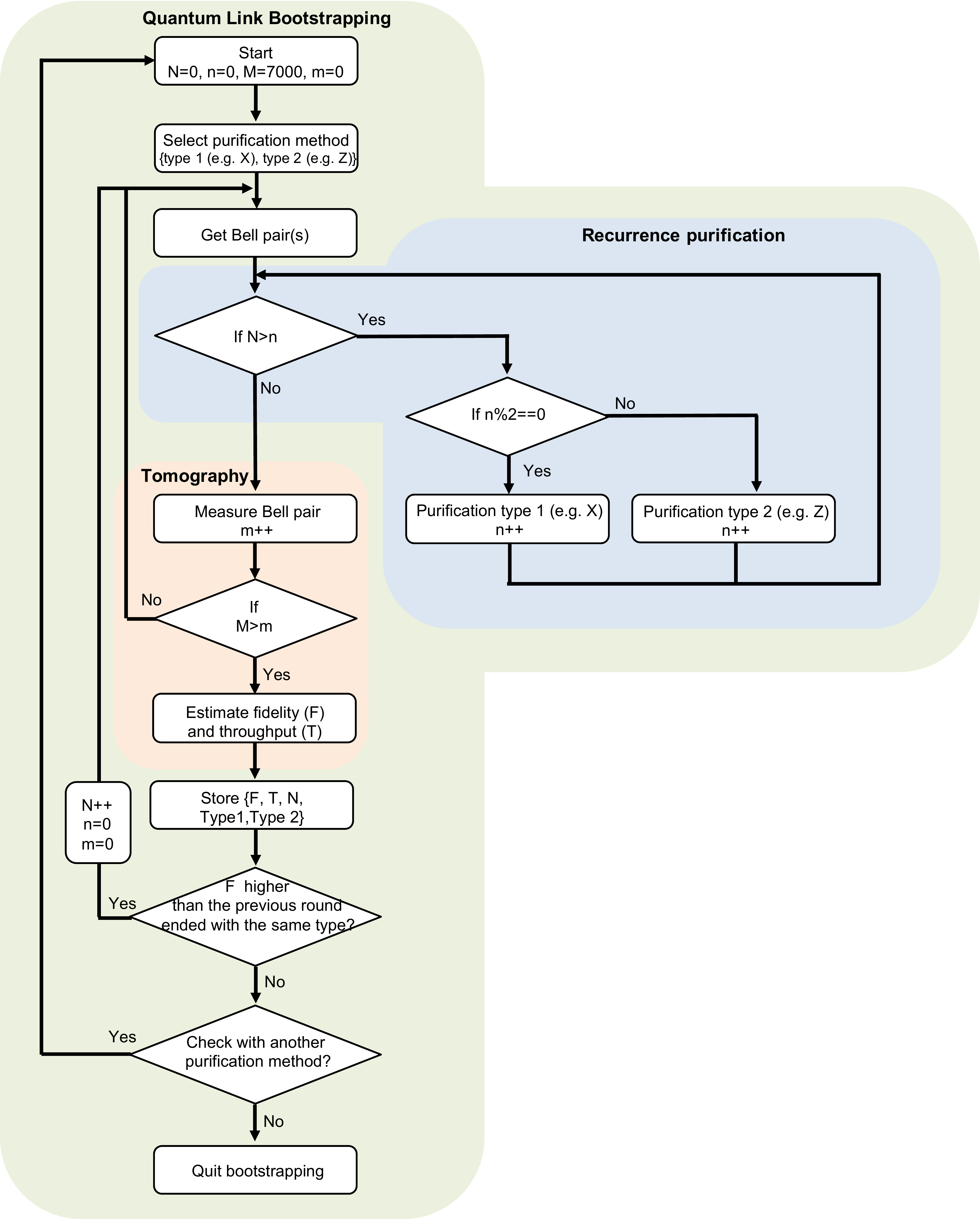}
  \caption{Flowchart of the quantum link bootstrapping. The termination process due to timeout is omitted from the chart.
  We need $1-F(\rho_{\mbox{reconstructed}},\rho_{\mbox{real}}) \ll 1-F(\rho_{\mbox{real}},\rho_{\mbox{ideal}})$ in order to make high-quality decisions about next operations in real time.
  }
  \label{FlowChartQLB}
\end{figure}

\subsubsection{Closely spaced repeater nodes}

\begin{figure*}
   \center
   \includegraphics[keepaspectratio,scale=0.24]{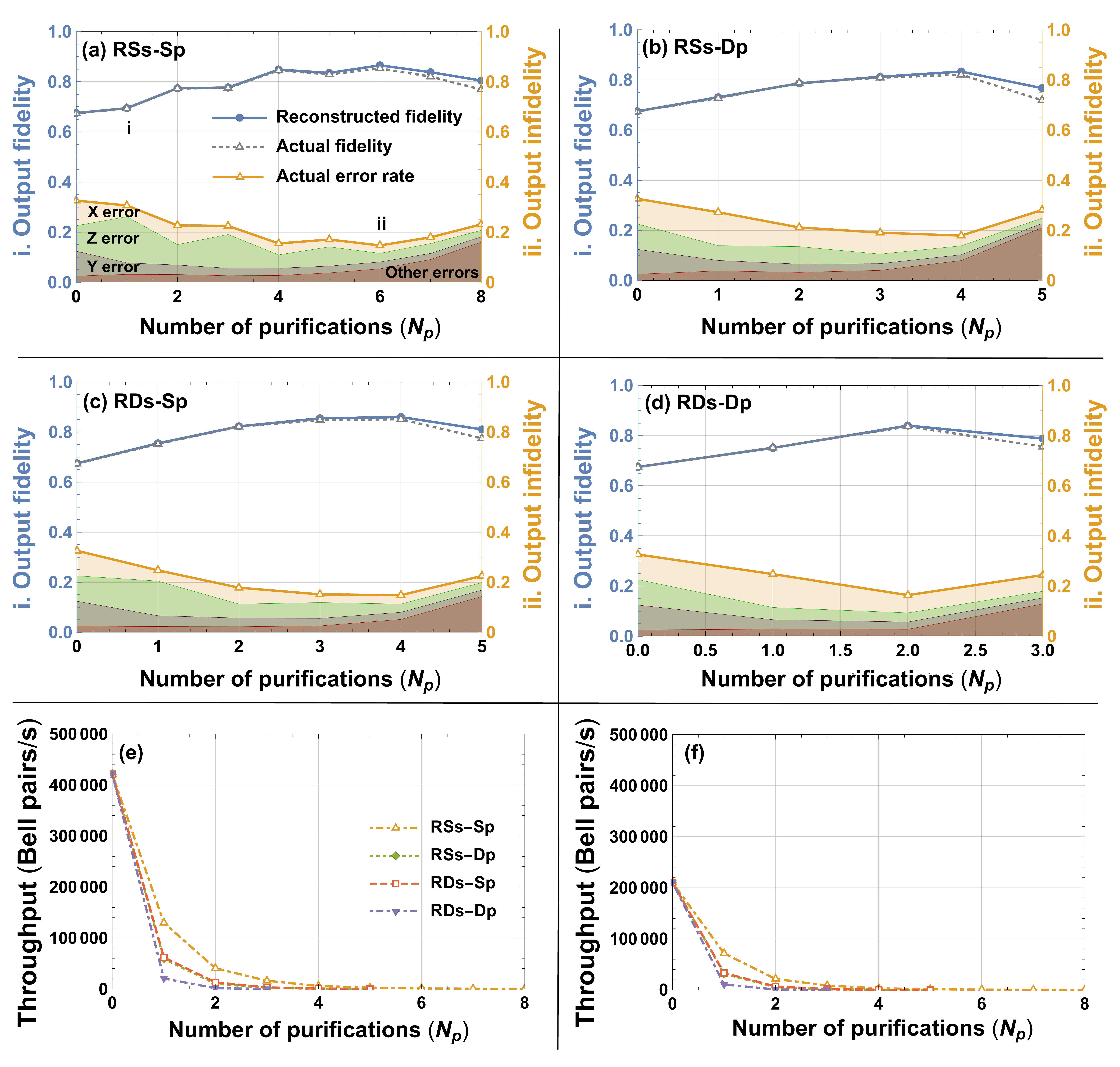}
   \caption[Reconstructed fidelity with actual fidelity and actual error rates, and estimated throughput of protocols over $L=10km$ MeetInTheMiddle link.]{
   Reconstructed fidelity with actual fidelity and actual error rates, and estimated throughput of protocols over $L=10km$ MeetInTheMiddle link.
   "Other errors" include memory excitation/relaxation error, and completely mixed error due to photon detector dark counts.
   (a) Simulation result of the RSs-Sp protocol.
   (b) Simulation result of the RSs-Dp protocol.
   (c) Simulation result of the RDs-Sp protocol.
   (d) Simulation result of the RDs-Dp protocol.
   (e) Protocol throughputs over the MeetInTheMiddle link.
   (e) Protocol throughputs over the SenderReceiver link.
   }
   \label{nTimesPurification}
 \end{figure*}

We start by investigating how recurrent purification benefits over a link between closely located repeater nodes, $L=10km$.
The simulated result over the MeetInTheMiddle link is shown in Fig.~\ref{nTimesPurification}.
Notice that each protocol has a different number of purification rounds performed.

As in the figure, the given system generates resources with an average fidelity $F_{r}\approx 0.675$,
but the RDs-Dp is capable of bringing up the quality up to $F_{r}\approx 0.840$ through a two round purification,
which is the fastest in terms of required rounds.
The RSs-Sp, on the other hand, starts with the slightest fidelity improvement
since applying X purification inherently results in a significant penalty to the Z error rate, as we discussed in section \ref{purifications}.
While this looks operationally impractical, by alternating X and Z purification mulitple rounds,
we can purify resources while minimizing error propagations --
this appears as the stairstep-like curve in Fig.~\ref{nTimesPurification}(a).
RSs-Sp gradually improves the fidelity as we perform more purifications,
and produces resources with optimal average fidelity, $F_{r} \approx 0.865$ ($F_{r} \approx 0.852$ for the SenderReceiver link) at $N_p = 6$.
About 3.1\% of the errors are X errors, 2.7\% are Y errors and 3.4\% are Z errors.
5.6\% of the outputs consists of unfixable errors,
which can mainly be avoided by improving the memory lifetime, shortening the distance between nodes
or adjusting the algorithm and the goal.
RDs-Sp is also capable of producing high fidelity resources, $F_{r} \approx 0.860$,
with more rounds of purifications than RDs-Dp but fewer than RSs-Sp.
The fidelity starts declining for all protocols in the end,
due to critical waiting time of memories relative to its own lifetime.
The reconstructed fidelity also gets overestimated,
because the randomness of measuring unentangled qubits due to errors,
stochastically contributes to the fidelity estimation.

The number of rounds does not necessarily determine how fast the protocol processes resources, especially with a limited number of qubits.
As shown in Fig.~\ref{nTimesPurification},
the RDs-Dp protocol over the MeetInTheMiddle link reaches its highest fidelity with a throughput of roughly 1565/s,
completing 7000 purified Bell pair measurements in around 37.1 seconds.
The RSs-Sp protocol produces 1106 resources per second with the local optimal fidelity,
but also achieves similar fidelity as RDs-Dp, $F_{r}\approx 0.848$,
at $N_p=4$, which obtains a higher generation rate of 5923/s.
The RDs-Sp protocol and the RSs-Dp protocol have a throughput of 816/s and 520/s respectively.
Although both protocols consume the same amount of resources per purification,
the RSs-Dp took longer in total because at each step, RSs-Dp faces higher error propagation probability,
resulting in greater purification failure rate.
As provided in Fig.~\ref{nTimesPurification}(e) and Fig.~\ref{nTimesPurification}(f), the throughput over the SenderReceiver link is approximately half of what the MeetInTheMiddle link achieves.
In this case, RSs-Dp shows no advantage over RDs-Sp.

\begin{figure}[!htbp]
   \center
   \includegraphics[keepaspectratio,scale=0.7]{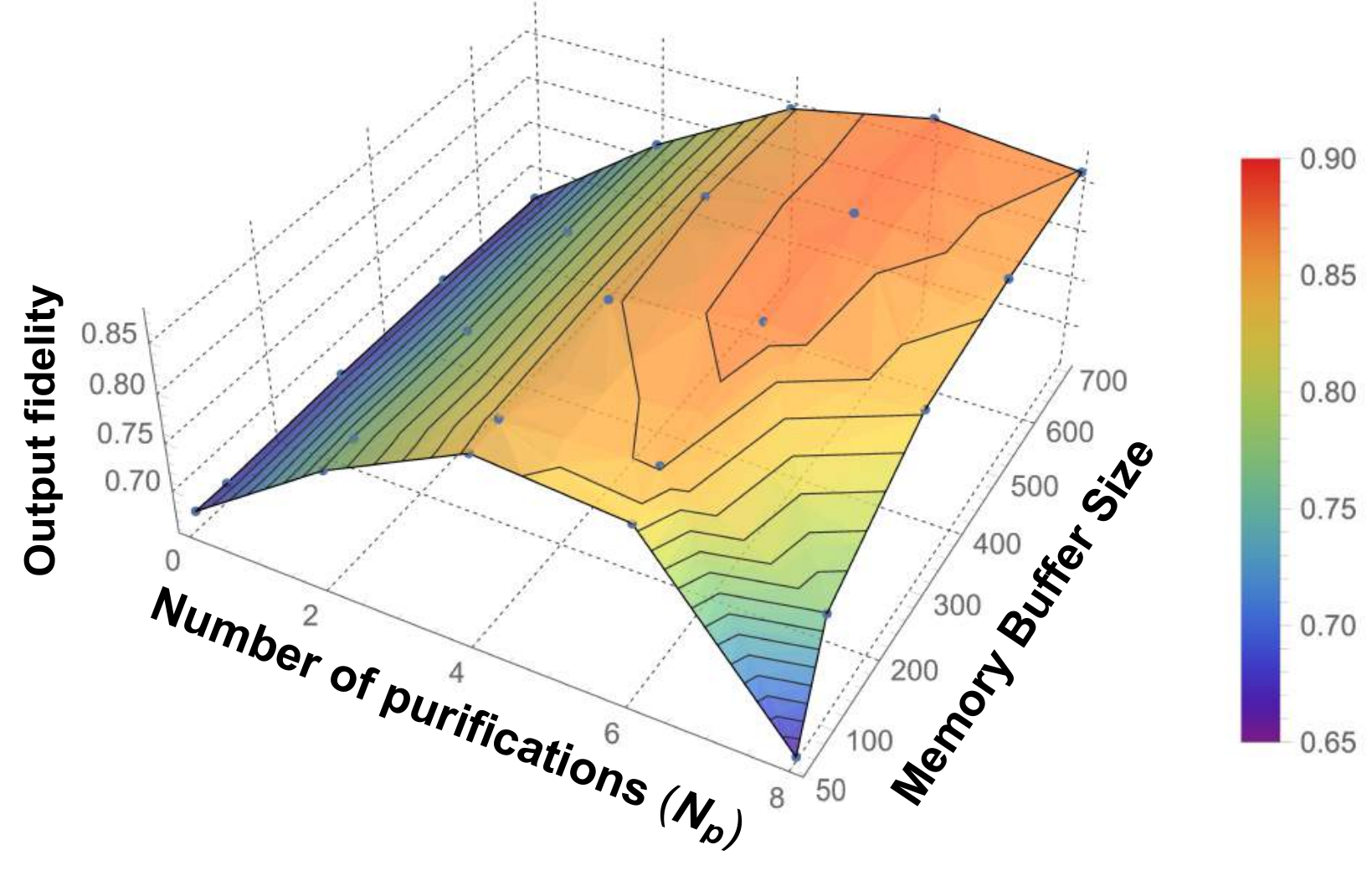}
   \caption{
   Impact of the memory buffer size to the maximum fidelity purified through the RSs-Sp protocol over a 10km MeetInTheMiddle link.
   }
   \label{memorysize}
 \end{figure}

The quality of a link can also be increased by installing larger sets of memory qubits.
As in Fig.~\ref{memorysize}, with 700 memory qubits, a six-round RSs-Sp can generate resources with $F_{r}\approx 0.879$.
Compared to the case with only 100 qubits, the fidelity at $N_{p}=4$ also increases around by 0.6\%.
For $N_{p}<4$, the fidelity stays unchanged because having 100 qubits is more than sufficient for a four-round RSs-Sp -- the throughput will, however, improve.
Notice that a linear increase in memory buffer size will not provide an effective solution to extend the performable rounds.
Over a 10km MeetInTheMiddle link, owning 700 memory qubits is still not enough for an eight-round RSs-Sp.

\subsubsection{Distantly spaced repeater nodes}

\begin{figure*}
  \center
  \includegraphics[keepaspectratio,scale=0.24]{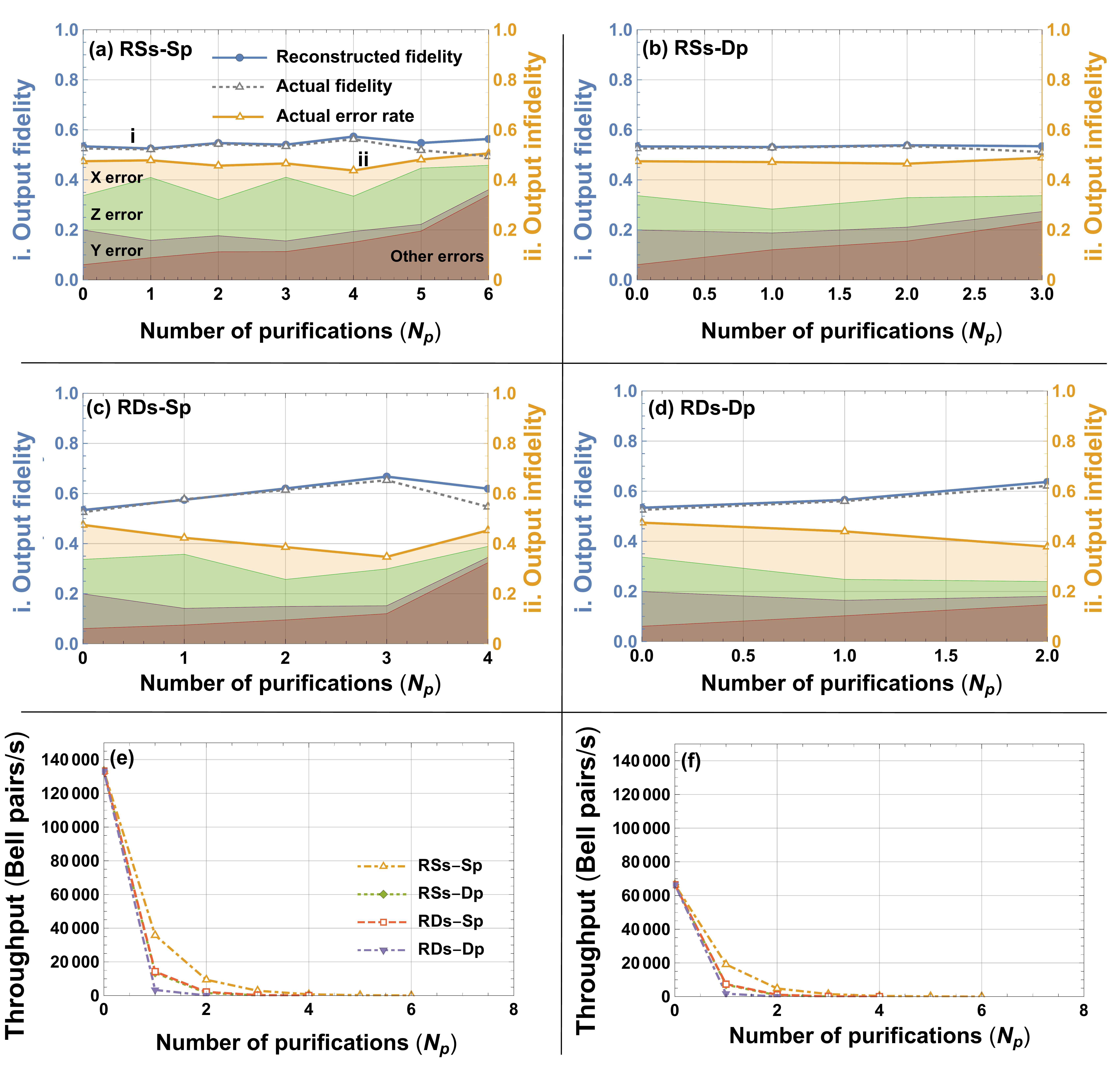}
  \caption[Reconstructed fidelity with actual fidelity and actual error rates, and estimated throughput of protocols over $L=20km$ MeetInTheMiddle link.]{
  Reconstructed fidelity with actual fidelity and actual error rates, and estimated throughput of protocols over $L=20km$ MeetInTheMiddle link.
  "Other errors" include memory excitation/relaxation error, and completely mixed error due to photon detector dark counts.
  (a) Simulation result of the RSs-Sp protocol.
  (b) Simulation result of the RSs-Dp protocol.
  (c) Simulation result of the RDs-Sp protocol.
  (d) Simulation result of the RDs-Dp protocol.
  (e) Protocol throughputs over the MeetInTheMiddle link.
  (e) Protocol throughputs over the SenderReceiver link.
  }
  \label{nTimesPurification20km}
\end{figure*}

Errors on qubits have more time to develop when two repeater nodes are located farther away.
Hence, how well errors are detected at each round, especially with noisy inputs, becomes an important factor for an effective recurrence purification protocol.
The simulation results over a repeater network with channel length $L=20km$ are is provided in Fig.~\ref{nTimesPurification20km}.

Unlike the case of Fig.~\ref{nTimesPurification},
neither RSs-Sp nor RSs-Dp improves the fidelity effectively,
regardless of the number of purification rounds.
As shown in the corresponding error distributions, performing purifications with the given input resources,
in such a way done by RSs-Sp and RSs-Dp, ends up failing because the loss lead by error propagations and memory errors is greater than the gain.
Resources used for RSs-Sp suffer more from memory errors as they go through more rounds of purifications, resulting in longer average idle time per resource.
The other two protocols implemented with double selection,
however, are capable of effectively purifying resources.
The RDs-Sp protocol produces the highest fidelity among all,
with an output value $F_{r} \approx 0.667$ and a rate of 325/s (168/s with $F_{r} \approx 0.651$ for the SenderReceiver link) at $N_{p}=3$,
whereas RDs-Dp also produces similar quality, $F_{r} \approx 0.637$, or $F_{r} \approx 0.626$ for the SenderReceiver link, within two rounds.
The corresponding throughput is 116/s for the MeetInTheMiddle link, and 60/s for the SenderReceiver link.

Under this scenario, consuming another Bell pair via the double selection shows advantage in terms of fidelity.
However, from the previous discussion, we know that RSs-Sp generates resources with higher fidelity and higher rate,
given slightly better average input resources and sufficiently long-lived memories relative to its average idle time.
Thus, for longer distances, we can also perform double selection-based purification at the beginning to raise the fidelity at a certain level,
then change to single selection purification afterwards to maximize the fidelity.
Below in Fig.~\ref{MX20km} are the simulation results of two cases;
{\emph Case A} that switches to RSs-Sp after performing a single round of RDs-Sp,
and {\emph Case B} that switches to RSs-Sp after performing two rounds of RDs-Sp.

\begin{figure}[!hbt]
  \center
  \includegraphics[keepaspectratio,scale=0.4]{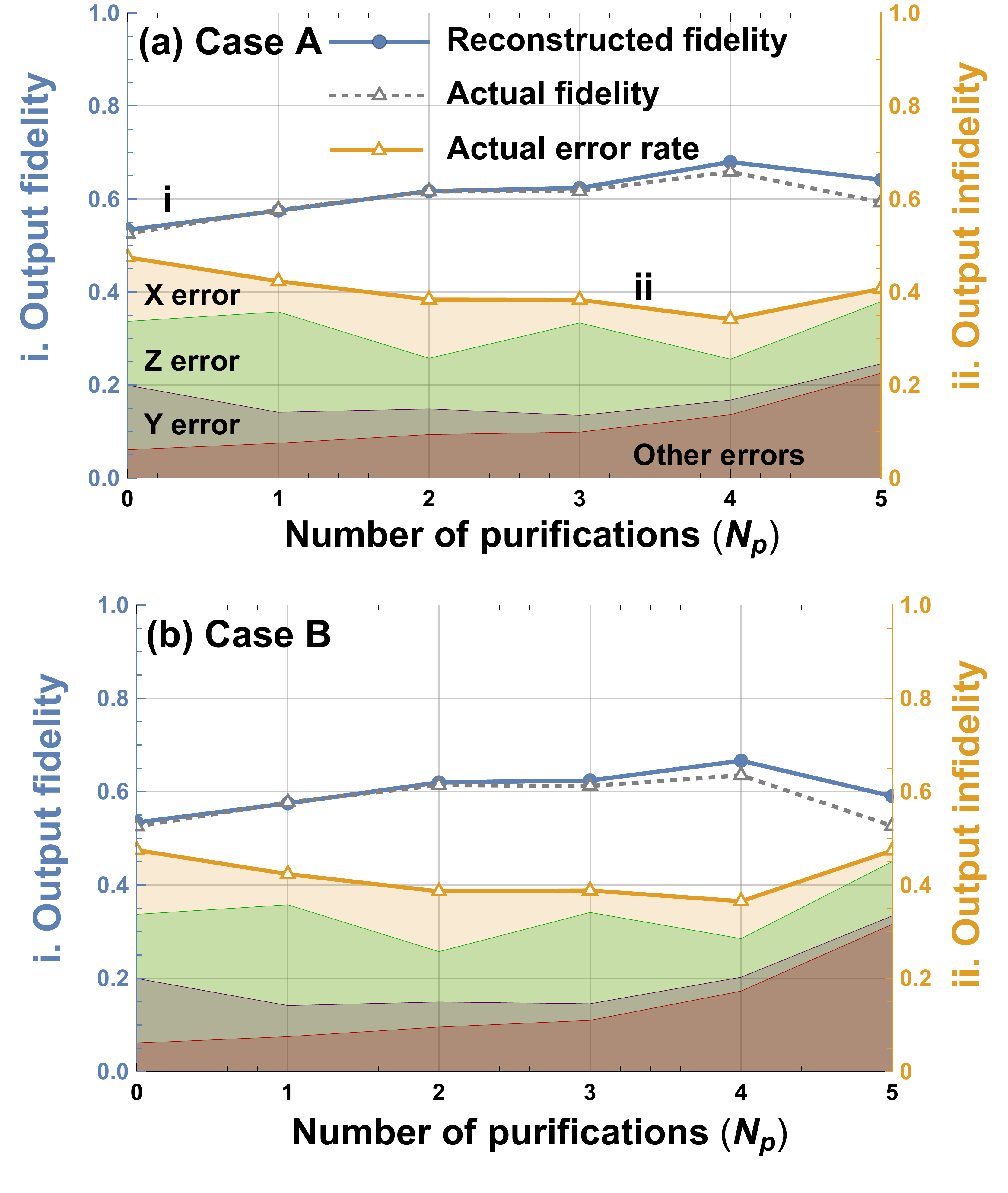}
  \caption[Switching RDs-Sp to RSs-Sp over $L=20km$]{Switching RDs-Sp to RSs-Sp over $L=20km$.
  a) Pattern 1 consists of one round RDs-Sp and four rounds of RSs-Sp. Purification methodology switches at $N_{p}=1$.
  b) Pattern 2 consists of two rounds RDs-Sp and three rounds of RSs-Sp. Purification methodology switches at $N_{p}=2$.}
  \label{MX20km}
\end{figure}

Whether this method is beneficial or not very much depends on the situation, and the switching timing as shown.
Under this scenario, {\emph Case A} produces higher fidelity, $F_{r}=0.680$, with a higher rate of 396 Bell pairs per second,
relative to the optimal case of RDs-Sp in Fig.~\ref{nTimesPurification20km}.
{\emph Case B} results in worsening the fidelity to $F_{r}=0.666$, and the throughput to 218/s.
The total bootstrapping time for completing the RSs-Sp, RSs-Dp, RDs-Sp, RDs-Dp, Case A and Case B with a single trial was 561 seconds, which is approximately 9.4 minutes.

\chapter{Conclusion}
\label{conslusion}

This thesis has introduced the RuleSet-based communication protocol,
 which is used to autonomously support coordinated decision making in quantum operations over a network.
 The RuleSet architecture is not only useful for the coordination,
 but also allows us to dynamically generate suitable operations based on link qualities,
 dynamically assign resources and manage network processes based on distinguishable connections.
The technology is adapted to conduct quantum link bootstrapping in a network simulation.
In total, four different recurrent purification schemes were simulated with the goal
to quantifying the achievable link fidelity with its corresponding throughput via link-level tomography,
 using Markov-Chain Monte-Carlo simulation with a noisy quantum repeater system.
Such information may be distributed across the network, and used for different purposes such as quantum routing.

The RSs-Sp requires more purification rounds to optimize the fidelity compared to other schemes,
which results in a longer average idle time for purified resources.
Given 100 link memory qubits each with a 50ms lifetime,
we found that RSs-Sp is still stronger than other simulated schemes in terms of the achievable fidelity and the throughput,
when the distance between nodes is kept short ($L \approx 10km$).
In contrast, because the system is noisier with longer channels ($L \approx 20km$), RSs-Sp alone may be incapable of purifying the link,
 as errors can evolve faster than the purification, in which case schemes such as the double selection becomes more preferable.
The simulations show that the bootstrapping process, therefore, must be able to search and choose the most effective purification method for that specific channel.
Changing from double selection to single selection in the middle indeed pushed the link limit,
but finding the optimal changeover timing may add an additional complexity to the process.
Other methods that detect errors with a higher probability requiring fewer rounds,
 such as RDs-Dp, may be more advantageous when a larger set of memory qubits is available for each link, relative to its demand.

While this work focuses on link fidelity optimization,
connecting source to destination multi-hops away also requires an adequately high generation rate, or possibly a synchronized generation timing of Bell pairs.
Such an optimization problem of the fidelity-throughput tradeoff remains an open question.

{\bf Future work}

A simulation of multi-hop quantum communication remains a future work.
Although not finished, OMNeT++ is capable of simulating a multi-hop quantum communication, and much preparatory work has been done for that.
Optimizing a link between adjacent nodes through bootstrapping, and optimizing multiple links for connecting non-adjacent nodes are two different tasks.
An interesting topic would be to investigate the minimum throughput required for connecting $n$ resources.
Discussion regarding the quantification of costs for quantum channels, required for routing, also remains an important open question.
The current version of the RuleSet-based communication protocol is also not flexible enough to accomplish a particular set of tasks,
such as performing purification through entanglement pumping - the current RuleSet does not have a shared resource list.
Improving the protocol also remains an important future work.

\chapter*{Acknowledgements}
\addcontentsline{toc}{chapter}{Acknowledgements}

First and foremost, I would like to express my deepest gratitude to Professor Rodney Van Meter at Keio University for his long term support through out my Bachelor's and Master's program.
I truly enjoyed working with you on research projects.
I have had countless amazing opportunities at the AQUA lab to learn, which I could have never gained without you.
When I joined AQUA, I never imagined myself presenting my own research in conferences, and publishing them as a paper.
It is, without doubt, the most valuable and irreplaceable experiences in my life.

I especially thank professor Jun Murai at Keio University for his kindness and for providing me all kinds of opportunities.
Working as your SA/TA was tough, but I am glad that I was capable of working with you throughout my university life.
I was lucky to have a lot of time to spend with you periodically inside and outside the campus.

I also acknowledge Project Research Associate Takahiko Satoh at Keio University for being always a supportive supervisor,
 and spending me a large fraction of your time without hesitation to share your knowledge.
I enjoyed talking with you, and I am fortunate to have you in the AQUA lab throughout my master's program.

I would also like to thank all the members of the AQUA lab.
Shin Nishio, Ryosuke Sato, Yasuhiro Okura, Shigetora Miyashita, Nozomi Tanetani, Sayyor Yusupov and Makoto Nakai, which I enjoyed frequently hanging out (bouldering, karaoke and eating) outside the campus.
I'm glad that I had you all.
I also thank all the new comers, Shuta Kobayashi, Liu Sitong, Yakabe Mahiro, Shu Toraketu and Nobuhide Hiranuma.
I can already imagine you all producing great research results in the future!!!

I also thank to those that used to be a member of the AQUA lab.
First, I have to thank Shota Nagayama for helping me out throughout my Bachelor's program, and for working with me as a SA/TA in many classes.
I shall thank Takafumi Oka, who is a good friend of mine.
You are very kind for always picking a place somewhere in between where I live and where you live, when we hang out!
I also thank Shinnosuke Ozawa for whom I enjoyed working in the AQUA lab for a long time.
I'm glad that I had you as a \emph{kohai} in AQUA.

I am grateful to all members in Murai, Kusumoto, Nakamura, Takashio, Uehara, Van Meter, Mitsugi, Nakazawa Laboratories for all the supports.

I must thank Nazmi Tanaka and Yuji Yamada for always being one of my best friends for more than 13 years.
I am grateful to have someone like you, and have never stopped enjoying to play games, watch movies, and do all other stuffs together.
I must also thank Kaho Yogi for always being closest to me.
Let's stay fit, and try to live together as long as possible!
I also thank Kennedy Ho, who always comes to Shonandai all the way from Tokyo just to hang out with me.
I truly enjoyed (and I'm still enjoying) spending time with you in the weekends, but please stop complaining how I never go out to Tokyo to meet you.
I did once!!!
Good for you I'm moving out from Shonandai!
I thank Peter Tien for frequently sending me messages, and sometimes advertising me your own company for no reason at all.
For about 3 years, I have been struggling to throw away the junk (or present, I shall say) you gave me for birthday.
Allow me to throw it away, and stop saying it is the symbol of our friendship.
I thank the \emph{AES family members}, Tomas Svitil, Petr Svitil, Jeremy Shih, Karanvir Chauhan, Sangwon Lee and Joshua Woods.
Lucky me to have you all to chit chat, and to spend time outside (and sometimes inside) Japan!

	\bibliographystyle{unsrt}
	\bibliography{main}

	\renewcommand{\thechapter}{\Alph{chapter}}
	\setcounter{chapter}{0}

	\appendix
	\chapter{Other details regarding the Quantum networking simulator}
	\label{appendix}

	\section{Simulator size}
	\label{SIMappendix}
	The Network Description File (NED) file specifies the connectivities of nodes and modules, and initialize parameters.
	A value may be assigned to a parameter simultaneously.

	The length of \emph{.ned} files, codes for describing the module connectivity within a single node, are as shown bellow.
	Length includes empty lines.

	\begin{itemize}
	 \item{138 {\tt QNode.ned}: \\Defines the module connectivity within a single quantum node. Modules in a quantum node include the Quantum Routing Software (QRSA) module, the classical routing software module, QNIC module, NIC module and the application module.}
	 \item{76 {\tt classical\_softwares.ned}: \\Defines classical software modules, including the queue module, the application module and the routing module.}
	 \item{184 {\tt qnic.ned}: \\Defines the module connectivity within a single QNIC. Modules in a QNIC include the stationary qubit module, the multiplexer module and the internal BSA module.}
	 \item{141 {\tt qrsa.ned}: \\Defines modules and their connectivity within a single Quantum Routing Software (QRSA) module. Modules include the routing daemon module, the hardware monitor module, the real time controller module, the rule engine module and the connection manager module.}
	 \item{29 {\tt channels.ned}: \\Defines channel modules. Channels include the quantum channel, and the classical channel.}
	 \item{180 {\tt HoM.ned}: \\Defines the BSA module, and the internal BSA module for a MeetInTheMiddle link and for a SenderReceiver link accordingly.}
	 \item{160 {\tt EPPS.ned}: \\Defines the SPDC module for a MidpointSource link.}
	\end{itemize}

	The length of an example \emph{.ned} file, codes for describing the node connectivity of a network, is as shown bellow.
	The length strongly depends on the number of simulated networks (including those used for testing).

	\begin{itemize}
	 \item 1108 {\tt topology\_realistic\_layer2\_experiments.ned}: \\Defines one or more network for a simulation. A network is also a module that is  composed of predefined nodes such as the QNode.
	\end{itemize}

	The following \emph{.cc} files specifies the behavior (functions) of modules defined in the NED files.
	The total length of the files is 6835 lines.

	\begin{itemize}
	 \item 191 {\tt Application.cc}
	 \item 388 {\tt BellStateAnalyzer.cc}
	 \item 266 {\tt ConnectionManager.cc}
	 \item 88 {\tt EntangledPhotonPairSource.cc}
	 \item 58 {\tt EntangledPhotonPairSource.h}
	 \item 936 {\tt HardwareMonitor.cc}
	 \item 148 {\tt HardwareMonitor.h}
	 \item 442 {\tt HoM\_Controller.cc}
	 \item 97 {\tt HoM\_Controller.h}
	 \item 47 {\tt QNIC.h}
	 \item 117 {\tt QNIC\_photonic\_switch.cc}
	 \item 53 {\tt QUBIT.h}
	 \item 160 {\tt Queue.cc}
	 \item 73 {\tt RealTimeController.cc}
	 \item 43 {\tt RealTimeController.h}
	 \item 24 {\tt ResourceManager.cc}
	 \item 38 {\tt ResourceManager.h}
	 \item 181 {\tt Router.cc}
	 \item 114 {\tt RoutingDaemon.cc}
	 \item 44 {\tt RoutingDaemon.h}
	 \item 1194 {\tt RuleEngine.cc}
	 \item 169 {\tt RuleEngine.h}
	 \item 247 {\tt SPDC\_Controller.cc}
	 \item 1332 {\tt stationaryQubit.cc}
	 \item 348 {\tt stationaryQubit.h}
	\end{itemize}

	A RuleSet is an object, which holds at least one Rule object.
	Each Rule object holds an Action object, and a Condition object that consist of one or more Clause object.
	Below are lengths of files defining the structure and the functions of a RuleSet, Rule, Action, Condition and Clause.

	\begin{itemize}
	 \item 686 {\tt Action.cc}
	 \item 436 {\tt Action.h}
	 \item 136 {\tt Clause.cc}
	 \item 151 {\tt Clause.h}
	 \item 66 {\tt Condition.cc}
	 \item 37 {\tt Condition.h}
	 \item 93 {\tt Rule.cc}
	 \item 70 {\tt Rule.h}
	 \item 20 {\tt RuleSet.cc}
	 \item 56 {\tt RuleSet.h}
	 \item 129 {\tt tools.h}
	\end{itemize}

	The \emph{.ini} files are used to initialize the simulation, and assigns values to all variables.
	Variables may be the channel cost, channel length, address, number of qubits installed in a particular QNIC and others.
	Values may also be assigned directly in the .ned file as a static value.
	Notice that each simulation will require one .ini definition.
	Some of the used files and their lengths are shown below.

	\begin{itemize}
	 \item 253 {\tt Fixed10\_No\_X\_XZ\_XXZPuri\_Low\_memErr\_.ini}
	 \item 288 {\tt Fixed10\_No\_X\_XZ\_XXZ\_DSPuri\_Low\_memErr\_.ini}
	 \item 250 {\tt Fixed15\_No\_X\_XZ\_XXZPuri\_Low\_memErr.ini}
	 \end{itemize}

	Simulated packet structures are defined in the \emph{.msg} files.

	\begin{itemize}
	 \item 217 {\tt classical\_messages.msg}
	 \item 70 {\tt PhotonicQubit.msg}
	\end{itemize}

	\section{How to build and run the simulator}
	The simulator is known to work with OmNET++ 5.4.1 and an external C++ library, Eigen 3.3.90.
	This thesis ran the simulator over Ubuntu 16.04.6 LTS using gcc 5.4.0.
	Building the simulation is done automatically when you hit the run button via the OMNeT++ GUI interface.
	Otherwise, use the {\tt opp\_makemake} command and type {\tt opp\_makemake -f --deep -O out -I/path\_to\_eigen/eigen3 -I. }.
	Once this is done, type {\tt make}. This will output an executable file {\tt quisp}.

	\subsection{Run-time configurable parameters}
	To run a simulation, you need to choose which simulation to run via specifying the .ini file for execution - all .ini files are stored in the {\tt network} directory.
	Suppose you want to run the {\tt Linear\_Single\_MIM} simulation, type {\tt ../quisp -m -u Cmdenv -n .. omnetpp.ini -c Linear\_Single\_MIM} inside the {\tt network} directory.
	All parameters values must be pre-specified in the corresponding .ini file.
	The current implementation supports the following parameters.
	Those parameters can also be specified per module (e.g. qubit).
	Details are also provided in the official OMNeT++ documentation.

	\begin{description}
	 \item[] {\tt **.Hgate\_error\_rate // Hadamard gate error rate. Scale from 0 to 1.}
	 \item[] {\tt **.Hgate\_X\_error\_ratio // The error rate will be divided in to three Pauli errors as in X:Y:Z, with P(X)+P(Y)+P(Z) = Hgate\_error\_rate}
	 \item[] {\tt **.Hgate\_Y\_error\_ratio}
	 \item[] {\tt **.Hgate\_Z\_error\_ratio}
	 \item[] {\tt **.Measurement\_error\_rate // Measurement gate error rate}
	 \item[] {\tt **.Measurement\_X\_error\_ratio}
	 \item[] {\tt **.Measurement\_Y\_error\_ratio}
	 \item[] {\tt **.Measurement\_Z\_error\_ratio}
	 \item[] {\tt **.Xgate\_error\_rate // X gate error rate}
	 \item[] {\tt **.Xgate\_X\_error\_ratio}
	 \item[] {\tt **.Xgate\_Y\_error\_ratio}
	 \item[] {\tt **.Xgate\_Z\_error\_ratio}
	 \item[] {\tt **.Zgate\_error\_rate // Z gate error rate}
	 \item[] {\tt **.Zgate\_X\_error\_ratio}
	 \item[] {\tt **.Zgate\_Y\_error\_ratio}
	 \item[] {\tt **.Zgate\_Z\_error\_ratio}
	 \item[] {\tt **.CNOTgate\_error\_rate // CNOT gate error rate}
	 \item[] {\tt **.CNOTgate\_IZ\_error\_ratio}
	 \item[] {\tt **.CNOTgate\_ZI\_error\_ratio}
	 \item[] {\tt **.CNOTgate\_ZZ\_error\_ratio}
	 \item[] {\tt **.CNOTgate\_IX\_error\_ratio}
	 \item[] {\tt **.CNOTgate\_XI\_error\_ratio}
	 \item[] {\tt **.CNOTgate\_XX\_error\_ratio}
	 \item[] {\tt **.CNOTgate\_IY\_error\_ratio}
	 \item[] {\tt **.CNOTgate\_YI\_error\_ratio}
	 \item[] {\tt **.CNOTgate\_YY\_error\_ratio}
	 \item[] {\tt **.memory\_X\_error\_rate // Memory X error rate per microsec.}
	 \item[] {\tt **.memory\_Y\_error\_rate // Memory Y error rate per microsec.}
	 \item[] {\tt **.memory\_Z\_error\_rate // Memory Z error rate per microsec.}
	 \item[] {\tt **.memory\_energy\_excitation\_rate // Memory excitation error rate}
	 \item[] {\tt **.memory\_energy\_relaxation\_rate // Memory relaxation error rate}
	 \item[] {\tt **.memory\_completely\_mixed\_rate // Memory completely mixed error rate}
	 \item [] {\tt **.channel\_Loss\_error\_rate //Channel photon loss error rate per km.}
	 \item [] {\tt **.channel\_X\_error\_rate //Channel X error rate per km.}
	 \item [] {\tt **.channel\_Z\_error\_rate //Channel Z error rate per km.}
	 \item [] {\tt **.channel\_Y\_error\_rate //Channel Y error rate per km.}
	 \item [] {\tt **.internal\_hom\_photon\_detection\_per\_sec //Internal BSA detection rate. This will determine the photon burst rate over the SenderReceiver link.}
	 \item [] {\tt **.internal\_hom\_darkcount\_probability //dark count probability per photon detector}
	 \item [] {\tt **.hom\_photon\_detection\_per\_sec //BSA detection rate. This will determine the photon burst rate over the MeetInTheMiddle link.}
	 \item [] {\tt **.hom\_darkcount\_probability //dark count probability per photon detector}
	 \item [] {\tt **.num\_measure //Number of qubits to be measured for the bootstrapping}
	 \item [] {\tt **.emission\_success\_probability //Photon emission probability from the qubit to optical fiber. If the emission fails, the error is treated the same way as photon loss in fiber.}
	 \item [] {\tt **.buffers //Number of qubits per QNIC}
	 \item [] {\tt **.tomography\_output\_filename //Output file name}
	 \item [] {\tt **.link\_tomography //A boolean identifying whether to perform link bootstrapping in the simulation. This may be turned off when simulating multi-hop quantum communication with predefined quantum link costs.}
	 \item [] {\tt **.initial\_purification //Integer identifying the number of purifications for the link bootstrapping.}
	 \item [] {\tt **.Purification\_type //Integer (ID) identifying the type of purification for link bootstrapping.}
	 \item[] {\tt **.Initial\_notification\_timing\_buffer // Extra buffer time for starting the entanglement generation (used when RuleSets need to be distributed before the entanglement generation)}
	\end{description}

	See the official OMNeT++ documentation for more details regarding building and running a simulation.

	\subsection{Output file format}

	In the .ini file, you have to specify the output file name for storing simulation data of the quantum link bootstrapping experiment:
	{\tt **.tomography\_output\_filename = "Output\_File\_Name"}.
	Once the simulation is done, it outputs two files: one as {\tt Output\_File\_Name\_dm} and another as {\tt Output\_File\_Name}.
	Those output files will be created in the {\tt network} directory.
	The {\tt Output\_File\_Name\_dm} is a raw ASCII text file with the reconstructed density matrix between all connected nodes.
	An example output of the {\tt Output\_File\_Name\_dm}  file is:

	\begin{description}
	  \item[] {\tt EndNode1[0]<--->Repeater1[0] \\
	REAL \\
	0.245333  0.00546737  0.00436971   0.239046\\
	0.00546737   0.250667   0.260954  0.0115033\\
	0.00436971   0.260954   0.253333  0.0104056\\
	0.239046  0.0115033  0.0104056   0.250667\\
	IMAGINARY\\
	0  0.00795211 0.00799364  -0.00275699\\
	-0.00795211 0 0.0163757 -0.00541632\\
	-0.00799364 -0.0163757  0  0.00106851\\
	 0.00275699 0.00541632 -0.00106851  0
	 }
	\end{description}

	The first line specifies which link the reconstructed density matrix corresponds to.
	In this case, the density matrix is reconstructed using resources between EndNode1[0] and Repeater1[0] (which are, of course, neighbors).
	The real part and the imaginary part of the density matrix is outputted separately as shown.

	The {\tt Output\_File\_Name}  file contains the important simulation data of related to the quantum link bootstrapping.
	An example output of the {\tt Output\_File\_Name}  file is:
	\begin{description}
	\item[] {\tt EndNode1[0]<-->QuantumChannel\{cost = 602856; distance = 5km; fidelity = 0.487046; bellpair\_per\_sec = 699.271; tomography\_time = 0.010420677295; tomography\_measurements = 7000; actualmeas = 7000; GOD\_clean\_pair\_total = 3525; GOD\_X\_pair\_total = 3475; GOD\_Y\_pair\_total = 0; GOD\_Z\_pair\_total = 0;\}<-->Repeater1[0]; F = 0.487046; X = 0.512954; Z = 0.00895361; Y = -0.00895361}
	\end{description}

	The line {\tt EndNode1[0]<-->QuantumChannel\{***\}<-->Repeater1[0];} is directly readable by OMNeT++ as a topology (node connectivity) definition in the .ned file.
	The {\tt cost} is the cost of the quantum channel. This is currently not in use, but needs to be defined properly later for routing.
	The {\tt distance} is the length of the quantum channel.
	The {\tt fidelity} is the reconstructed fidelity, and the {\tt bellpair\_per\_sec} is its corresponding throughput calculated as measured Bell pairs per second.
	The {\tt tomography time} shows how much simulated time it took for completing the link bootstrapping from the start of the simulation.
	The {\tt tomography\_measurements} is the expected number of measured resources, where {\tt actualmeas} was used for debugging to identify how many of them were actually measured.
	The {\tt GOD\_clean\_pair\_total} is the total amount of measured resources with no error.
	The {\tt GOD\_X\_pair\_total}, {\tt GOD\_Y\_pair\_total}, {\tt GOD\_Z\_pair\_total} are the total amount of measured resources with a Pauli error, X, Y and Z accordingly.
	Those {\tt GOD parameters} are not accessible in the real world, but is there for practical reasons (e.g. for evaluation).
	The {\tt F},{\tt X},{\tt Y},{\tt Z} correspond to the reconstructed fidelity (which is actually redundant because it is technically the same as {\tt fidelity}), and the reconstructed Pauli error rates.
	As you may have noticed by now, this output needs to be cleaned up a little bit more in the future.

	\section{How to extend the simulator}

	\subsection{Adding a new Action and a new Clause}
	The Action already has an abstract class defined as {\tt class Action}, which serves as an interface.
	This class provides a common function named {\tt  virtual cPacket* run( cModule *re ) = 0;}.
	As shown, the function may return a cPacket, which is a classical packet (e.g. for sharing measurement results) that will be passed to the RuleEngine, and takes the cModule RuleEngine as an argument because it may require some of the functions belonging to the RuleEngine (e.g. freeing resources).
	If you wish to add a new Action to the simulation, you need to write a constructor for the new Action (which extends the Action class) in the {\tt Action.h} file.
	The behavior of the new Action must be written inside the overriding {\tt run()} function in the {\tt Action.c} file.
	The Action currently only supports operations requiring a single list of allocated resources, which is physically located at a single QNIC.
	Therefore, when you need to simulate entanglement swapping, you have to rewrite and extend the code so that it supports two lists of allocated resources for each local QNIC.

	The Clause works similarly to the Action.
	The interface is provided as {\tt class Clause} in the {\tt Clause.h} file, with common functions  {\tt virtual bool check( std::map< int, stationaryQubit* > ) const = 0;} and {\tt virtual  bool checkTerminate( std::map< int, stationaryQubit* > ) const = 0;}.
	The function {\tt check( )} is used for checking whether the clause satisfies to invoke the corresponding Action, and the {\tt checkTerminate( std::map< int, stationaryQubit* > )} checks a when to terminate and delete the whole RuleSet.
	Those functions only return a boolean for identifying whether the clause has satisfied or not.
	The argument, {\tt std::map< int, stationaryQubit* >}, is a list of resources allocated for the Rule.

	\subsection{Adding a new quantum gate}
	Quantum gate operations in the simulation track Pauli error propagations, instead of doing the actual matrix calculation (except for measurement gates used for tomography).
	A gate may also be noisy, where the  error is stochastically applied to the qubit based on a random selection with a given probability distribution.
	All gate operations are defined inside the {\tt stationaryQubit.h} and {\tt stationaryQubit.cc}.
	If necessary, you can also add gate operations that actually do the math using vectors and matrices.
	In that case, you need to correctly update and manage the distributed state.
	Those function should look similar to the measurement function {\tt measurement\_outcome stationaryQubit::measure\_density\_independent()}.

	\subsection{How to add a new protocol (classical packet)}

	The packet must be defined in the {\tt classical\_messages.msg} file as: {\tt packet ExampleProtocol extends header\{ *** \}}.
	The header holds three basic parameters, which are {\tt int srcAddr; //Source node's address}, {\tt  int destAddr; //Destination node's address} and {\tt int hopCount; // Hop count for time-to-live, which is currently not implemented.}
	OMNeT++, by default, distinguishes protocols by trying to cast the received packet as {\tt if(dynamic\_cast<ExampleProtocol *>(msg) != nullptr)}, where {\tt msg} is the received message/packet.
	In the current implementation, the {\tt Router.cc} must recognize the protocol because every classical packet from another node goes through this module.

	\section{Current state of the simulator}

	The simulator meets the primary goals establishing for this project.
	It is also a platform for future experiments.
	The list below should help those who wish to extend the simulator.
	The simulator has been developed while prototyping simultaneously.
	In reading the code, you may find some parameters in modules have been used for debugging purposes, or are already unused.

	\begin{itemize}
	  \item The current simulator is capable setting a dynamic size of memory buffer per QNIC.
	  Each memory will transmit one photon, unless it is reinitialized after some operation.
	  \item Entangled qubits have a pointer pointing to each other.
	  \item The RuleSet is programmable, but Actions other than purifications (Ss-Sp, Ds-Sp, Ss-Dp, Ds-Dp) and measurements (based on error propagation or based on density matrix) are not implemented in the current version -- e.g. entanglement swapping.
	  \item The current Rule assumes a single allocated resource list. We may need two for operations requiring two resources from two different local QNICs.
	  \item The RuleEngine assigns all resources to the first RuleSet stored because my thesis assumes a single RuleSet per link (like circuit switching). You may improve this to simulate multiplexing schemes.
	  \item The real-time controller has not yet been implemented. Operations are directly called by the RuleEngine through a function in the RuleSet.
	  \item The hardware monitor is capable of generating a routing table for both classical and quantum based on Dijkstra's algorithm. This, however, assumes a single simulated network.
	  To simulate quantum Internet, you need to make each node recognize its own network, and implement other routing protocols to connect those (e.g. BGP).
	  \item For a multi-hop quantum communication, the connection manager can break down the path to decide where to perform entanglement swapping. This, however, does not consider the link quality at all.
	  \item The output file of the link bootstrapping simulation does include a \emph{cost} parameter, but the calculation process isn't carefully designed yet.
	  Once this is taken care, the value is used by the hardware monitor to create the routing table.
	\end{itemize}

	\subsection{Limitations}

	First of all, the developed simulator only tracks Pauli error propagations through gates.
	This will simplify the calculation process, but approximate the output fidelity to a pessimistic value because non-Pauli errors will be ignored -- the throughput, therefore, will be an optimistic value.

	The simulation assumes a Bell pair with exactly two qubits entangled across arbitrary nodes.
	You can extend the simulator to make it recognize some other types of entangled states, but making this dynamic is very hard.

	\subsection{Example run}

	An example configuration is shown below.

\begin{figure}[!hbt]
	  \center
	  \includegraphics[keepaspectratio,scale=0.6]{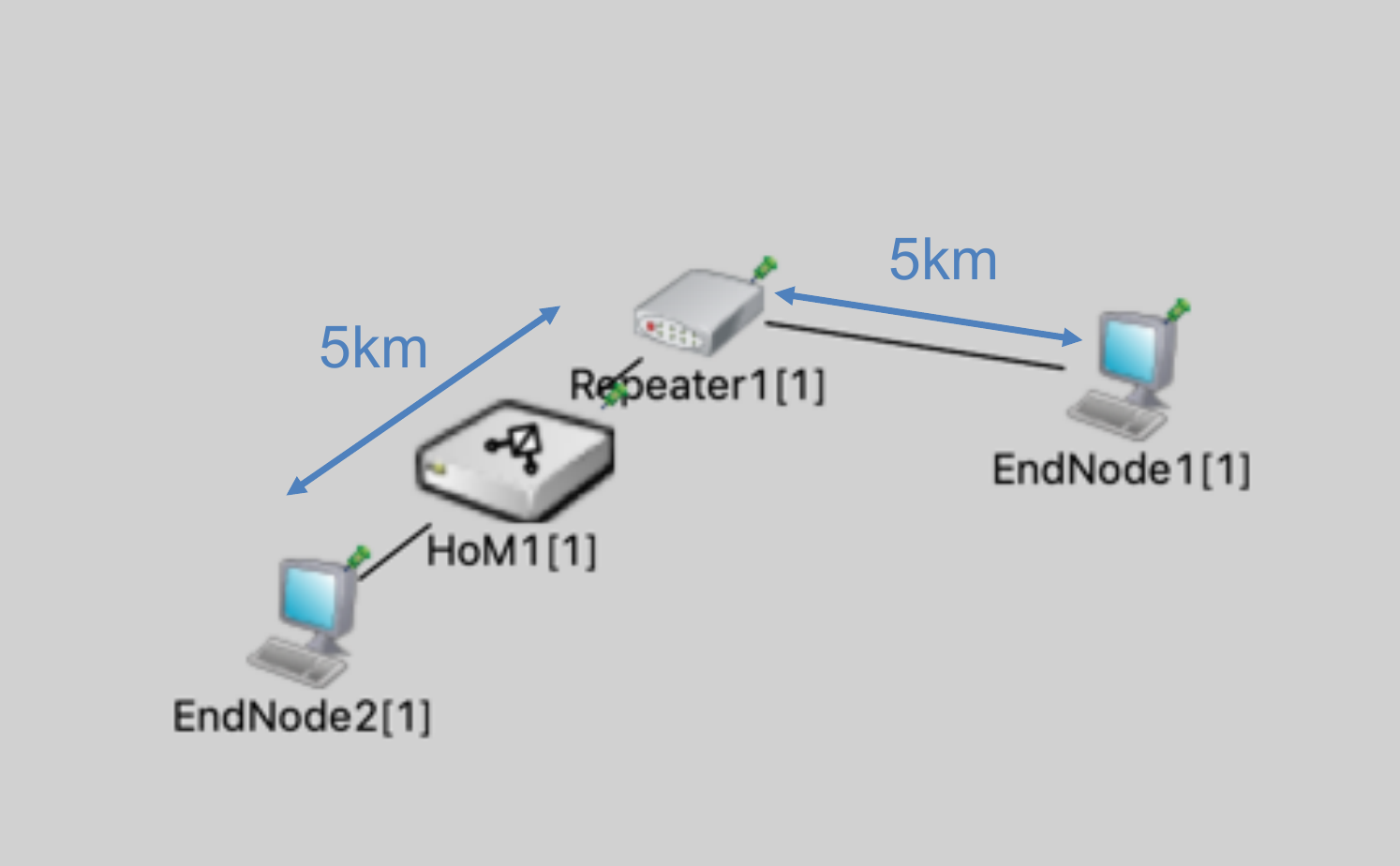}
	  \caption{Network topology defined as Realistic\_Layer2\_Simple\_MIM\_MM\_5km.}
	  \label{networktopo}
	\end{figure}

	{\tt
	[General]\\
	**.Hgate\_error\_rate = 1/2000\\
	**.Hgate\_X\_error\_ratio = 0\\
	**.Hgate\_Y\_error\_ratio = 0\\
	**.Hgate\_Z\_error\_ratio = 0\\
	\\
	**.Measurement\_error\_rate = 1/2000\\
	**.Measurement\_X\_error\_ratio = 1\\
	**.Measurement\_Y\_error\_ratio = 1\\
	**.Measurement\_Z\_error\_ratio = 1\\
	\\
	**.Xgate\_error\_rate = 1/2000\\
	**.Xgate\_X\_error\_ratio = 0\\
	**.Xgate\_Y\_error\_ratio = 0\\
	**.Xgate\_Z\_error\_ratio = 0\\
	\\
	**.Zgate\_error\_rate = 1/2000\\
	**.Zgate\_X\_error\_ratio = 0\\
	**.Zgate\_Y\_error\_ratio = 0\\
	**.Zgate\_Z\_error\_ratio = 0\\
	\\
	\#Error on Target, Error on Controlled\\
	**.CNOTgate\_error\_rate = 1/2000\\
	**.CNOTgate\_IZ\_error\_ratio = 1 \\
	**.CNOTgate\_ZI\_error\_ratio = 1 \\
	**.CNOTgate\_ZZ\_error\_ratio = 1 \\
	**.CNOTgate\_IX\_error\_ratio = 1 \\
	**.CNOTgate\_XI\_error\_ratio = 1 \\
	**.CNOTgate\_XX\_error\_ratio = 1 \\
	**.CNOTgate\_IY\_error\_ratio = 1 \\
	**.CNOTgate\_YI\_error\_ratio = 1 \\
	**.CNOTgate\_YY\_error\_ratio = 1 \\
	\\
	**.memory\_X\_error\_rate = 1.11111111e-7 \#ratio. Not the error rate!!\\
	**.memory\_Y\_error\_rate = 1.11111111e-7 \#ratio. Not the error rate!!\\
	**.memory\_Z\_error\_rate = 1.11111111e-7 \#ratio. Not the error rate!! By default the ratio is 1:1:1\\
	**.memory\_energy\_excitation\_rate = 0.000198\\
	**.memory\_energy\_relaxation\_rate = 0.00000198\\
	**.memory\_completely\_mixed\_rate = 0\\
	\\
	**.Initial\_notification\_timing\_buffer = 10 s \#when to start the BSA timing notification.
	}
	\\ \\ \\
	{\tt
	[Config Example\_run\_Ss-Sp]\\
	network= Realistic\_Layer2\_Simple\_MIM\_MM\_5km\\
	seed-set = \${0}\\
	**.num\_measure = 7000\\
	**.buffers = 100\\
	**.tomography\_output\_filename = "Example\_run\_Ss-Sp\_5km"\\
	**.emission\_success\_probability = 0.46*0.49\\
	\\
	\# Error on optical qubit in a channel\\
	**.channel\_Loss\_error\_rate = 0.04500741397 \# per km\\
	**.channel\_X\_error\_rate = 0.01\\
	**.channel\_Z\_error\_rate = 0.01\\
	**.channel\_Y\_error\_rate = 0.01\\
	\\
	\# Internal HoM in QNIC\\
	**.internal\_hom\_photon\_detection\_per\_sec = 1000000000 \\
	**.internal\_hom\_darkcount\_probability = 10e-8 \#10/s\\
	\\
	\#Stand-alone HoM in the network\\
	**.hom\_photon\_detection\_per\_sec = 1000000000\\
	**.hom\_darkcount\_probability = 10e-8\\
	\\
	**.link\_tomography = true \# true activates bootstrapping.\\
	**.initial\_purification = 1 \# Number of purification to be performed before tomography. Since it is once, only Ss-Sp X purification will be performed.\\
	**.Purification\_type = 3003 \# ID=3003 refers to Ss-Sp. Switches Ss-Sp X purification and Ss-Sp Z purification every attempt.
	}
	\\ \\
	This simulation will run quantum link bootstrapping, based on Ss-Sp with only $n=1$ round.
	The output, therefore, will only purify the X error before performing tomography.
	The network topology used for the simulation is shown in Fig.~\ref{networktopo}.
	This particular simulation took about 3.5 minutes over a server with OS {\tt Ubuntu 16.04.6 LTS} including 24 processors,
	with a maximum total memory size of 491520 MB.
	The \emph{processor 0}'s specification is summarized in Tab.\ref{table_sys}.

	\begin{table}[H]
	  \begin{center}
	    \caption{Major Processor 0 specification.}
	    \label{table_sys}
	    \begin{tabular}{|c|c|}
	      \hline
	      processor	& 0 \\
	      vendor\_id	& GenuineIntel\\
	      cpu family	& 6\\
	      model		& 85\\
	      model name	& Intel(R) Xeon(R) Gold 6136 CPU @ 3.00GHz\\
	      stepping	& 4\\
	      microcode	& 0x2000043\\
	      cpu MHz		& 2999.999\\
	      cache size	& 25344 KB\\
	      physical id	& 0\\
	      siblings	& 1\\
	      core id		& 0\\
	      cpu cores	& 1\\
	      apicid		& 0\\
	      initial apicid	& 0\\
	      fpu		& yes\\
	      fpu\_exception	& yes\\
	      cpuid level	& 22\\
	      wp		& yes\\
	      bugs		& cpu\_meltdown spectre\_v1 spectre\_v2 spec\_store\_bypass l1tf\\
	      bogomips	& 5999.99\\
	      clflush size	& 64\\
	      cache\_alignment	& 64\\
	      address sizes	& 43 bits physical, 48 bits virtual\\
	      \hline
	    \end{tabular}
	  \end{center}
	\end{table}

	The output {\tt Example\_run\_Ss-Sp\_5km} is:
	\begin{description}
	\item[] {\tt EndNode1[0]<-->QuantumChannel\{cost=209924;  distance=5km;  \\fidelity=0.850604;  bellpair\_per\_sec=658.389; \\ tomography\_time=10.63201768213;   tomography\_measurements=7000;  actualmeas=7000;   GOD\_clean\_pair\_total=5996;   GOD\_X\_pair\_total=44;   GOD\_Y\_pair\_total=30;   GOD\_Z\_pair\_total=691; \\ \}<-->Repeater1[0];   F=0.850604;   X=0.0292881;   Z=0.110106;  \\ Y=0.0100021}
	\item[] {\tt EndNode2[0]<-->QuantumChannel\{cost=202156;  distance=2.5km; \\ fidelity=0.854019;  bellpair\_per\_sec=678.232; \\ tomography\_time=10.320951596033;  tomography\_measurements=7000;  actualmeas=7000;   GOD\_clean\_pair\_total=6025;   GOD\_X\_pair\_total=36;   GOD\_Y\_pair\_total=31;   GOD\_Z\_pair\_total=687; \\ \}<-->HoM1[0];   F=0.854019;   X=0.0185928;   Z=0.11239;  \\ Y=0.0149989}
	\item[] {\tt Repeater1[0]<-->QuantumChannel\{cost=209924;  distance=5km;  \\ fidelity=0.850604;  bellpair\_per\_sec=658.389; \\ tomography\_time=10.63201768213;  tomography\_measurements=7000;  actualmeas=7000;   GOD\_clean\_pair\_total=5996;   GOD\_X\_pair\_total=44;   GOD\_Y\_pair\_total=30;   GOD\_Z\_pair\_total=691;\\  \}<-->EndNode1[0];   F=0.850604;   X=0.0292881;   Z=0.110106; \\  Y=0.0100021}
	\item[] {\tt Repeater1[0]<-->QuantumChannel\{cost=202156;  distance=2.5km; \\ fidelity=0.854019;  bellpair\_per\_sec=678.232; \\ tomography\_time=10.320951596033;  tomography\_measurements=7000;  actualmeas=7000;   GOD\_clean\_pair\_total=6025;   GOD\_X\_pair\_total=36;   GOD\_Y\_pair\_total=31;   GOD\_Z\_pair\_total=687;\\  \}<-->HoM1[0];   F=0.854019;   X=0.0185928;   Z=0.11239; \\  Y=0.0149989}
	\end{description}

	The output {\tt Example\_run\_Ss-Sp\_5km\_dm} is:
	\begin{description}
	\item[] {\tt EndNode1[0]<--->Repeater1[0]\\
	REAL\\
	   0.462611  0.00723047   0.0170436    0.370249\\
	 0.00723047   0.0202788  0.00964299 -0.00152404\\
	  0.0170436  0.00964299   0.0190114  0.00154146\\
	   0.370249 -0.00152404  0.00154146    0.498099\\
	IMAGINARY\\
	          0  -0.0146235   0.0182761 -0.00885229\\
	  0.0146235           0   0.0129395 -0.00312463\\
	 -0.0182761  -0.0129395           0  0.00553257\\
	 0.00885229  0.00312463 -0.00553257           0}
	 \item[] { \tt EndNode2[0]<--->HoM1[0]\\
	REAL\\
	  0.488372  0.0162139 0.00510361   0.370815\\
	 0.0162139  0.0142119 0.00179692 0.00827218\\
	0.00510361 0.00179692  0.0193798 0.00990076\\
	  0.370815 0.00827218 0.00990076   0.478036\\
	IMAGINARY\\
	           0  0.000422923  3.21287e-05   -0.0176171\\
	-0.000422923            0  -0.00498175  -0.00506991\\
	-3.21287e-05   0.00498175            0  -0.00294181\\
	   0.0176171   0.00506991   0.00294181            0}
	\item[]{\tt Repeater1[0]<--->EndNode1[0]\\
	REAL\\
	   0.462611   0.0170436  0.00723047    0.370249\\
	  0.0170436   0.0190114  0.00964299  0.00154146\\
	 0.00723047  0.00964299   0.0202788 -0.00152404\\
	   0.370249  0.00154146 -0.00152404    0.498099\\
	IMAGINARY\\
	          0   0.0182761  -0.0146235 -0.00885229\\
	 -0.0182761           0  -0.0129395  0.00553257\\
	  0.0146235   0.0129395           0 -0.00312463\\
	 0.00885229 -0.00553257  0.00312463           0}
	 \item[]{\tt Repeater1[0]<--->HoM1[0]\\
	REAL\\
	  0.488372 0.00510361  0.0162139   0.370815\\
	0.00510361  0.0193798 0.00179692 0.00990076\\
	 0.0162139 0.00179692  0.0142119 0.00827218\\
	  0.370815 0.00990076 0.00827218   0.478036\\
	IMAGINARY\\
	           0  3.21287e-05  0.000422923   -0.0176171\\
	-3.21287e-05            0   0.00498175  -0.00294181\\
	-0.000422923  -0.00498175            0  -0.00506991\\
	   0.0176171   0.00294181   0.00506991            0\\}
	\end{description}

	As shown in the output, the reconstructed fidelity is about $F_r = 0.85$.
	As the simulation only performs X purification, the X error rate is lower than the Z error rate.
	The actual process time for the link bootstrapping in this case is {\tt tomography\_time} (in the {\tt Example\_run\_Ss-Sp\_5km}) - {\tt Initial\_notification\_timing\_buffer} (in the configuration).
	To check the impact of purification, you can run the same simulation with a different {\tt initial\_purification} value.

\end{document}